\documentclass[twocolumn,notitlepage,aps,prb,10pt,superscriptaddress,floatfix,letterpaper,reprint
]{revtex4-1}	

\usepackage{hyperref}
\usepackage[usenames,dvipsnames]{color}
\usepackage{amsmath}
\usepackage[english]{babel}
\usepackage{amssymb}
\usepackage{graphicx}
\usepackage{epstopdf}
\usepackage[normalem]{ulem}
\usepackage{subfigure}
\usepackage{braket}
\usepackage{bbold}
\usepackage{placeins}
\usepackage[para]{threeparttable}
\usepackage{lipsum}
\usepackage{multirow}
\usepackage{bm}
\usepackage{siunitx}

\usepackage{todonotes}

\definecolor{linkcolor}{HTML}{399B03}
\definecolor{urlcolor}{HTML}{399B03}
\hypersetup{pdfstartview=FitH, linkcolor=linkcolor,urlcolor=urlcolor,colorlinks=true}
\hypersetup{
    unicode=false,          
    pdftoolbar=true,        
    pdfmenubar=true,        
    pdffitwindow=false,     
    pdfstartview={FitH},    
    pdfauthor={},         
    colorlinks=true,        
    linkcolor=blue,         
    citecolor=red,          
}

\newcommand{\lati}{\bm{i}}
\newcommand{\latj}{\bm{j}}
\newcommand{\latk}{\bm{k}}
\newcommand{\latl}{\bm{l}}

\newcommand{\tk}{\mathbf{k}}
\newcommand{\tr}{\mathbf{r}}
\newcommand{\tR}{\mathbf{R}}
\newcommand{\tq}{\mathbf{q}}

\begin{document}

\title{Ab-Initio self-energy embedding for the photoemission spectra of NiO and MnO}

\author{Sergei Iskakov}
\affiliation{%
 Department of Physics, University of Michigan, Ann Arbor, Michigan 48109, USA
}%
\author{Chia-Nan Yeh}%
\affiliation{%
 Department of Physics, University of Michigan, Ann Arbor, Michigan 48109, USA
}%
\author{Emanuel Gull}%
\affiliation{%
 Department of Physics, University of Michigan, Ann Arbor, Michigan 48109, USA
}%
\author{Dominika Zgid}%
\affiliation{%
 Department of Chemistry, University of Michigan, Ann Arbor, Michigan 48109, USA
}%
\affiliation{%
 Department of Physics, University of Michigan, Ann Arbor, Michigan 48109, USA
}%

\date{\today}

\begin{abstract}
    The accurate ab-initio simulation of periodic solids with strong correlations is one of the grand challenges of condensed
    matter.
    While mature methods exist for weakly correlated solids, the ab-initio description of strongly correlated systems is an active field of research.
    In this work, we show results for the single particle spectral function of the two correlated $d$-electron solids NiO and MnO
    from self-energy embedding theory.
    Unlike earlier work, the theory does not use any adjustable parameters and is fully ab-initio, while being able to treat
    both the strong correlation and the non-local screening physics of these materials.
    We derive the method, discuss aspects
    of the embedding and choices of physically important orbitals, and compare our results to x-ray and angle-resolved photoemission
    spectroscopy as well as bremsstrahlung-isochromat spectroscopy.
\end{abstract}

\maketitle

\section{Introduction}
The ab-initio simulation of periodic solids with strong correlations is an important problem in condensed matter physics.
While reliable computational methods exist for weakly correlated solids, they tend to be less suitable where the underlying
independent electron approximation fails, such as in systems with d-electrons. Where strong correlations are important,
the condensed matter community has historically resorted to the construction of low-energy effective models, such as single-
or multi-orbital Hubbard models, in order to describe effective low-lying degrees of freedom.

For systems where a treatment of the electronic structure in addition to strong correlation physics is desired, embedding methods
such as a combination of the dynamical mean field theory (DMFT)~\cite{DMFT_infinite_dim_Georges_Kotliar_1992,Hubbard_inf_dim_Vollhardt_Metzner,Georges96}
with density functional theory (DFT) electronic structure codes~\cite{LDAplusDMFT_Anisimov1997, Lichtenstein98,Kotliar06}, led to a combination of both approaches.
These methods are very successful in their region of applicability. However, they suffer fundamentally from the need to determine free parameters,
such as the double counting correction or the values of screened interaction parameters, at the interface between the electronic structure and strong correlation calculations.

Diagrammatic perturbation theory provides an alternative route to standard electronic structure methods such as DFT. Perturbative methods
are free from adjustable parameters, and solutions of the (bare or self-consistent) second-order perturbation theory~\cite{GF2_Alexander16, GF2_Sergei19}
and several variants of Hedin's GW approximation~\cite{G0W0_Pickett84, G0W0_Hybertsen86, GW_Aryasetiawan98, QSGW_Kotani07, scGW_Andrey09}
can be performed for realistic solids. However, due to their perturbative nature, these methods are not able to access the strong correlation
regime. Nevertheless, the diagrammatic language in which these theories are formulated lends itself ideally to embedding methods,
which aim to selectively enhance the solution of a weakly correlated problem with non-perturbative strong-correlation answers in a
small but potentially strongly correlated subset of orbitals. Moreover, the Green's function language in which they are formulated
allows one to calculate experimentally observable quantities such as the (momentum- and energy-resolved) spectral function, making
them ideal candidates for studying condensed matter systems.

A combination of extended DMFT (EDMFT)~\cite{EDMFT_Qimiao_2000,PhysRevB.63.115110,PhysRevLett.77.3391} with a perturbative
method such as GW lead to a formulation of the GW+EDMFT approach~\cite{GWplusEDMFT_Sun02,PhysRevLett.90.086402,PhysRevB.88.235110,
Tomczak_2012,PhysRevB.87.125149,PhysRevB.90.195114,TMO_MIT_Leonov16,PhysRevB.95.245130,haule_lee_h2_2017, Boehnke16,multitier_GW+DMFT_werner_2017,
ComDMFT_2019, chan_zhu_2020_gw_dmft,PhysRevB.93.165106}, where the weakly correlated electrons are treated at the GW level and the strongly correlated electrons
are handled by an accurate non-perturbative approach.

In this paper, we focus on the discussion and performance assessment of another diagrammatic  ab-initio embedding theory -
the self-energy embedding theory (SEET)~\cite{Kananenka15,Zgid17,Lan17}. This theory combines the GW approximation~\cite{Hedin1965}
with the non-perturbative solution of quantum impurity models. SEET was extensively tested on molecular
problems~\cite{Tran_jcp_2015,Tran_jctc_2016,Lan17,Tran_Shee_2017,Tran_useet,PhysRevX.7.031059,PhysRevX.10.011041}
and very simple solids~\cite{doi:10.1021/acs.jctc.8b00927}. However, this paper presents first tests for fully realistic solids.

The two antiferromagnetic compounds NiO and MnO are ideal materials for testing the capabilities of SEET. Correlation effects
in those materials are believed to be strong, and Mott~\cite{Mott_1949} considered NiO as a paradigmatic example of a `Mott' insulator.
The NiO solid has been carefully studied with a wide range of experiments, including angle-integrated and angle-resolved photoemission
and bremsstrahlung-isochromat spectroscopy for NiO~\cite{NiO_expt_Powell70, NiO_expt_Sawatzky84, NiOCoO_expt_Shen90, NiO_expt_Shen91,
NiO_expt_Tjernberg96, NiO_expt_AFMandPM_Tjernberg96, NiO_expt_Jauch04, NiO_expt_Schuler05, NiO_expt_MIT_Gavriliuk12, NiO_expt_MIT_Potapkin16,
MnONiO_expt_Eastman75, PE_review_SHEN95} and MnO~\cite{MnONiO_expt_Eastman75, MnO_expt_Lad88, MnO_expt_Elp91, PE_review_SHEN95,
MnO_PT_Kondo00, MnO_MIT_Patterson04}. The material has also been studied with a wide array of theoretical methods, including the
Hartree-Fock (HF) approximation~\cite{NiOMnO_HF_Towler94}, configuration interactions within the metal-ligand clusters~\cite{NiO_Fujimori84},
density functional theory(DFT)~\cite{MnO_DFT_PT_Fang99}, LDA+U~\cite{LDAplusU_Anisimov97}, different variants of the GW
approximation~\cite{NiO_GW_Aryasetiawan95,GW_Aryasetiawan98, Sergey04, NiO_QSGW_Li05, Rodl09}, the variational cluster approximation
(VCA)~\cite{NiO_VCA_Eder15},  LDA+DMFT~\cite{LDADMFT_Ren06, Kunes07_prb,Kunes07_prl}, and Linearized QSGW+DMFT~\cite{ComDMFT_2019}.

This paper proceeds as follows.  Sec.~\ref{sec:method} introduces the GW approximation and presents the self-energy embedding theory.
Sec.~\ref{sec:compasp} describes the computational details necessary for reproducing our calculations. Sec.~\ref{sec:results} shows
theoretical photoemission results as compared to experiment, and Sec.~\ref{sec:conclusions} presents our conclusions.

\section{Method}\label{sec:method}
We model a solid as an arrangement of atoms in a Bravais lattice with periodicity in all three directions. We employ the Born-Oppenheimer approximation and
choose a basis of single-particle wave functions. In this work we use Bloch waves constructed from Gaussian basis functions as 
\begin{equation}\label{Gaussian}
\phi_{\tk_i,i}(\tr) = \sum_{\tR} \phi^{\tR}_{i}(\tr)e^{i\tk\cdot\tR},
\end{equation}
where $\phi^{\tR}_{i}(\tr)$ is a Gaussian atomic orbital centered in Bravais lattice cell $\tR$. These states are not orthogonal and define the overlap matrix
\begin{equation}
\mathbf{s}_{\lati\latj} = \int_{\Omega} d\tr \phi^{*}_{\tk_i,i}(\tr) \phi_{\tk_j,j}(\tr)\delta_{\tk_i,\tk_j} \label{Eqn:Ovlp}.
\end{equation}
The electronic structure Hamiltonian in second quantization is
\begin{align}
H = \sum_{\lati\latj,\sigma} h^{0}_{\lati\latj} c^{\dagger}_{\lati\sigma} c_{\latj\sigma} + 
\frac{1}{2} \sum_{\substack{\lati\latj\latk\latl\\\sigma\sigma'}} 
v_{\lati\latj\latk\latl} c^{\dagger}_{\lati\sigma} c_{\latk\sigma'}^\dagger c_{\latl\sigma'} c_{\latj\sigma}. \label{Eqn:Hamiltonian}
\end{align}
Where $c_{\lati\sigma}$ ($c_{\lati\sigma}^{\dagger}$) are annihilation (creation) operators corresponding to the single
particle state $\phi_{\tk_i,i}(\tr)$, with spin $\sigma$ and index $\lati (\latj, \latk, \latl)$
denotes the combined orbital-momenta index $\lati = (i, \tk_i)$.
The single-particle operator $h^{0}_{\lati\latj}$ and two-particle operator $v_{\lati\latj\latk\latl}$ are defined respectively as
\begin{subequations}
\begin{align}
h^{0}_{\lati\latj} &= \int_{\Omega} d\tr 
\phi^{*}_{\tk_i,i}(\tr)
\left[-\frac{1}{2}\nabla^{2}_{\tr} - \sum_{\alpha} \frac{Z_\alpha}{r_{\alpha,\tr}}  \right]
\phi_{\tk_j,j}(\tr)\label{Eqn:H0}, \\
v_{\lati\latj\latk\latl} &= \frac{1}{V} \int_{\Omega} d\tr \int_{\mathbb{R}^3} d\tr'
\frac{\phi^{*}_{\tk_i,i}(\tr)\phi_{\tk_j,j}(\tr)\phi^{*}_{\tk_k,k}(\tr')\phi_{\tk_l,l}(\tr')}
{|\tr-\tr'|},
\label{Eqn:V}
\end{align}
\end{subequations}
where $Z_{\alpha}$ is the nuclear charge of atom $\alpha$, $r_{\alpha,\tr}= |\tr-\tr_\alpha|$ is the distance to nucleus $\alpha$ at $r_\alpha$,
$\Omega$ is the volume of the unit cell and $V$ is the volume of the system.

The primary object of interest in this paper is in the single-particle imaginary time Green's function $G^{\mathcal{H},\sigma}_{\lati\latj}(\tau)$
for Hamiltonian $\mathcal{H}$ and indices $\lati$ and $\latj$,
\begin{align}
G^{\mathcal{H},\sigma}_{\lati\latj}(\tau)  = -\frac{1}{\mathcal{Z}} Tr\left[e^{-(\beta-\tau) (\mathcal{H} -\mu N)} c_{\lati,\sigma} e^{-\tau (\mathcal{H} -\mu N)} c^{\dagger}_{\latj,\sigma}\right].
\end{align}
Here $\mathcal{Z} = Tr\left[e^{-\beta(\mathcal{H} -\mu N)}\right]$ is the grand partition function, $\mu$ is the chemical potential, $\beta$ is the inverse
temperature and $N$ is the number of particles in the system. We define the non-interacting Green's function as
$G^{0,\sigma}_{\lati\latj}(\tau) = G^{H^{0},\sigma}_{\lati\latj}(\tau)$, where
$H^{0} = \sum_{\lati\latj,\sigma} h^{0}_{\lati\latj} c^{\dagger}_{\lati\sigma} c_{\latj\sigma}$, and the interacting one as $G^{\sigma}_{\lati\latj}(\tau) = G^{H,\sigma}_{\lati\latj}(\tau)$.
Translation symmetry implies that Green's functions are diagonal in reciprocal space but dense in orbital space and can be defined as
\begin{align}
G^{\tk\sigma}_{ij}(\tau) = G^{\sigma}_{\lati\latj}(\tau),
\end{align}
with $\tk = \tk_i = \tk_j$.

The Matsubara frequency Green's function is defined through the Fourier transform
\begin{align}
G^{\sigma}_{\lati\latj}(\omega_n) = \int_{0}^{\beta} d\tau G^{\sigma}_{\lati\latj}(\tau) e^{i\omega_n\tau},
\end{align}
where $\omega_n = (2n + 1)\frac{\pi}{\beta}$ is the fermionic Matsubara frequency with $n$ integer. The self-energy is defined by the Dyson equation
\begin{align}
\Sigma^{\sigma}_{\lati\latj}(\omega_n) = \left(G^{0,\sigma}_{\lati\latj}(\omega_n)\right)^{-1} - \left(G^{\sigma}_{\lati\latj}(\omega_n)\right)^{-1} \label{Eqn:Dyson}.
\end{align}
Knowledge of the single particle Green's function allows the computation of the spectral function or density of states as
\begin{align}
    G^{\sigma}_{ij}(\tau)= \int d\omega \frac{A^{\sigma}_{ij}(\omega)e^{-\tau\omega}}{1+e^{-\beta\omega}}.\label{eqn:dos_tau}
\end{align}

\subsection{GW approximation}
\begin{figure}[tbh]
\includegraphics[width=\linewidth]{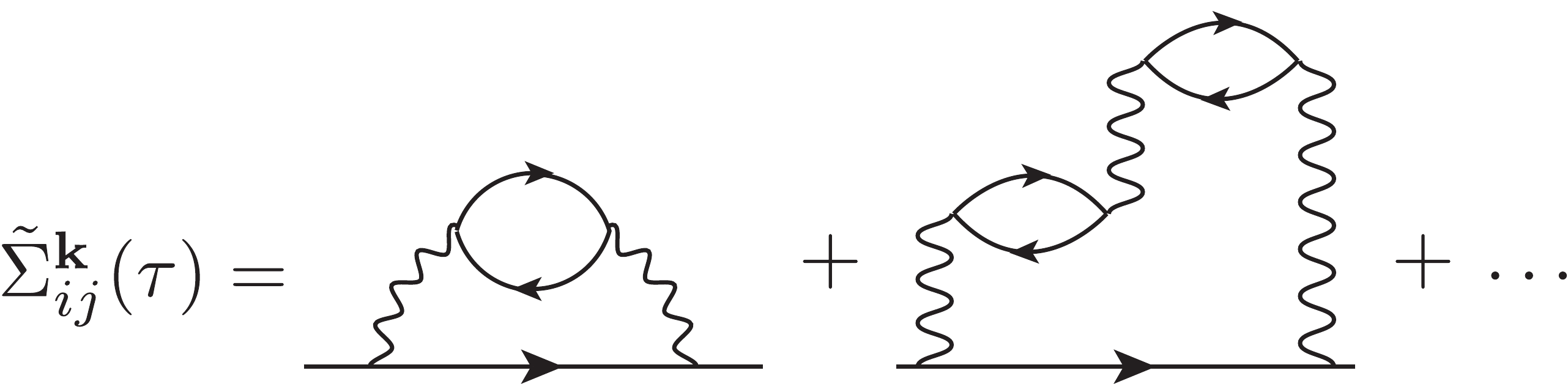}
\caption{Diagrams beyond the Hartree diagram in the self-consistent GW approximation. Wiggly lines denote bare interactions $v$, lines with arrow dressed Green's functions $G$.} \label{fig:bubbles}
\end{figure}

In a first step, we solve the system in the fully self-consistent finite temperature GW approximation introduced by Hedin~\cite{Hedin1965}.
This approximation is thermodynamically consistent and conserving but neglects second-order and higher exchange terms. The GW self-energy is given by
\begin{align}
\Sigma^{\tk,\sigma}_{ij}(\omega_n) &= - \frac{1}{\beta V}\sum_{\substack{m\\\tk',kl}} \Big[G^{\tk',\sigma}_{lk}(\omega_n + \Omega_m)
W^{\tk\tk'\tk'\tk}_{ilkj}(\Omega_m) \nonumber \\ & - \sum_{\sigma'}G^{\tk',\sigma'}_{lk}(\omega_m) v^{\tk\tk\tk'\tk'}_{ijkl}
\Big], \label{eq:sigma_gw}
\end{align} 
where $\Omega_m = \frac{2m\pi}{\beta}$ are the bosonic Matsubara frequencies and the `screened interaction' $W^{\tk\tk'\tk'\tk}_{ilkj}$ is defined as
\begin{align}
W_{\lati_1\lati_2\lati_3\lati_4}(\Omega_n) &= v_{\lati_1\lati_2\lati_3\lati_4} + \tilde{W}_{\lati_1\lati_2\lati_3\lati_4}(\Omega_n)
\nonumber \\ 
\tilde{W}_{\lati_1\lati_2\lati_3\lati_4}(\Omega_n) &= \frac{1}{V}\nonumber \\  \sum_{\lati_5\lati_6\lati_7\lati_8}&v_{\lati_1\lati_2\lati_5\lati_6}
\Pi_{\lati_5\lati_6\lati_7\lati_8}(\Omega_n)W_{\lati_7\lati_8\lati_3\lati_4}(\Omega_n), \label{eqn:secondhedin}
\end{align}
with the approximate polarization operator
\begin{align}
\Pi_{\lati_1\lati_2\lati_3\lati_4}(\Omega_n) &= \frac{1}{\beta}\sum_{m}G^{\sigma}_{\lati_1\lati_3}(\omega_m)G^{\sigma}_{\lati_4\lati_2}(\omega_m + \Omega_n).
\end{align}
Eq.~\ref{eq:sigma_gw} can be written as
\begin{subequations}
\begin{align}
\Sigma^{\tk,\sigma}_{ij}(\omega_n) &= (\Sigma^{\text{GW}}_{\infty})^{\tk,\sigma}_{ij} + (\Sigma^{\text{GW}})^{\tk,\sigma}_{ij}(\omega_n)  \\
(\Sigma^{\text{GW}})^{\tk,\sigma}_{ij}(\omega_n)  &= - \frac{1}{\beta V}\sum_{\substack{m\\\tk',kl}} G^{\tk',\sigma}_{l,k}(\omega_n + \Omega_m) \tilde{W}^{\tk\tk'\tk'\tk}_{ilkj}(\Omega_m),\label{eq:gwcorr_sigma}
\end{align}
\end{subequations}
where $(\Sigma^{\text{GW}}_{\infty})^{\tk,\sigma}_{ij}$ is the Hartree-Fock self-energy. The self-consistent GW correction to the  Hartree-Fock self-energy,
$(\Sigma^{\text{GW}})^{\tk,\sigma}_{ij}(\omega_n)$, contains an infinite series of `bubble' diagrams as shown in Fig. ~\ref{fig:bubbles}.

\begin{figure}[thb]
\includegraphics[width=0.6\linewidth]{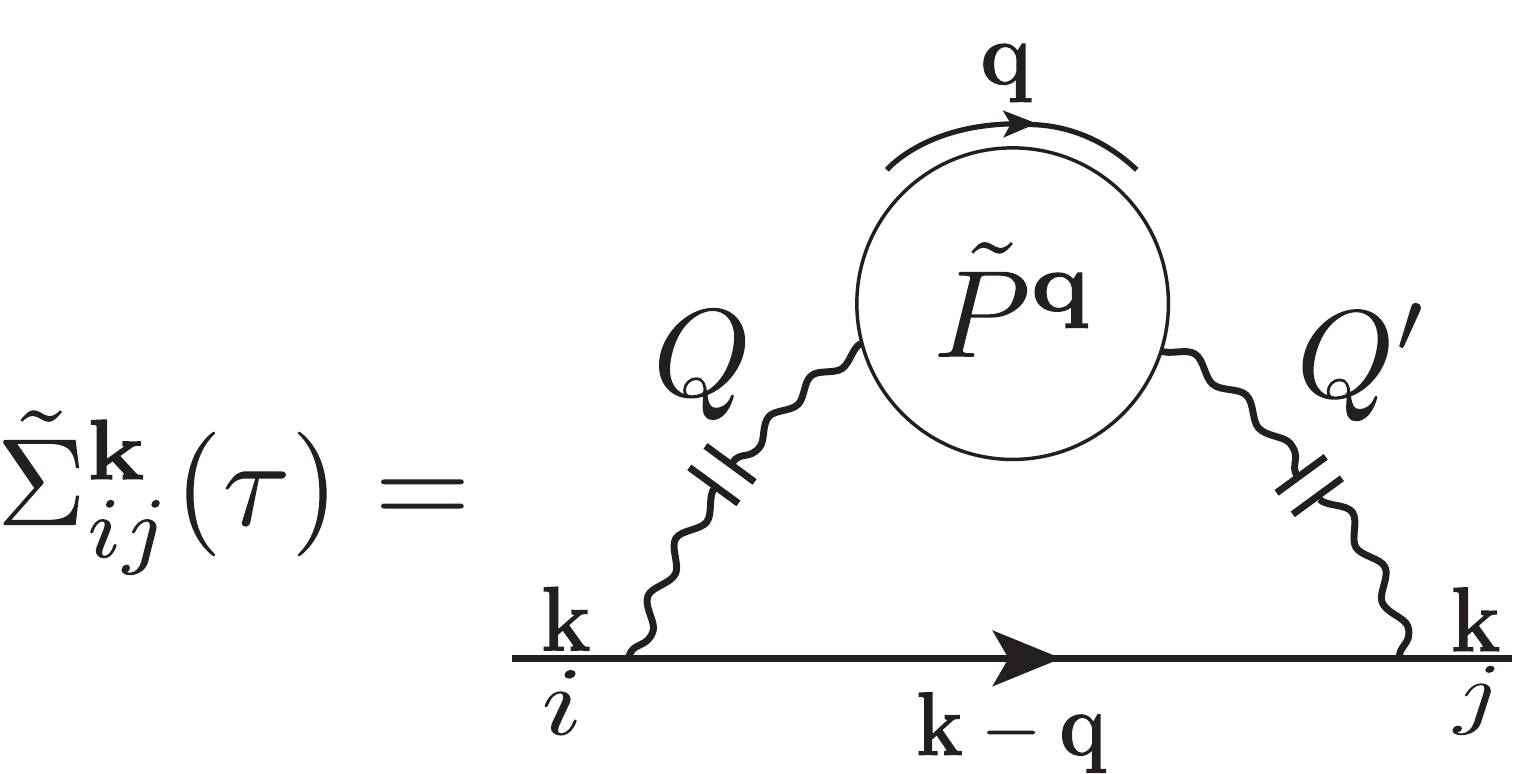}
\caption{Diagrams of Fig.~\ref{fig:bubbles} expressed with the decomposition of Eq.~\ref{eq:gwcorr_sigma}. Interrupted wiggly lines denote the auxiliary basis decomposition indices $Q$ and $Q'$.}\label{fig:self_contract}
\end{figure}

In our GW implementation, we use a Coulomb integral decomposition since 
due to its size, it is not practical to store the full four-index Coulomb integral. 
Several ways to employ its symmetry to
decompose it are known, such as Cholesky decomposition~\cite{Cholesky2008} or the resolution of identity (also known as
density fitting)~\cite{Werner2003,Ren2012,Sun2017}.
Here, we write $v_{\lati_1\lati_2\lati_3\lati_4} = V^{Q}_{\lati_1\lati_2}V^{Q}_{\lati_3\lati_4}$ where
$Q$ is an auxiliary index and $V^{Q}_{\lati_1\lati_2}$ is a three-point integral defined as
\begin{align}\label{eqn:VQ}
V^{Q}_{\lati_1\lati_2}= \sum_{P}\int_{\Omega} d\tr d\tr'
\frac{\phi^{*}_{\lati_1}(\tr)\phi_{\lati_2}(\tr)\chi^{\tq}_{P}(\tr')} {|\tr-\tr'|} \mathbf{J^{-\frac{1}{2}}}^{\tq}_{PQ},
\end{align}
with momentum transfer $\tq = \tk_{i_1}-\tk_{i_2} = \tk_{i_3} - \tk_{i_4}$, $\chi^{\tq}_{P}(\tr')$ an auxiliary basis function and $\mathbf{J}^{-1} = \mathbf{J}^{-\frac{1}{2}}\mathbf{J}^{-\frac{1}{2}}$ the inverse of
\begin{align}
J^{\tq}_{PQ} = \int_{\Omega} d\tr d\tr' \frac{\chi^{\tq*}_{P}(\tr)\chi^{\tq}_{Q}(\tr')}{|\tr-\tr'|}.
\end{align}
This allows to simplify Eq.~\ref{eqn:secondhedin} to
\begin{align}
\tilde{W}_{\lati_1\lati_2\lati_3\lati_4}(\Omega_n) = -\sum_{Q,Q'} V^{Q}_{\lati_1\lati_2}
\tilde{P}_{QQ'}^{\tq}(\Omega_n) V^{Q'}_{\lati_3\lati_4},
 \label{eqn:screened_series}
\end{align}
\\
where the renormalized polarization matrix $\tilde{P}^{\tq} (\Omega_n)$ is
\begin{align}\label{eqn:polar}
\tilde{P}^{\tq} (\Omega_n) &= [ \mathbb{1} - \tilde{P}_{0}^{\mathbf{q}}(\Omega_{n})]^{-1} \tilde{P}_{0}^{\mathbf{q}}(\Omega_{n}),
\end{align}
and
\begin{align}\label{eqn:barepolar}
\tilde{P}_{0,Q,Q'}^{\mathbf{q}}(\Omega_{n}) &= \frac{1}{V}\sum_{\substack{\tk,m,\sigma\\i_1,i_2,i_3,i_4}} \nonumber\\
V^{Q,\tk,\tk+\tq}_{i_1i_2}  &G^{\mathbf{k},\sigma}_{i_1,i_4}(\omega_m) G^{\mathbf{k+q},\sigma}_{i_3,i_2}(\omega_m + \Omega_n)  V^{Q'\tk+\tq,\tk}_{i_3i_4}.
\end{align}
Eq.~\ref{eq:gwcorr_sigma} then simplifies to
\begin{align} \label{eqn:GWselfenergy}
\tilde{\Sigma}^{\tk,\sigma}_{i_1i_2}(\tau)\nonumber &= \\-\frac{1}{V}\sum_{\substack{\tq,i_3,i_4\\Q,Q'}}&
 V^{Q,\tk\tk-\tq}_{i_1,i_4} 
 G^{\tk-\tq,\sigma}_{i_3,i_4}(\tau)\tilde{P}^{\tq}_{Q, Q'}(\tau)
 V^{Q',\tk-\tq,\tk}_{i_3i_2}.
\end{align}
We diagrammatically represent this decomposition in Fig.~\ref{fig:self_contract}.

\subsection{Self-energy embedding method}\label{sec:seet}
GW is an approximate method with well known limitations. To capture correlation effects beyond the GW approximation,
either high-order diagrammatic methods or quantum embedding methods can be used.
Embedding theories that are $\Phi$-derivable and based on diagrammatic expansions such as DMFT, GW+EDMFT, SEET, or
self-energy functional theory aim to systematically improve low-order perturbative results.
These embedding theories satisfy conservation laws and are thermodynamically consistent.

Here, we briefly summarize the SEET equations used by in this paper. In this section we assume that all quantities
are expressed in an orthogonal basis, which we will discuss later.
The real space Green's function and the lattice (k-space) Green's function are related by the Fourier transform
\begin{align}
    G_{ij}^{\tR\tR'}(\omega_n)=\frac{1}{V} \sum_k e^{i \tk \tR}G_{ij}^{\tk}(\omega_n) e^{-i \tk\tR'}.
\end{align}
The GW momentum resolved Green's function of the entire lattice is defined as 
\begin{align}
(G^\text{GW}(\omega_n))^\tk=\big[(\omega_n+\mu)\mathbb{1} -h^{0,\tk}-(\Sigma^\text{GW})^{\tk}\big]^{-1},
\end{align}
where $(\Sigma^\text{GW})^{\tk}=(\Sigma^\text{GW}_\infty)^{\tk}+(\Sigma^\text{GW}(\omega))^{\tk}$. 
As a result of embedding procedure, we define a lattice Green's function in the following way
\begin{align}\label{eq:embedding_lattice_gf}
(G(\omega_n))^\tk=\big[(\omega+\mu)\mathbb{1} -h^{0,\tk}-\Sigma^\tk \big]^{-1},
\end{align}
where 
\begin{align}\label{eq:embedding_lattice_se}
\Sigma^\tk_{ij}=(\Sigma^{GW})_{ij}^{\tk}+\sum_A\left((\Sigma^\text{imp}_A)_{ij}-(\Sigma_A^\text{DC-GW})_{ij}\right)\delta_{(ij)\in A }
\end{align}
with $\Sigma^\text{imp}=\Sigma^\text{imp}_\infty+\Sigma^\text{imp}(\omega_n)$ containing non-perturbatively added self-energy
diagrams and $\Sigma^\text{DC-GW}=\Sigma^\text{DC-GW}_\infty+\Sigma^\text{DC-GW}(\omega_n)$ subtracting those diagrams that
are contained both in the GW solution and the non-perturbative construction.
Subsets $A$ of impurity orbitals with indices $ij \in A$, sometimes also called active orbitals, are defined as groups of the most
physically relevant orbitals for the problem that have correlations that are necessary to be included at a higher than perturbative  level.

To define the self-consistency condition used in SEET we perform Fourier transform of $(G(\omega))^\tk $, $\Sigma^\tk$, and $h^{0,\tk}$
from momentum to real space obtaining $G^{\tR\tR'}$, $\Sigma^{\tR\tR'}$, and $h^{0,\tR\tR'}$. The Fourier transform results in the
following structure of the self-energy matrix in the real space
\begin{align}
    \Sigma_{ij}^{\tR\tR'} &= (\Sigma^{GW})_{ij}^{\tR\tR'} \nonumber \\ &+\sum_A\left((\Sigma^\text{imp}_A)_{ij}-
    (\Sigma_A^\text{DC})_{ij}\right)\delta_{\tR\tR'}\delta_{(ij)\in A},\label{eq:SigmaWithA}
\end{align}
for unit cells away from central cell ($\tR \neq \tR'$) the self-energies are treated at the weakly correlated level $\Sigma^{\tR\tR'}_{ij}=(\Sigma^{GW})_{ij}^{\tR\tR'}$ 
while the local, central cell self-energy for $\tR = \tR'$ includes non-perturbative corrections $(\Sigma^\text{imp}_A)_{ij}$ for every orbital group $A$.

This leads us to a definition of an embedding condition in SEET, where we apply the block-matrix inversions of real space quantities and absorb all terms
containing contributions connecting orbitals in $A$ to the remainder of the system in the matrix $\Delta^A_{ij}(\omega)$ in the following way
\begin{align}\label{eq:seet_embedding_condition}
(G(\omega_n))^{\tR\tR}_{ij\in A}=\big[ (\omega_n +\mu)\mathbb{1}-h^{0,\tR\tR}_{ij\in A}-\Sigma^{\tR\tR}_{ij\in A} - \Delta^A_{ij}(\omega_n)\big]^{-1}.
\end{align}
The hybridization matrix $\Delta^A_{ij}(\omega_n)$ arises since an inverse of a subset is not equal to a subset of an inverse, namely
$(G(\omega_n))^{\tR\tR}_{ij\in A} \ne [(G(\omega_n))^{\tR\tR'})^{-1}]^{\tR\tR}_{ij\in A}= \big[ (\omega_n +\mu)\mathbb{1}-h^{0,\tR\tR}_{ij\in A}-\Sigma^{\tR\tR}_{ij\in A}\big]^{-1}$.
Note that Eq.~\ref{eq:seet_embedding_condition} can further be rewritten as
\begin{align}\label{eq:Gloc}
    [(G(\omega_n))^{\tR\tR}_{ij\in A}]^{-1}
    &=(i\omega_n+\mu)\mathbb{1}-\tilde{h}^{0,\tR\tR}_{ij\in A} +\\
    &- \Sigma^{\text{corr},\tR\tR}_{ij\in A}(\omega_n) 
    -\Sigma^{\text{imp}}_{ij\in A} - \Delta^A_{ij}(\omega_n),\nonumber 
\end{align}
where $\tilde{h}^{0,\tR\tR}_{ij} = h^{0,\tR\tR}_{ij} + (\Sigma^\text{GW})^{\tR\tR}_{\infty,ij} - \Sigma^\text{DC}_{\infty,ij}$ is the renormalized non-interacting Hamiltonian,
and $\Sigma^{\text{corr},\tR\tR}_{ij}(\omega_n) = (\Sigma^\text{GW})^{\tR\tR}_{ij}(\omega_n) - \Sigma^\text{DC}_{ij}(\omega_n)$ is the local correction from the weakly correlated method.

We emphasize that, in SEET, the substantial contribution of $\Sigma^{\text{corr},\tR\tR}_{ij\in A}(\omega_n) $ to the local correlated orbitals
is included explicitly in the real space self-consistency condition in Eq.~\ref{eq:seet_embedding_condition} and is not included as a part of
hybridization as done in the GW+DMFT schemes described in Ref.~\onlinecite{haule_lee_h2_2017,chan_zhu_2020_gw_dmft}.
These contributions stem from GW diagrams that have both external legs $i$ and $j$ in the active space but contain one or
more internal indices on the remaining orbitals. Furthermore, the explicit treatment of $\Sigma^{\text{corr},\tR\tR}_{ij\in A}(\omega_n) $
prevents us from observing non-causality problems with hybridization as described in Ref.~\onlinecite{haule_lee_h2_2017} since
$\Delta^A_{ij}(\omega_n)$ as defined in Eq.~\ref{eq:seet_embedding_condition} is always causal.

We also emphasize that, while the total chemical potential is adjusted to give a fixed number of particles in the unit cell, each impurity subspace $A$ may have any non-integer
occupancy. In addition, the number of particles in each subspace may change substantially during the iterative
procedure as electrons shift from the subspaces to the rest of the system and back, while maintaining the total number of
particles.

To evaluate $\Sigma^\text{imp}_{ij\in A}$, we define the auxiliary propagator
\begin{align}\label{eq:Gimp}
   \mathcal{G}_{A}^{-1}(\omega_n)=\mathcal{G}_{A}^{0,-1}(\omega_n) -\Sigma^\text{imp}_{ij\in A},
\end{align}
where the zeroth order $\mathcal{G}_{A}^{0,-1}(\omega_n)$ is defined as 
\begin{align}\label{eq:Gimp0}
    \mathcal{G}_{A}^{0,-1}(\omega_n)
    =(i\omega_n+\mu)\delta_{ij}-\tilde{h}^{0,\tR\tR}_{ij\in A} - \Delta^A_{ij}(\omega_n).
\end{align}
As realized in the context of DMFT~\cite{Georges96}, a propagator of the form of Eq.~\ref{eq:Gimp} can be obtained by solving the quantum
impurity model with impurity orbitals defined as the active orbitals from a space $A$. In SEET, the two-body interactions in the impurity
remain the bare, unchanged interactions of the original lattice Hamiltonian, since screening is included by the explicit treatment
of $\Sigma^{\text{corr},\tR\tR}_{ij\in A}(\omega_n)$ at the level of the embedding condition and Eq.~\ref{eq:Gloc}.

The fact that the bare interactions do not need to be adjusted in the impurity model is a major difference to formulations of
GW+EDMFT, as implemented {\it e.g.} in Ref.~\cite{Boehnke16}. The GW+EDMFT double counting correction due to the presence of screened $W^\text{imp}(\omega_n)$ removes local
correction to the self-energy from the weakly correlated method, therefore $\Sigma^{corr,\tR\tR}_{ij}(\omega_n) \equiv 0$~\cite{PhysRevLett.90.086402}.
This GW+EDMFT construction containing $W^\text{imp}(\omega_n)$ leads to an impurity model with a different hybridization and noninteracting Hamiltonian and, as the model needs
to take into account correlations outside the active space accordingly, to a rescaling of the interactions. However,
while operationally different, both GW+EDMFT and SEET are consistent, conserving, and contain RPA screening by GW diagrams.

In practice, our method starts from a self-consistent finite temperature GW solution of the lattice problem. It then proceeds by solving all independent
impurity problems for the different disjoint subspaces $A$ independently. The non-perturbative solution of $\Sigma^\text{imp}_{ij}$
is used to update the lattice self-energy and the Green's function from Eq.~\ref{eq:embedding_lattice_se} and ~\ref{eq:embedding_lattice_gf},
followed by a new calculation of the real space Green's function and hybridization (Eq.~\ref{eq:seet_embedding_condition}) and a subsequent solution of the impurity model. In principle, after obtaining the self-consistent solution of
Eq.~\ref{eq:seet_embedding_condition}, the GW solution would need to be iterated again. This has not been done in this work.

\section{Computational aspects}\label{sec:compasp}
\subsection{Basis and lattice structure}
We study the electronic properties of antiferromagnetic fcc NiO and MnO with lattice constant $a=4.1705$\si{\angstrom}~\cite{PhysRevB.3.1039}
and $4.4450$\si{\angstrom}~\cite{MnO_a} at temperature $T\sim 451$ K($\beta = 700$ Ha$^{-1}$). In order to capture the type-II anti-ferromagnetic
ordering we double the unit cell along the [$111$] direction. The resulting unit cell is rhombohedral with two transition metal
atoms and two oxygen atoms. 
Any small rhombohedral distortion below the Ne\'{e}l temperature is neglected.
For both systems we use the \emph{gth-dzvp-molopt-sr} basis~\cite{GTHBasis} with \emph{gth-pbe} pseudopotential~\cite{GTHPseudo}. The
\emph{def2-svp-ri} basis is chosen as the auxiliary basis for the Coulomb integral decomposition~\cite{RI_auxbasis}.
The finite-size errors of the GW exchange diagram are corrected by the Ewald probe-charge
approach~\cite{EwaldProbeCharge, CoulombSingular}.
The Coulomb integrals (Eq.~\ref{eqn:VQ}) and non-interacting matrix elements (Eq.~\ref{Eqn:H0}) are prepared by \texttt{PySCF}~\cite{PySCF}.

The use of a finite basis of Gaussian orbitals introduces an error which is difficult to assess independently.
We therefore compared results of simple DFT calculations of our systems in this basis to those obtained in a
plane wave code~\cite{Kresse1999} and found satisfactory agreement.

\subsection{Imaginary-time/Matsubara frequency grid}
All dynamical functions, such  as the Green's function, polarization, or self-energy, are computed  in an imaginary time formalism.
We use the compact intermediate representation (IR)~\cite{PhysRevB.96.035147} with sparse frequency sampling~\cite{PhysRevB.101.035144} for their storage and manipulation.
The IR has one dimensionless  parameter $\Lambda$ that should be chosen larger than $\beta \omega_\text{max}$, where $\omega_\text{max}$
is the bandwidth of the system (difference between highest and lowest single particle energy).
In this work we use $\Lambda = 10000$ and generate the IR basis functions using the \texttt{irbasis}~\cite{CHIKANO2019181} open-source software package.
Other representations such as
Legendre~\cite{Boehnke2011,dong2020legendrespectral} or Chebyshev polynomials~\cite{PhysRevB.98.075127} and other sparse grids~\cite{kaltak2019minimax}
could be used instead.

\subsection{Orthogonalization}

We solve the GW approximation in the basis of atomic orbitals. In this basis, obtaining analytically continued results is
difficult as the spectral functions of Eq.~\ref{eqn:dos_tau} are not strictly positive, nor normalized, and straightforward application of the
Maximum Entropy Method~\cite{Jarrell1996} is not possible. In addition, most impurity solvers (including the exact
diagonalization solver used in this work~\cite{ISKAKOV2018128}) require orthogonal orbitals.
Finally, to perform the SEET embedding procedure the Green's functions have the be in an orthogonal basis.
It is therefore convenient to orthogonalize the basis and express the GW Green's function in it before performing further analysis.

In this paper we use two types of orbital orthogonalization, $G^\text{orth} = X G X^{*}$, which differ in the transformation matrix $X$ employed.
Symmetrical orbital orthogonalization~\cite{LOWDIN1970185} uses $X = S^{\frac{1}{2}}$, with $\mathbf{s} = S^{\frac{1}{2}} S^{\frac{1}{2}}$,
and $\mathbf{s}$ defined in the Eq.~\ref{Eqn:Ovlp}.
In the canonical orthogonalization~\cite{LOWDIN1970185} the transformation matrix is $X = (V_S s^{\frac{1}{2}})^{-1}$, where  $V_S$ is the matrix constructed
from the eigenvectors of the overlap matrix $\mathbf{s}$ and $s^{\frac{1}{2}}$ is the diagonal matrix constructed from the square-roots of the corresponding eigenvalues of the overlap matrix~\cite{LOWDIN1970185}.

\subsection{Analytical continuation}
The $\tk$-space spectral function measured in ARPES,
\begin{align}
    A^{\tk}(\omega) = \sum_{j} A_{jj}^{\tk}(\omega)
\end{align}
is the trace of the orbitally resolved spectral functions $A_{jj}^{\tk}(\omega)$ which are determined by the Green's
functions $G^{\tk}_{ij}(\tau)$ according to Eq.~\ref{eqn:dos_tau}.
In the orthogonal basis $A_{jj}^{\tk}(\omega)$ is normalized to one and strictly positive. It can therefore be obtained from a
maximum entropy continuation~\cite{Jarrell1996}. 
We use the open-source  \texttt{ALPS}~\cite{Gaenko2017} \texttt{Maxent} package~\cite{Levy2017} with a truncated
continuation kernel, with the Green's functions defined on the grid points of the IR basis~\cite{PhysRevB.101.035144}.
We have verified for select data points that our results are consistent with the Pad\'{e} continued fraction method.
Alternative methods for continuation exist, including the stochastic optimization method~\cite{Mishchenko2000} and the Sparse Modelling~\cite{Otsuki2017} approach.
In addition, continuations of derived quantities, such as the cumulant~\cite{Stanescu2006} or the self-energy~\cite{Wang2009}, are possible.
We have not explored these methods.

In order to obtain the local spectral function we first perform the summation over momenta and then continue the resulting
orbitally-resolved local Green's function $G^\text{loc}_{ij}(\tau) = \frac{1}{V}\sum G^{\tk}_{ij}(\tau) $ as
\begin{align}
G^\text{loc}_{ii}(\tau)= \int d\omega \frac{A^\text{loc}_{ii}(\omega)e^{-\tau\omega}}{1+e^{-\beta\omega}}.\label{eqn:ldos_tau}
\end{align}
While continuation and linear transforms, such as basis change and transforms to real space, commute in principle,
in practice analytically continued data will depend on the order of these operations due to the ill conditioned nature of the analytical continuation kernel.
The total local spectral function is defined as
\begin{align}
A^\text{loc}(\omega) = \sum_{i} A^\text{loc}_{ii}(\omega).\label{eq:ldos}
\end{align}

\subsection{Attribution of the orbital character}

In order to gain additional understanding of the spectral function, it is useful to ascribe orbital character to
analytically continued function. The basis transformation to the orthogonal orbitals allows such an identification by writing
the spectral function of an orthogonalized orbital as a sum of contributions from various atomic orbitals.

We find that the symmetrical orthogonal basis provides an
almost unique correspondence between orthogonal and atomic states.
Basis functions in the canonical orthogonal basis typically mix several atomic states, such that the
attribution to a single atomic state is more difficult. However, we find that in some
cases orbitals of similar type, such as Ni $t_{2g}$ states, are grouped together.

While the attribution of the orbital character is quite straightforward in reciprocal space,
where each k-point can be analyzed independently, it may be
problematic in real space, since the basis transformation will mix
 Gaussian basis functions centered in different unit cells. However, we found that in the symmetrical orthogonal basis
the configuration (contribution from different atomic orbitals) of each orthogonal orbital remains the same for different k-points.
In the local unit cell, each orthogonal orbital can therefore be uniquely traced back to its corresponding atomic orbital.

\subsection{Solution of the impurity model}
SEET is based on the embedding of a non-perturbative impurity model into a self-consistently adjusted hybridization with the environment.
Solving impurity models is a computationally difficult problem and requires a quantum impurity solver such as QMC~\cite{Gull2011},
NRG~\cite{Bulla2008}, Exact Diagonalization (ED)~\cite{doi:10.1063/1.4823192}, CI~\cite{Zgid_2011,Zgid2012}, or Coupled Clusters~\cite{Shee2019,Zhu2019}.

SEET requires the solution of impurity problems with general off-diagonal interactions and hybridizations at potentially
strong interaction. However, the ability to treat multiple active spaces keeps the size of the impurities to be treated
relatively moderate. We found ED to be an ideal impurity solver for SEET problems with $2-5$ orbitals.

ED requires discretization of the continuous hybridization function $\Delta(\omega_n)$ in Eq.~\ref{eq:Gimp0}
and its approximation by a finite, typically small, number of discrete bath sites.

In the symmetric orthogonal basis, off-diagonal elements of the hybridization function are in our experience 1-2 orders
of magnitude smaller than diagonal elements. This allows us to neglect them entirely and fit $\Delta^{\sigma}_{ii}(\omega_n)$ by
minimizing the fit residue
\begin{align}
\chi^2_{\sigma i} = \sum_n f(n)\lVert\Delta^{\sigma}_{ii}(\omega_n) - \sum_{b=1}^{N_b}
\frac{\mathcal{V}^{\sigma}_{ib} \mathcal{V}^{\sigma*}_{ib}}{i\omega_n - \epsilon^{\sigma}_b}\rVert,
\label{eqn:delta_ed}
\end{align}
with weight function $f(n)$ chosen to suppress high frequency contributions to $\Delta^{\sigma}_{ii}(\omega_n)$ (we usually choose $f(n)=1/\omega_n$).
Using a bound-constrained nonlinear least square method~\cite{Voglis_arectangular,Branch1999} we enforce the constraint
that $\mathcal{V}_{ib}$  be positive and $\epsilon_b$ in the vicinity of the Fermi energy.
For two-orbital problems we use 5 bath sites per orbital; for three orbitals we use 3, and for four orbitals we use 3. We solve impurity problems using the open-source ED impurity solver of Ref.~\cite{ISKAKOV2018128}.

\begin{figure}[tbh]
\includegraphics[width=0.47\textwidth]{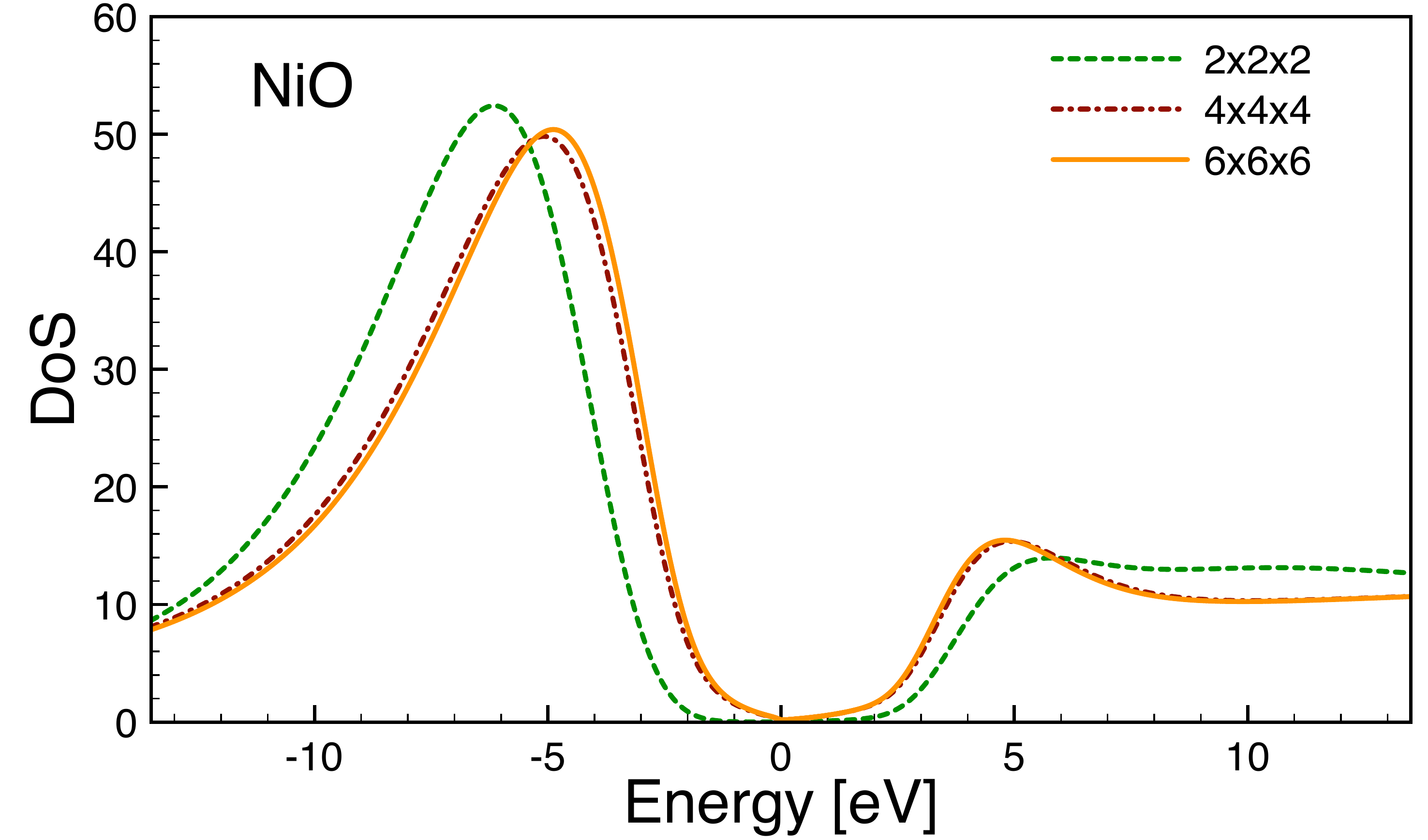}
\includegraphics[width=0.47\textwidth]{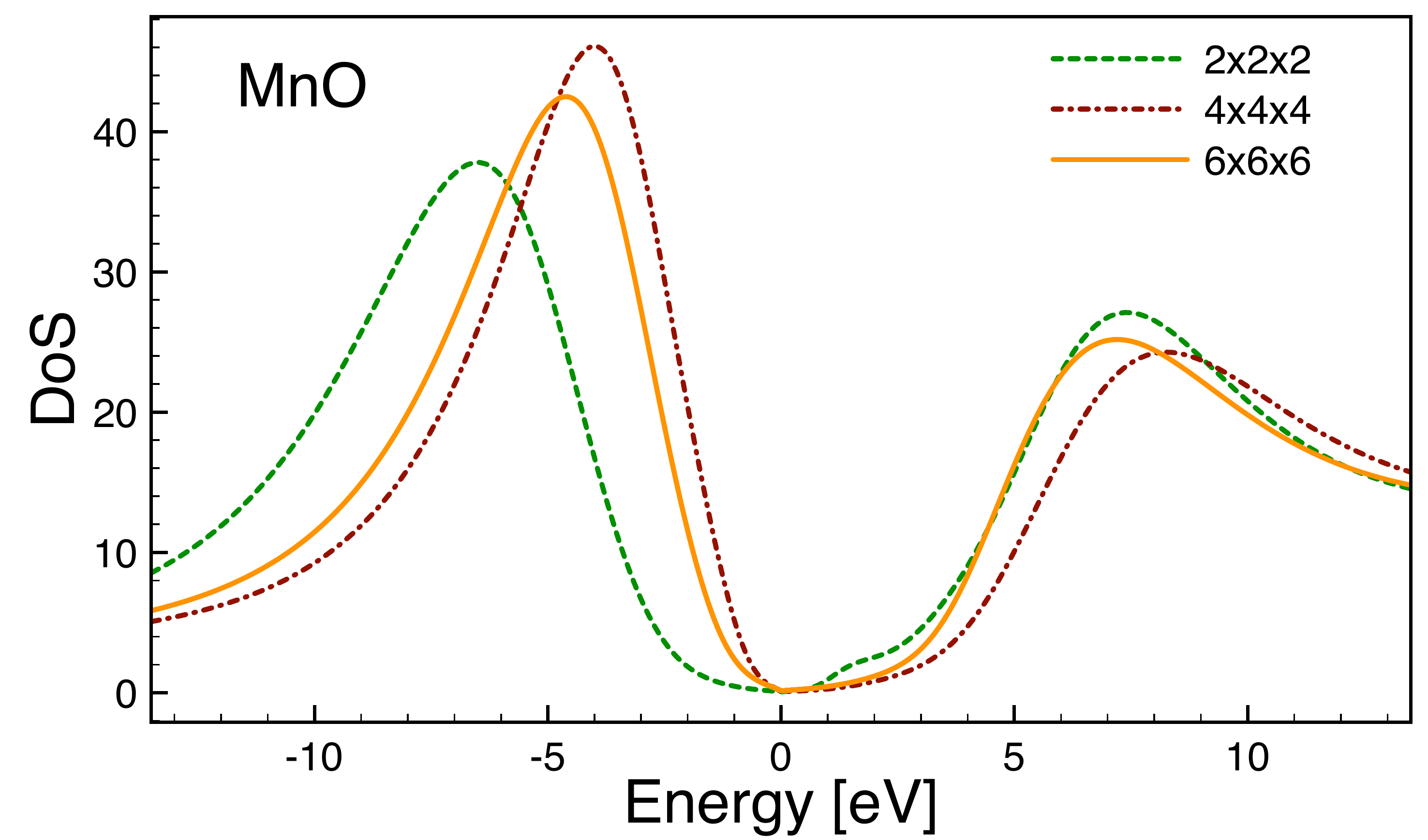}
\caption{Local total density of states of NiO (top panel) and MnO (bottom panel), for systems of size $2$~$\times$~$2$~$\times$~$2$ (dashed green), $4$~$\times$~$4$~$\times$~$4$ (dash-dotted dark red)
    and $6$~$\times$~$6$~$\times$~$6$ (orange), obtained with self-consistent GW.}
\label{Fig:FiniteSize}
\end{figure}

\section{Results}\label{sec:results}

\subsection{Finite size effects in NiO and MnO}

Fig.~\ref{Fig:FiniteSize} shows the self-consistent GW approximation to the local spectral function $A^\text{loc}(\omega)$,
as defined in Eq.~\ref{eq:ldos}, obtained for NiO (top panel) and MnO (bottom panel).
We show curves for three different momentum discretizations; $2$~$\times$~$2$~$\times$~$2$, $4$~$\times$~$4$~$\times$~$4$ and
$6$~$\times$~$6$~$\times$~$6$, to examine finite size effects.
These appear to be substantial between $2$~$\times$~$2$~$\times$~$2$ and
$4$~$\times$~$4$~$\times$~$4$ lattices. Increasing lattice size further to $6$~$\times$~$6$~$\times$~$6$ shows the saturation of the local density of states
In particular, in both NiO and MnO the size of the gap shrinks substantially ($\sim$~2~eV)  as the system size is enlarged from $2$~$\times$~$2$~$\times$~$2$ to $4$~$\times$~$4$~$\times$~$4$,
with an additional correction of $0.5$~eV as the system size is enlarged to $6$~$\times$~$6$~$\times$~$6$.
An extrapolation in the inverse linear size to the infinite system size limit suggests that an additional reduction of the gap size by
$0.5$~$-$~$0.7$~eV is present when comparing the  $6$~$\times$~$6$~$\times$~$6$ lattice to the thermodynamic limit.
Consequently, when analyzing all our results from the embedding procedure presented in the subsequent sections one should be aware
that they will be affected by the presence of the finite size effects and the resulting gaps should be ``rescaled''
by an additional $0.5$~$-$~$0.7$~eV. Additional basis truncation effects due to the incompleteness of the basis set are possible in addition. These effects have not been
assessed in this work due to the prohibitive computational cost and linear dependency issues in Gaussian basis sets.

Spectral function from the self-consistent GW method do not consist of sharp peaks but rather exhibit smooth, broad features.
This smoothness is a result of the analytical continuation procedure which leads to a significant broadening of features, especially at energies far from the Fermi level.

The self-consistent GW response function in Fig.~\ref{Fig:FiniteSize} corresponds to photoemission (occupied part) and inverse photoemission (unoccupied part).
Note that this theoretical assignment is done while neglecting the effect of element- and energy-dependent photoemission cross-sections present when collecting
experimental data. (In angle-resolved photoemission spectroscopy (ARPES), the registered spectrum is equal to a sum of orbital photocurrents multiplied by cross-section matrix elements.)

\begin{figure}[tbh]
\includegraphics[width=0.47\textwidth]{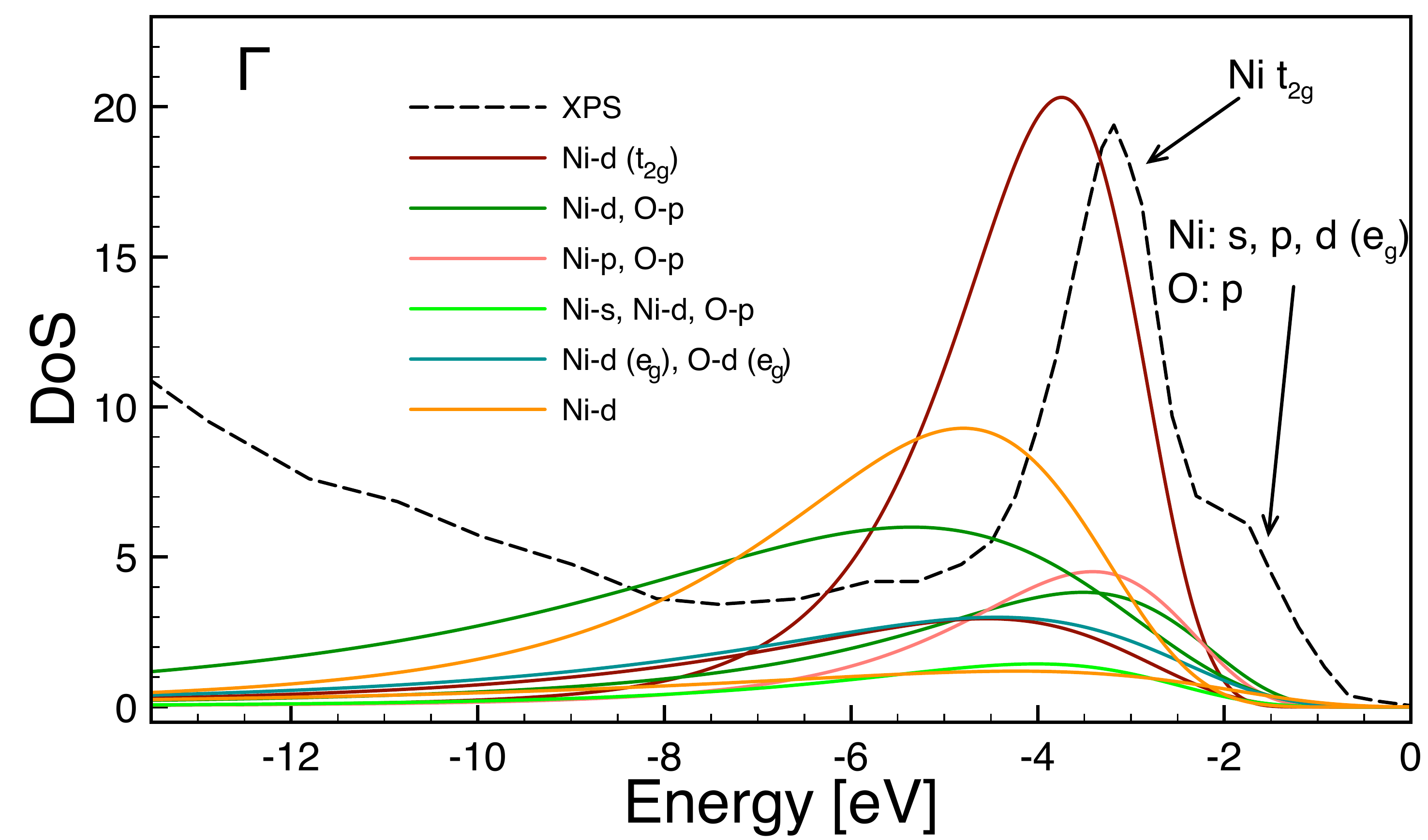}
\caption{Orbitally resolved $\tk$-space spectral function obtained with GW in the canonical orthogonal basis at the $\Gamma$ point.
Dashed black line: experimental ARPES data near $\Gamma$, Ref.~\cite{NiOCoO_expt_Shen90}.}
\label{Fig:GammaVsArpes}
\end{figure}

\begin{figure}[tbh]
\includegraphics[width=0.47\textwidth]{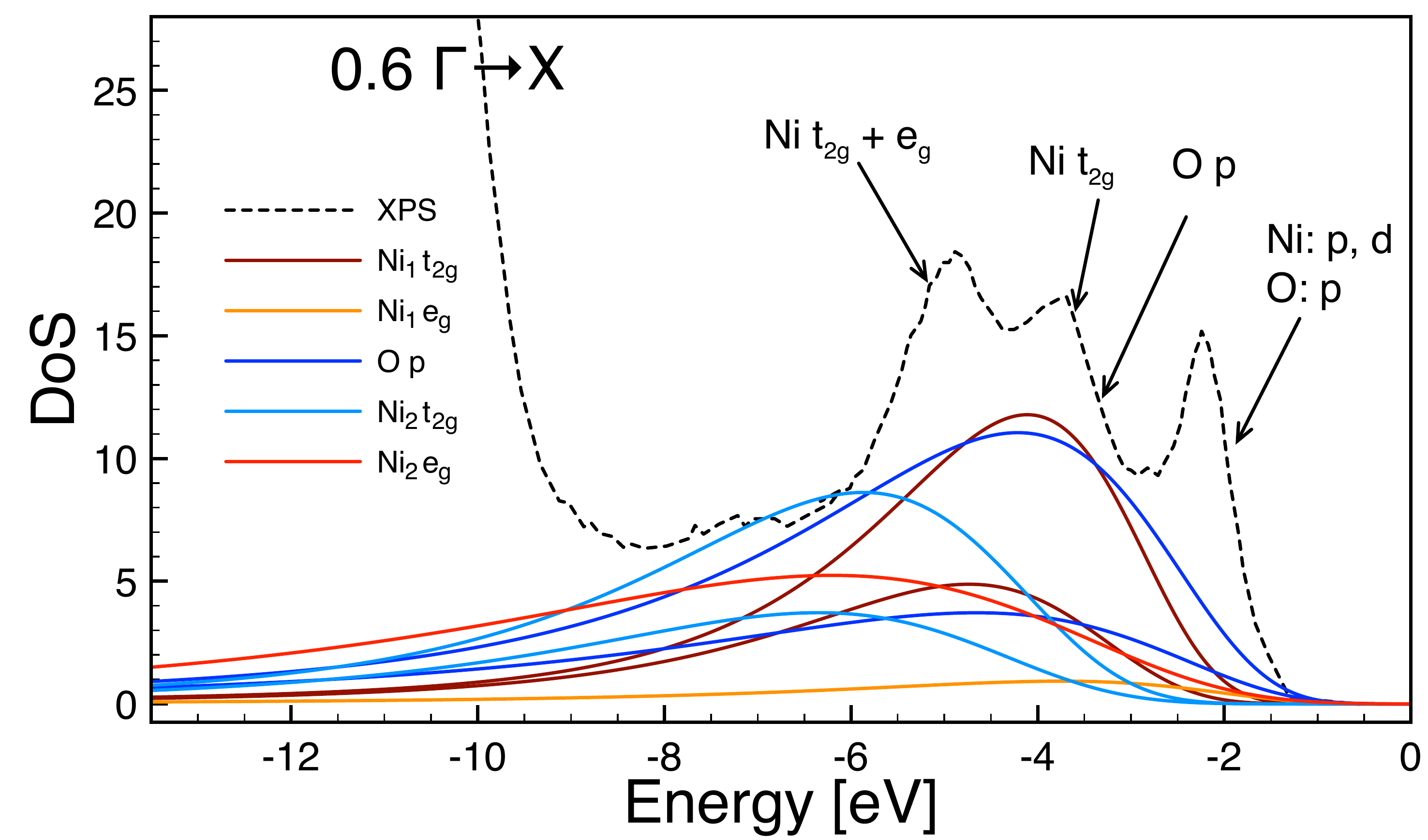}
\caption{Orbitally resolved $\tk$-space spectral function obtained with GW in the symmetrical orthogonal basis at $\frac{2}{3}$ of the distance between $\Gamma$ and $X$ (i.e. closer to $X$).
Dashed black line: experimental ARPES data, Ref.~\cite{NiOCoO_expt_Shen90}.}
\label{Fig:05GammaXVsArpes}
\end{figure}

\subsection{NiO}

\begin{figure*}[bth]
\includegraphics[width=0.47\textwidth]{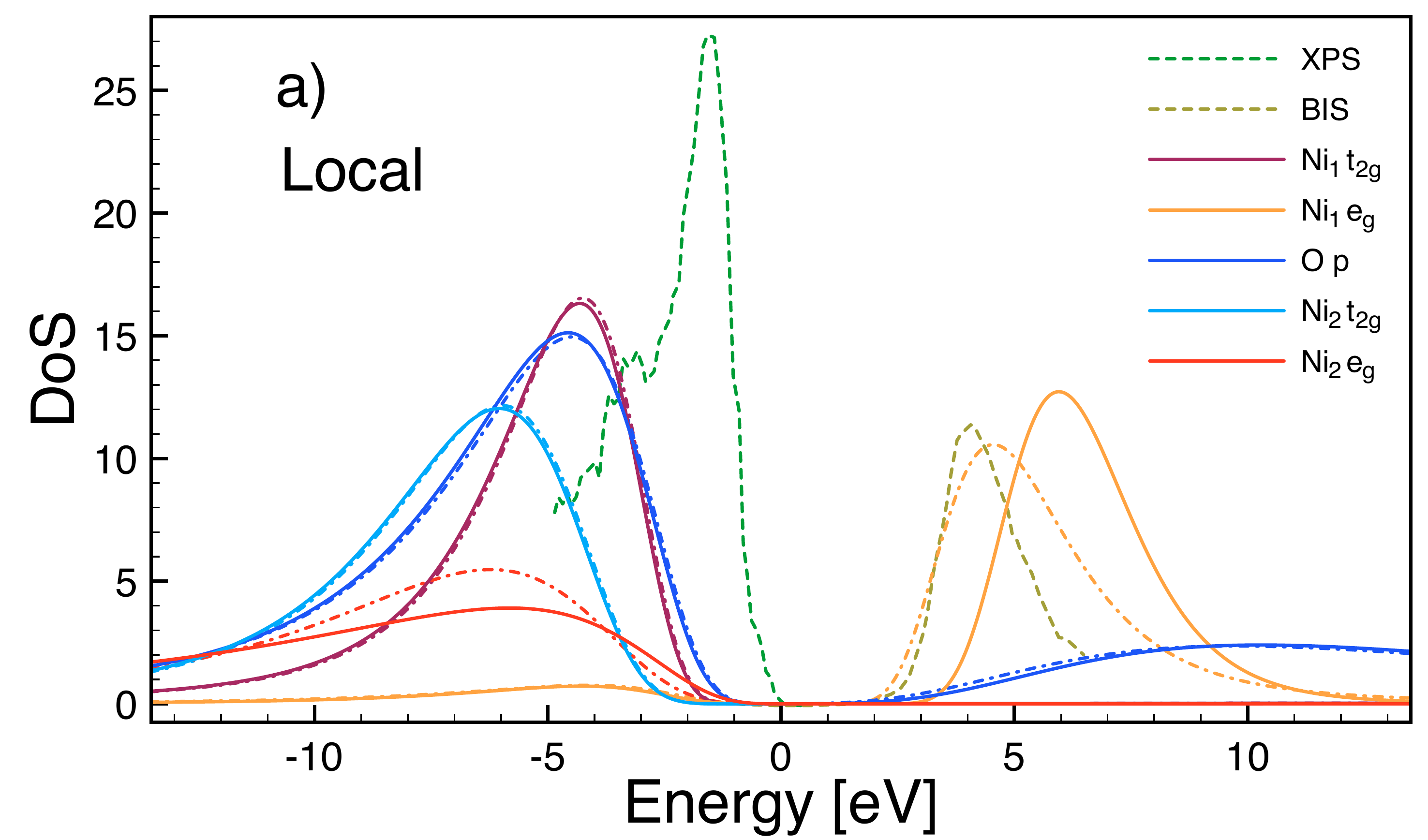}
\includegraphics[width=0.47\textwidth]{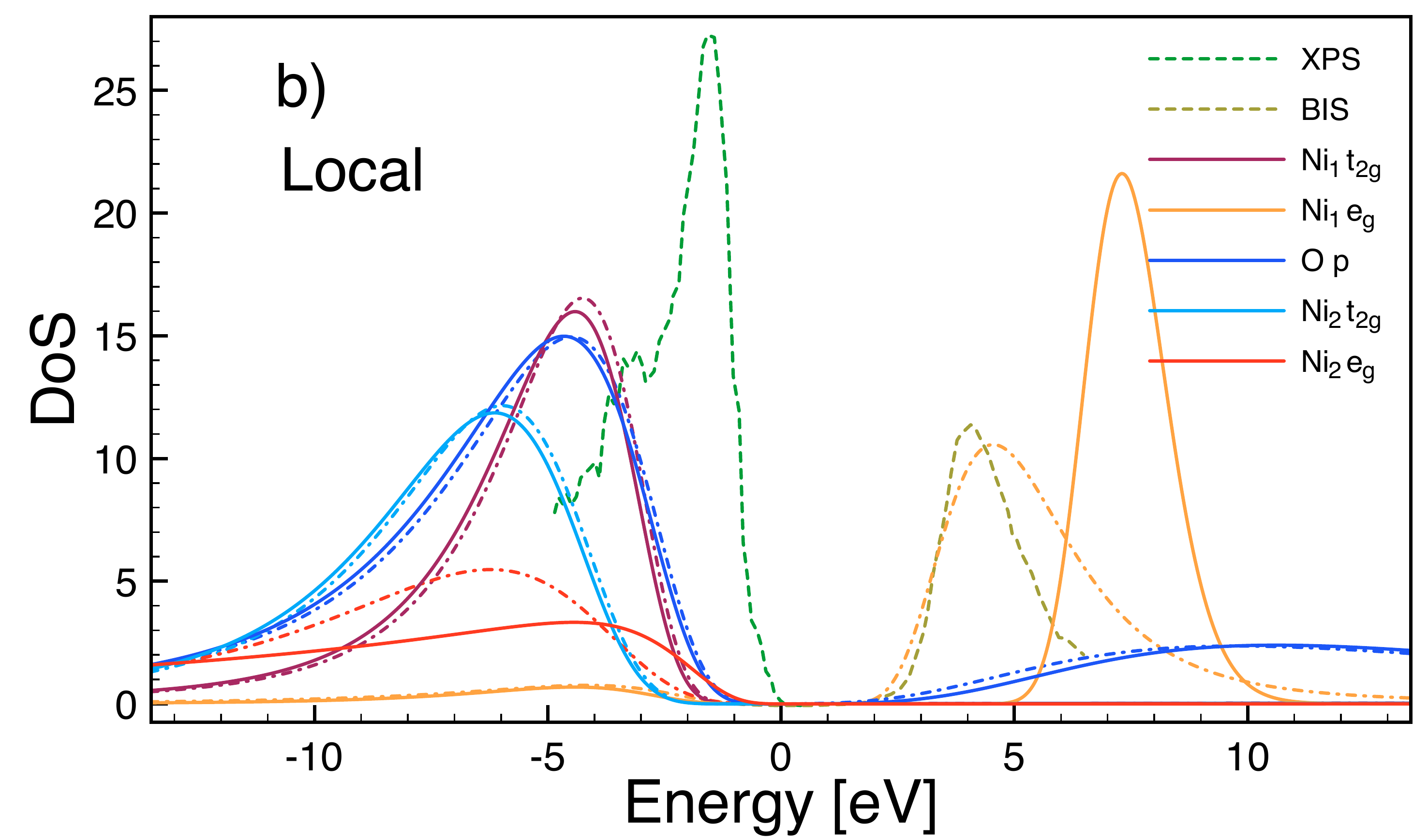}

\includegraphics[width=0.47\textwidth]{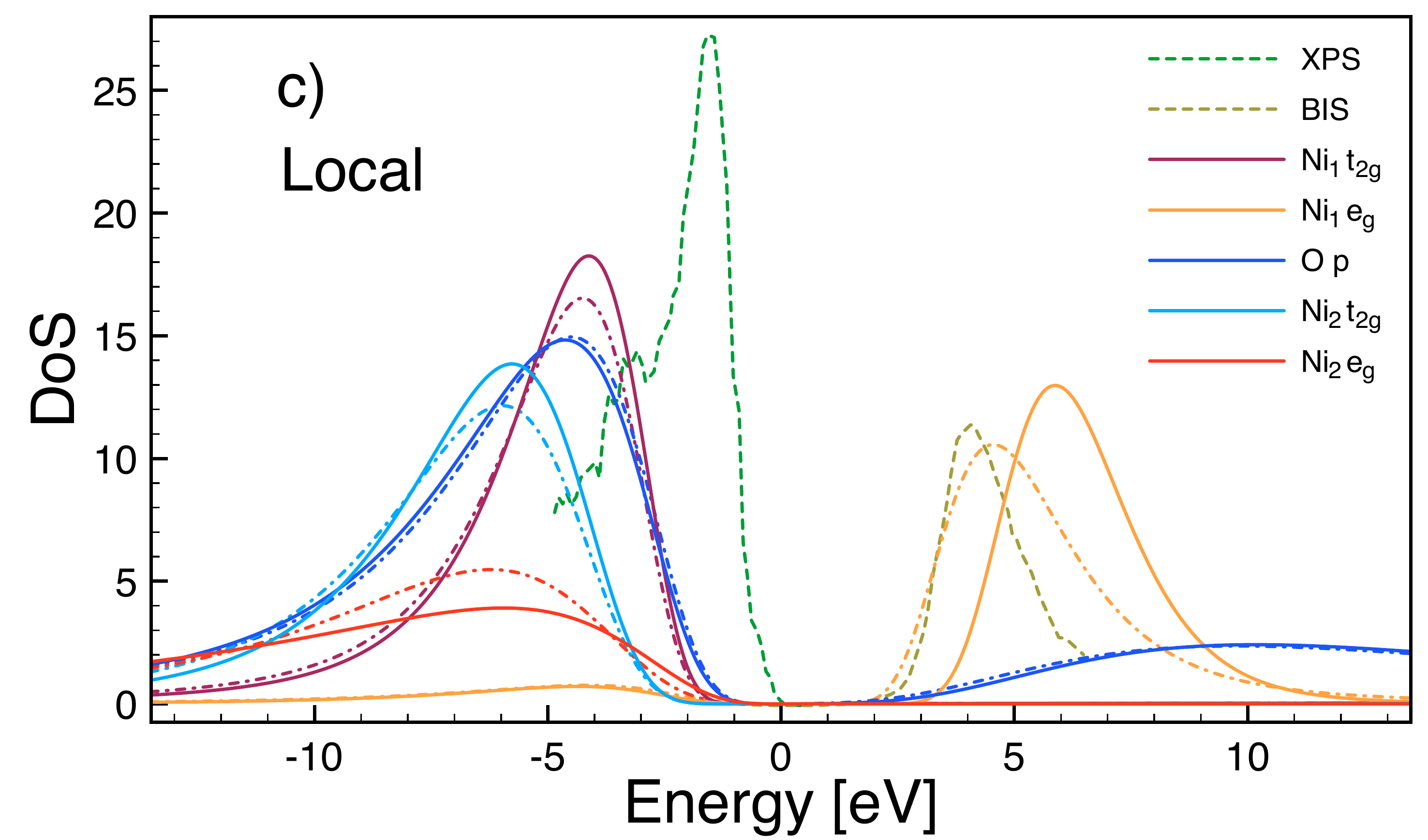}
\includegraphics[width=0.47\textwidth]{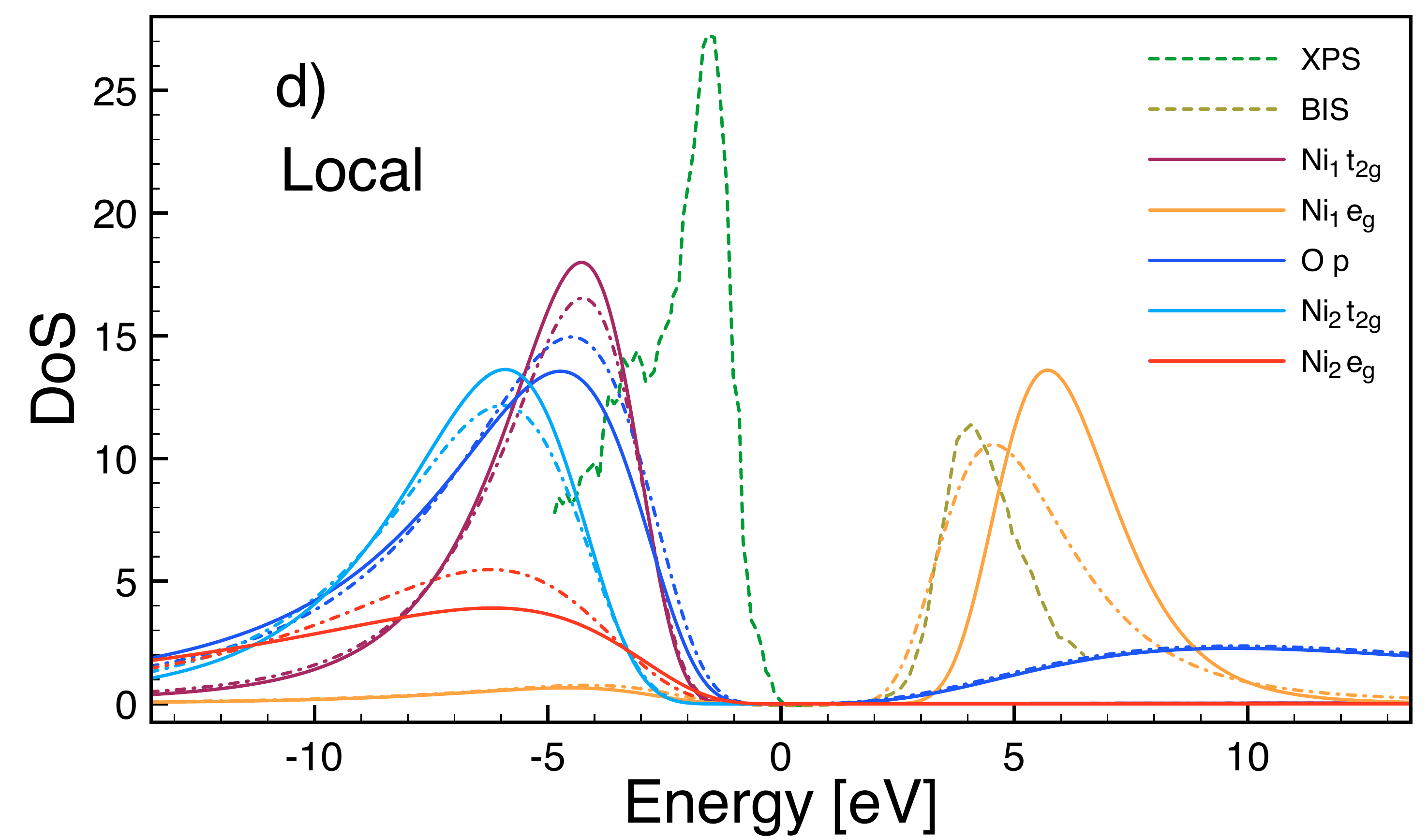}
\caption{
Orbitally resolved SEET local spectral function for NiO. Panels in the plots correspond to impurity choices from
Table~\ref{tab:impurities_NiO}. (Specifically, the panel a) corresponds to impurity choice a), etc.)
Dash-dotted lines: GW,
solid lines: SEET; Dashed lines: experimental data (see Ref.~\cite{NiO_expt_Sawatzky84}).}
\label{Fig:NiOLocal_SEETvsGW}
\end{figure*}

\begin{figure*}[htp]
\includegraphics[width=0.47\textwidth]{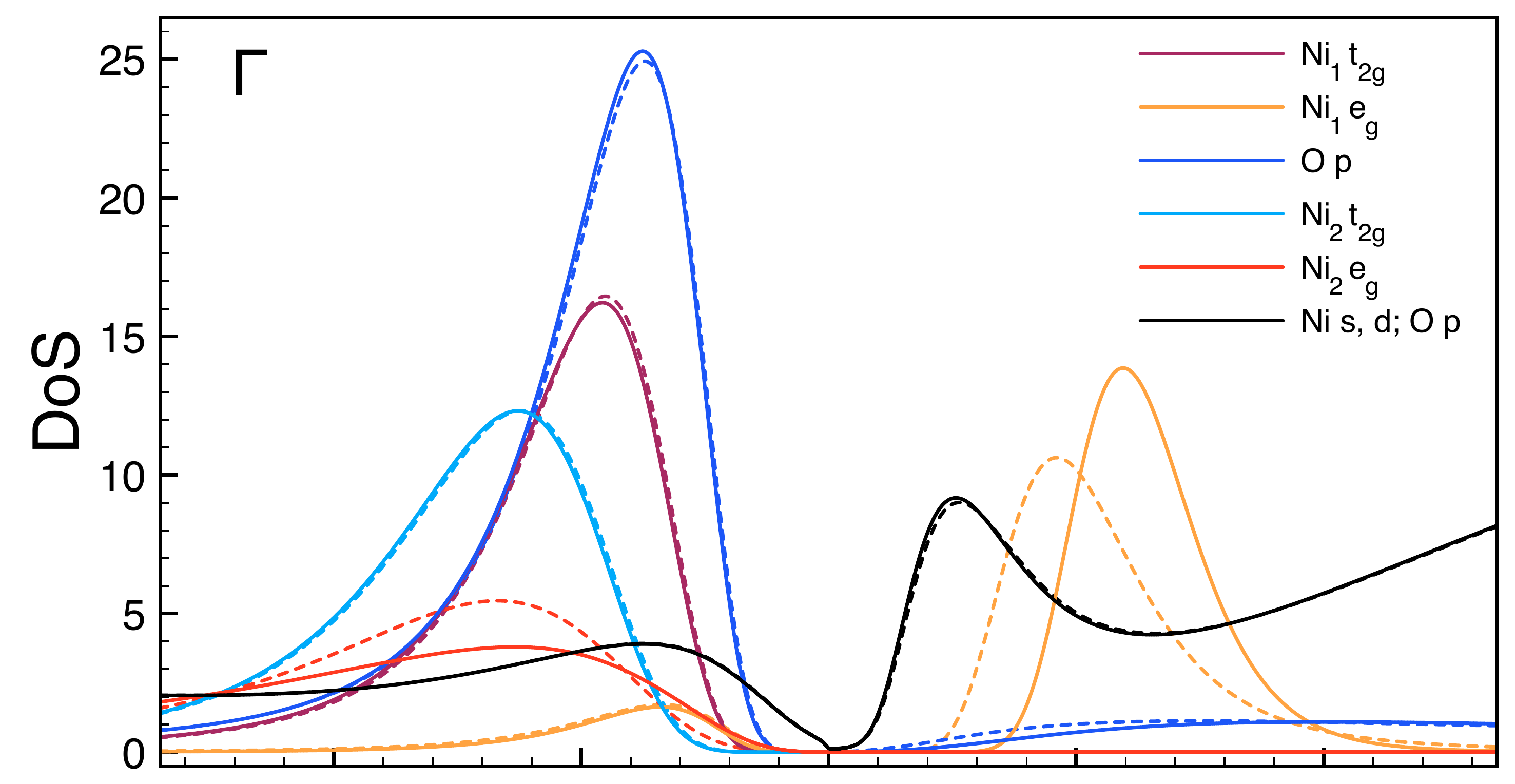}
\includegraphics[width=0.47\textwidth]{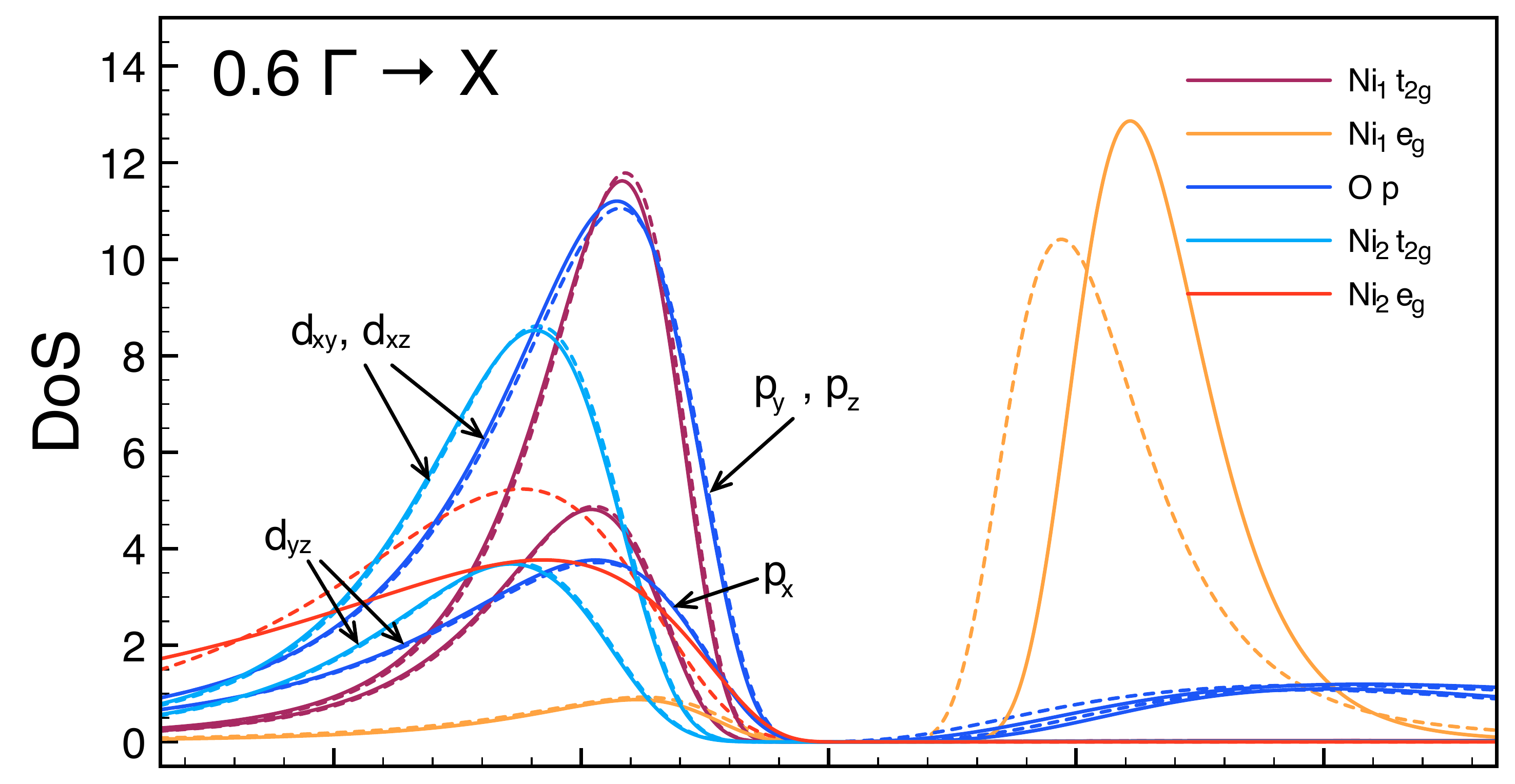}
\includegraphics[width=0.47\textwidth]{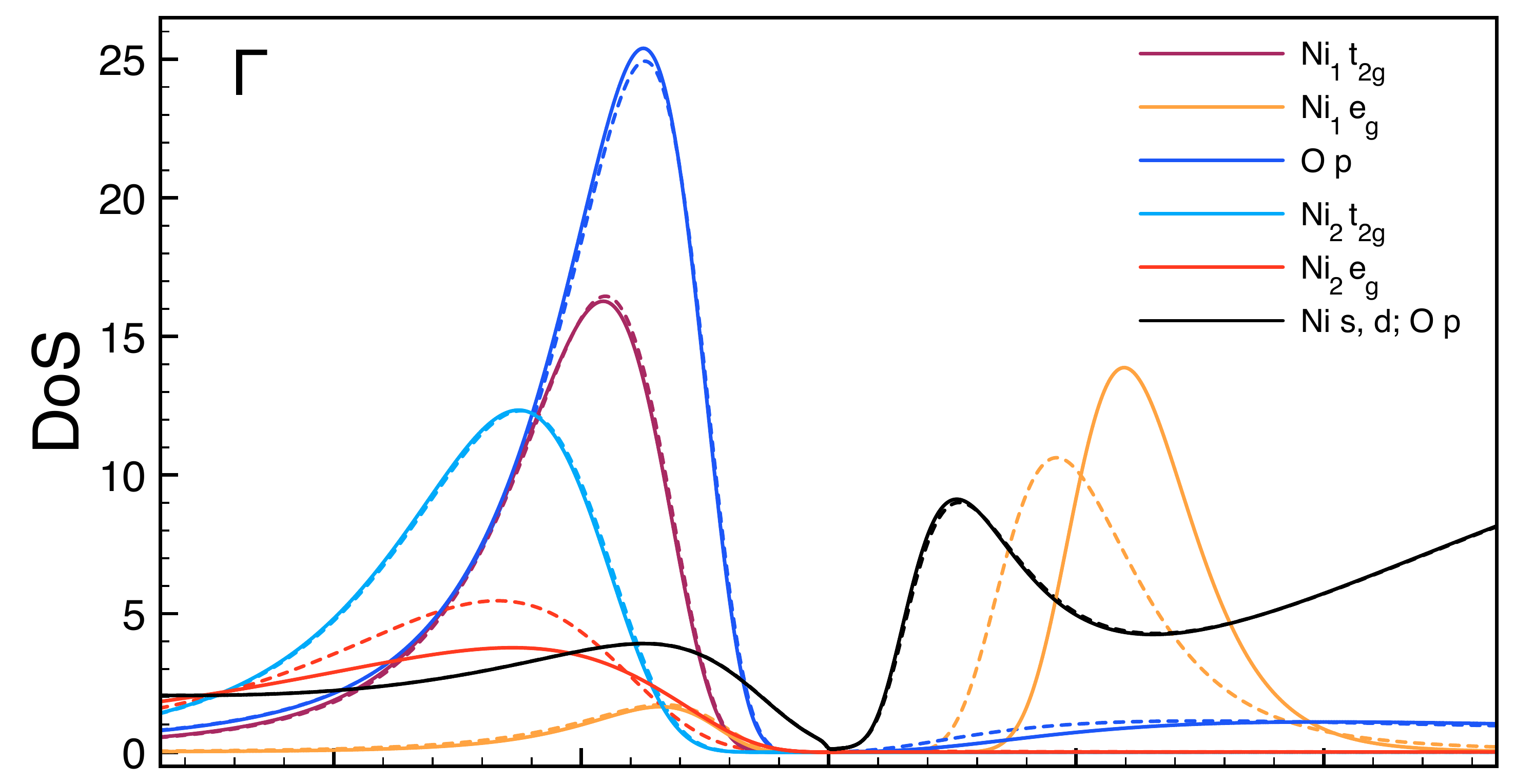}
\includegraphics[width=0.47\textwidth]{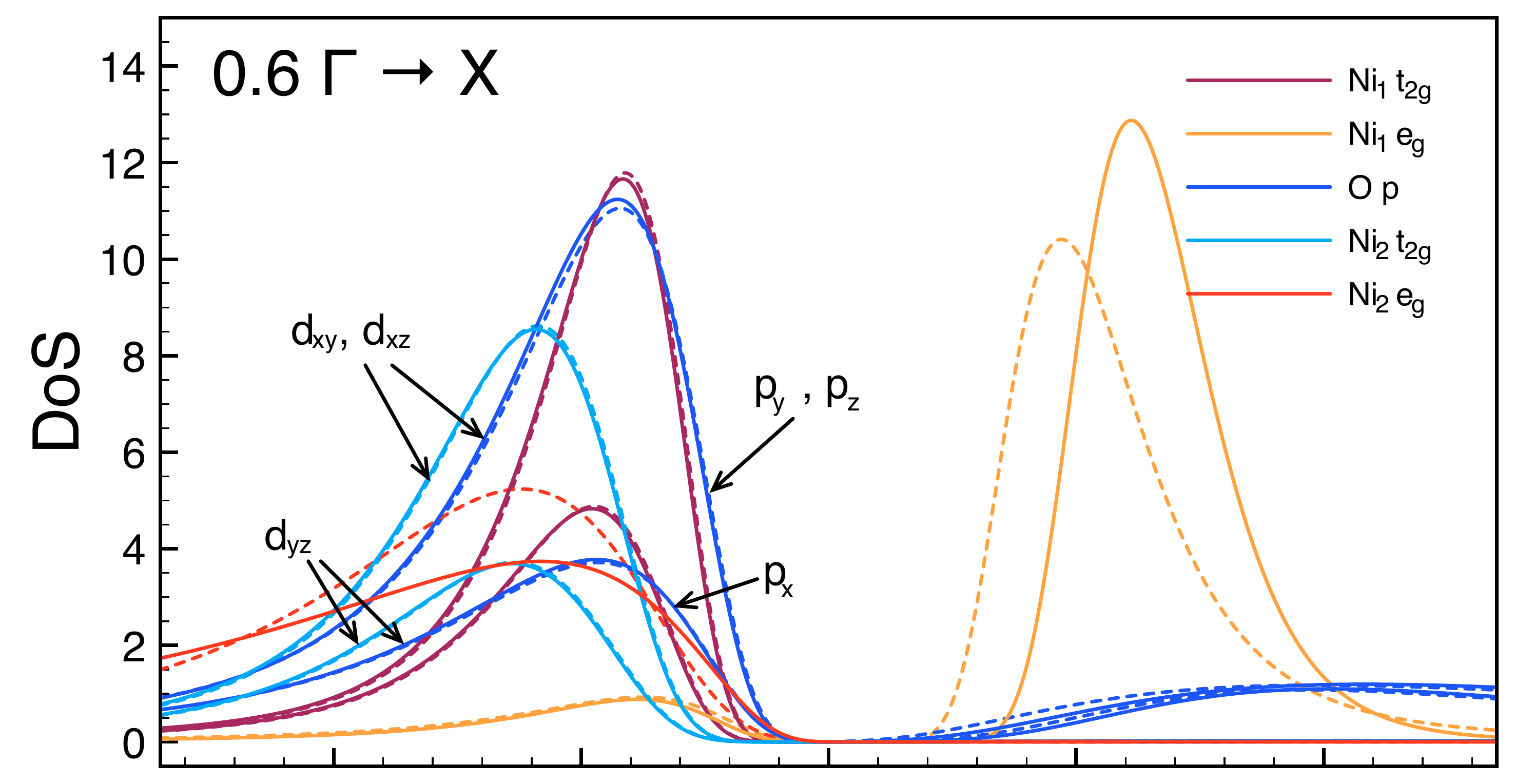}
\includegraphics[width=0.47\textwidth]{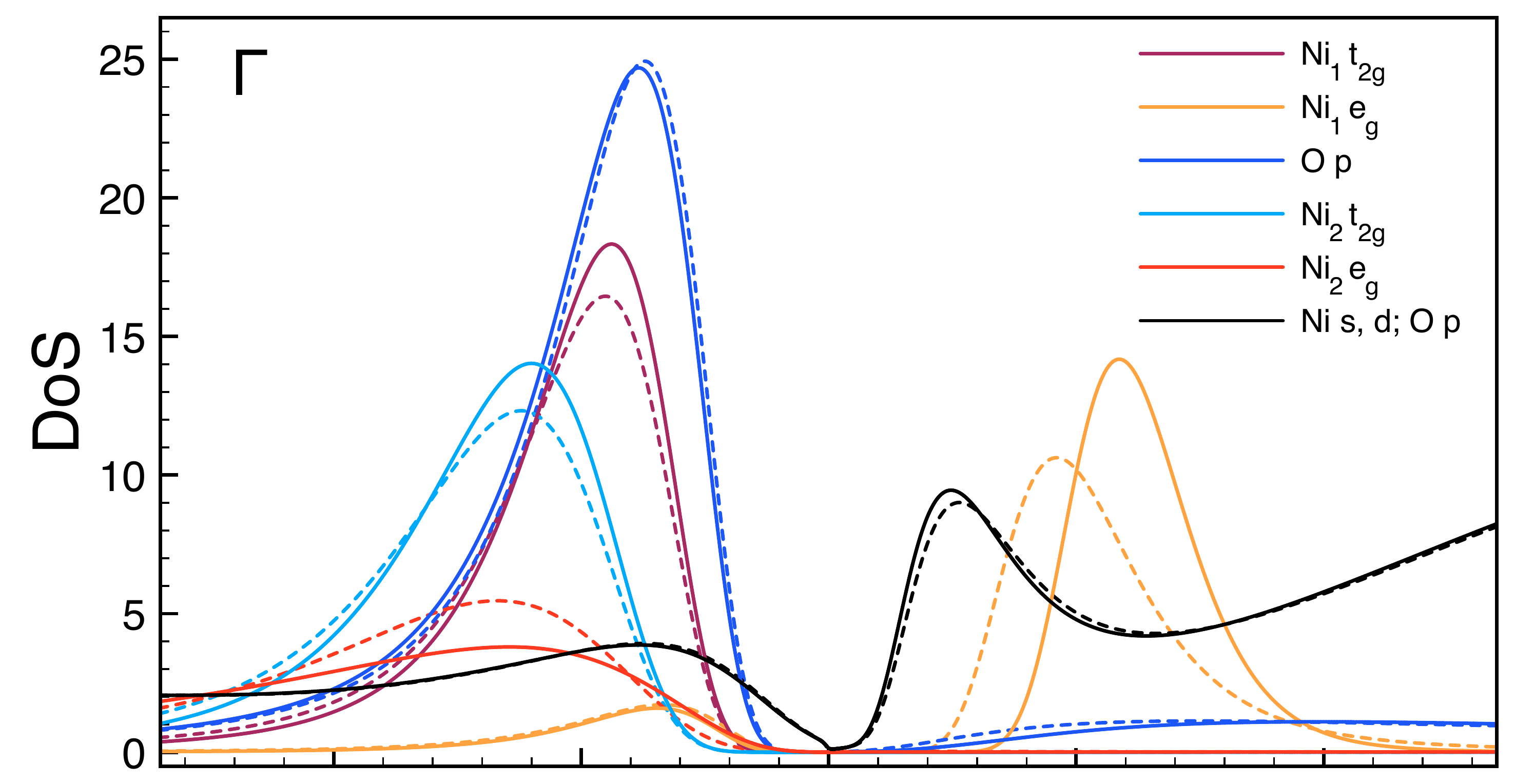}
\includegraphics[width=0.47\textwidth]{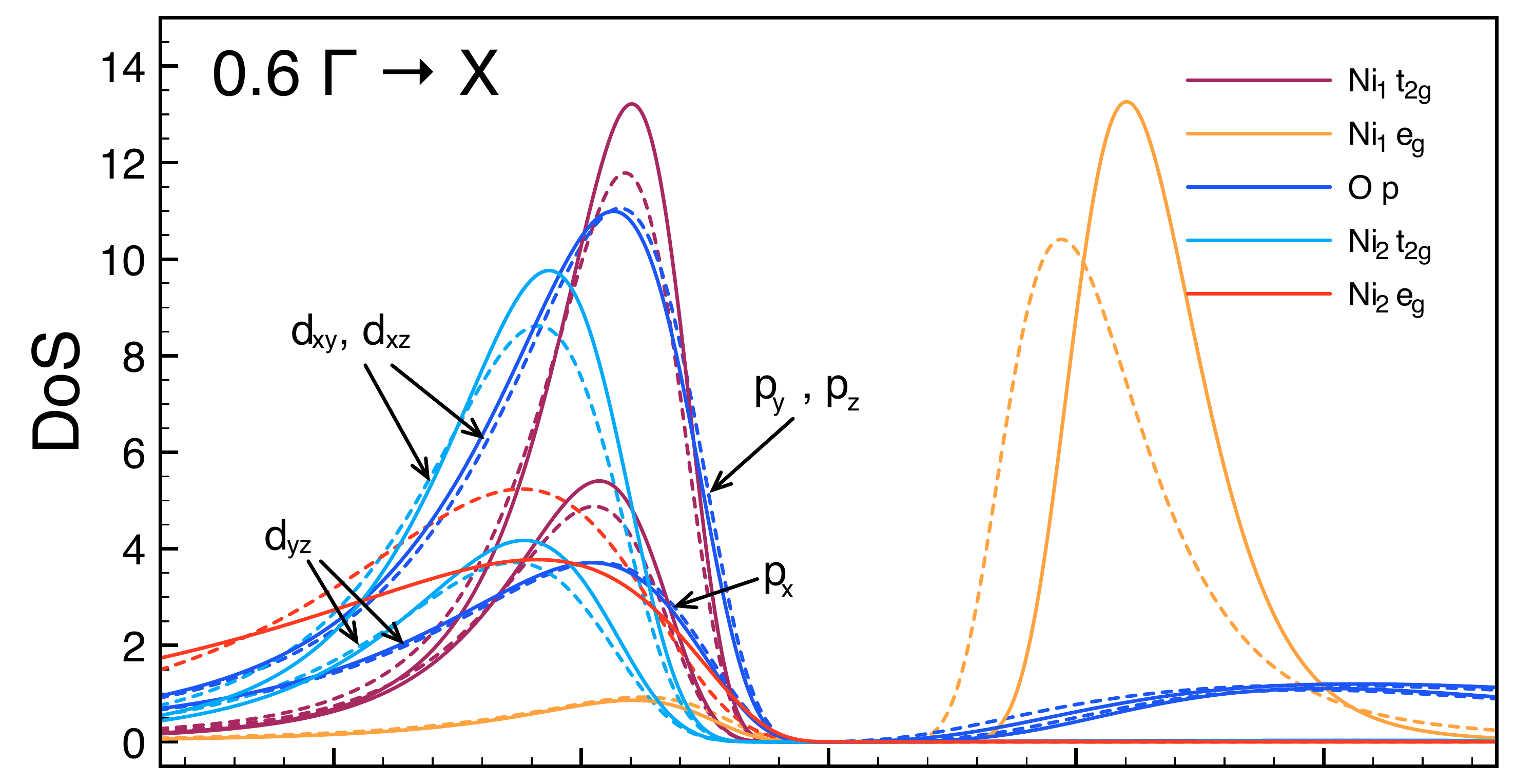}
\includegraphics[width=0.47\textwidth]{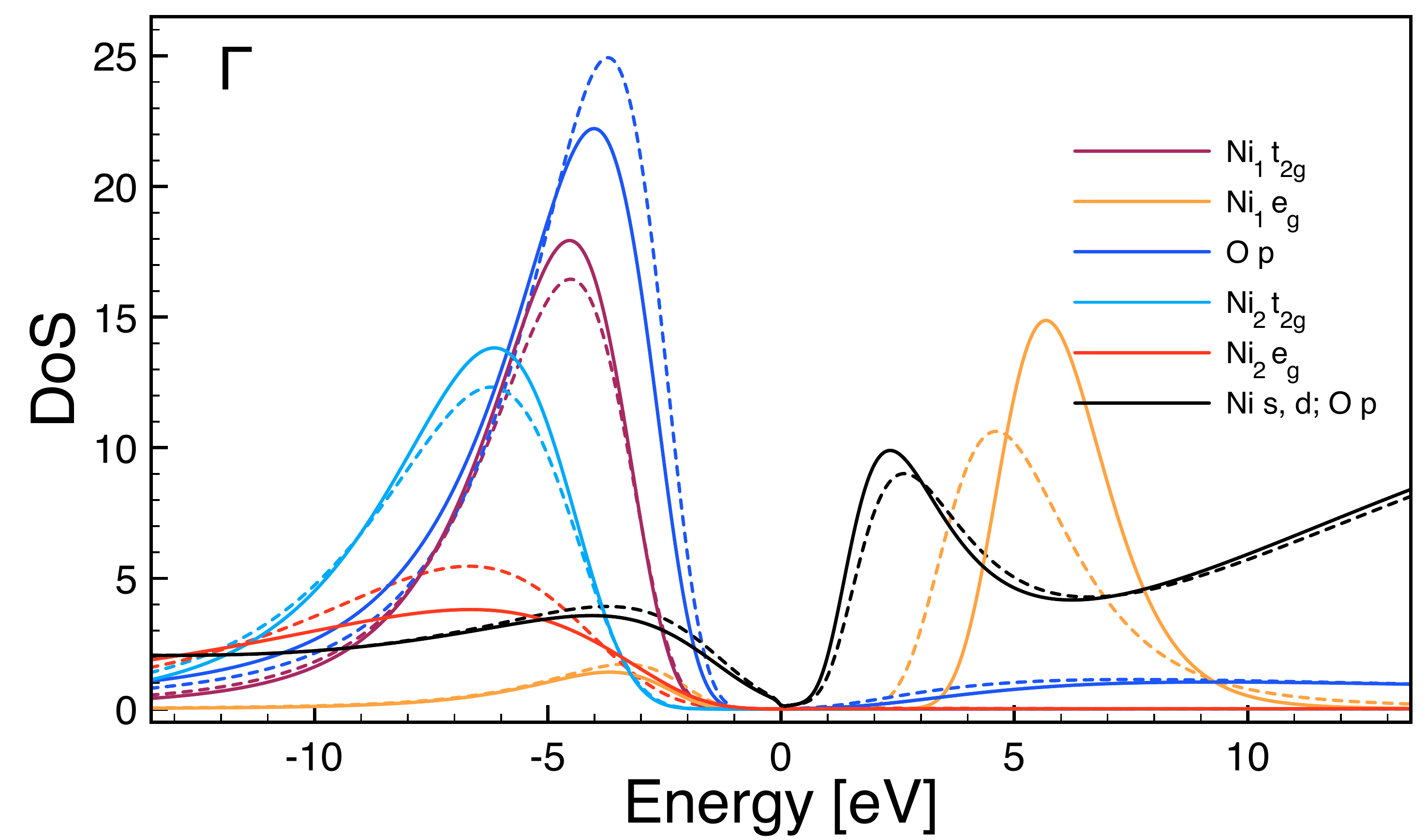}
\includegraphics[width=0.47\textwidth]{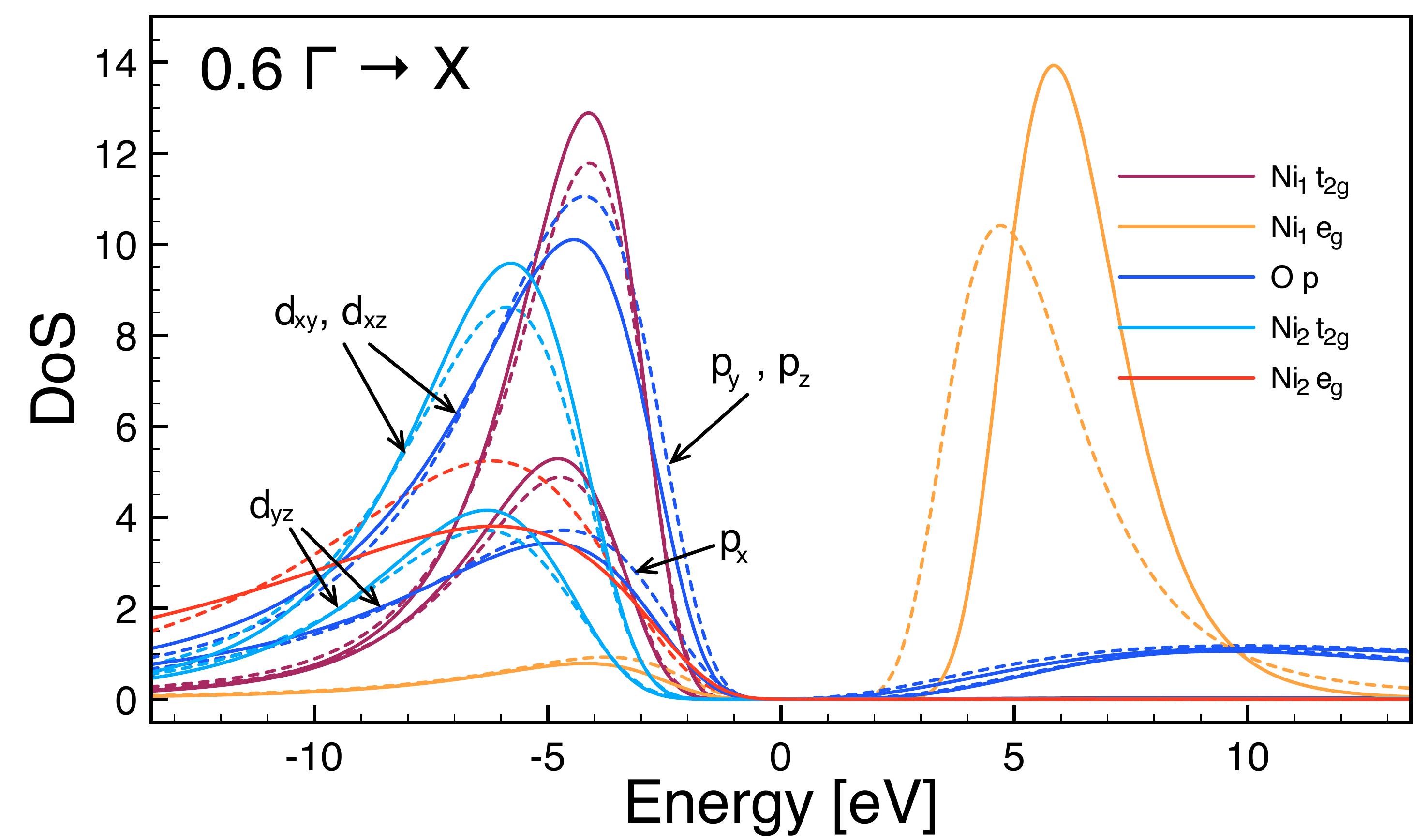}
\caption{$\tk$-resolved SEET (solid lines) and GW (dashed lines) spectral functions for NiO at $\Gamma$ (left column) and at $\frac{2}{3}$ distance
between $\Gamma$ and $X$  (right column).  Different impurities were chosen in each of the rows. These impurity choices correspond to the rows
of Table~\ref{tab:impurities_NiO}, with first-to-fourth rows corresponding to impurities chosen in a-d rows of Table~\ref{tab:impurities_NiO}, respectively.
}
\label{Fig:NiOSEETvsGW}
\end{figure*}

ARPES obtains the  $\tk$-resolved spectral function of materials. Traditional band-structure simulations can then be used to attribute features in the spectral function to their atomic origin.
In the case of moderately to strongly correlated materials, where band-structure methods may become unreliable,
this attribution may break down, due to both broadening effects and shifts of the spectral functions caused by the presence of higher level correlations.

GW is expected to remain reliable for stronger correlation strengths than standard band-structure calculations. Consequently, for NiO
in Figs.~\ref{Fig:GammaVsArpes}~and~\ref{Fig:05GammaXVsArpes}, we attempt to assign an atomic orbital character to the features present in the GW spectrum. 
Fig.~\ref{Fig:GammaVsArpes} presents the orbitally resolved spectral function in canonical orthogonal orbitals at the $\Gamma$ point while
Fig.~\ref{Fig:05GammaXVsArpes} shows results in symmetric orthogonal orbitals for a point in the $\Delta$ direction at $\frac{2}{3}$ distance between $\Gamma$
and $X$.

Note that in the canonical orthogonal basis, orbitals do not correspond to a single atomic state but to a linear combination of atomic states. However,
in our case they are dominated (for the curves shown on the level of 70-80\%)
by a specific atomic state. We label the important orbitals near the Fermi energy by the dominant atomic orbital character.
Fig.~\ref{Fig:GammaVsArpes} superimposes the experimental ARPES data from Ref.~\cite{NiOCoO_expt_Shen90}. Experimental data is
shifted such that the highest occupied state lies at zero energy. As discussed earlier, our GW calculations suffer from the finite size effects
identified in Fig.~\ref{Fig:FiniteSize} estimated to be of $0.5$~$-$~$0.7$~eV, introducing an additional relative shift. Nevertheless, even with this shift
a clear identification of the main feature with atomic orbitals is possible.

We find that the dominant peak stems from the Ni $t_{2g}$ orbitals, whereas the states closest to the Fermi energy contain
a mixture of both nickel and oxygen $p$-states.

Fig.~\ref{Fig:05GammaXVsArpes} shows data for the point at $\frac{2}{3}$ distance along the $\Delta$ direction, between the $\Gamma$ and $X$ points, along with
identifications of dominant atomic contributions. Ni$_1$ and Ni$_2$ denote the two antiferromagnetically ordered Nickel contributions.
Data is obtained in the symmetric orthogonal orbital basis, where we find
that atomic orbital character can be attributed almost uniquely ($\sim$ 95\%) for each of the linear combinations present in this basis.
While ARPES shows a sequence of clearly distinct peaks, our results are smoother and only allow a general attribution of the dominant contribution
in a broad energy window, which we indicate in the plot. 

We emphasize here that these GW results, when accounting for the systematic error due to  finite size effects, are qualitatively
correct for NiO. This means that as a result of embedding procedure, we only expect small improvements and we predict that SEET results should remain mostly unchanged when compared to GW.

\begingroup
\squeezetable
\begin{table}[tbh]
\begin{ruledtabular}
\begin{tabular}{c|c|c|p{6cm}}
Name & Imp & Orb & Description \\
\hline
a & 2 & 2 & Ni$_1$ $e_g$; Ni$_2$ $e_g$ \\
b & 1 & 4 &Ni$_1$ $e_g$ + Ni$_2$ $e_g$ \\
c & 4 & 3 &Ni$_1$ $e_g$; Ni$_2$ $e_g$; Ni$_1$ $t_{2g}$; Ni$_2$ $t_{2g}$ \\
d & 6 & 3 &Ni$_1$ $e_g$; Ni$_2$ $e_g$; Ni$_1$ $t_{2g}$; Ni$_2$ $t_{2g}$; O$_1$ $p$; O$_2$ $p$ \\
\end{tabular}
\end{ruledtabular}
\caption{Choice of the active space for NiO. Imp denotes the number of distinct disjoint impurity problems.
Orb stands for the number of impurity orbitals in the largest impurity problem.\label{tab:impurities_NiO}}
\end{table}
\endgroup

\subsection{Effect of strong electron correlations in NiO}

We now turn our attention to results from our embedding construction. The identification of the orbitals near the Fermi level
in Figs.~\ref{Fig:GammaVsArpes}~and~\ref{Fig:05GammaXVsArpes} suggests a choice of active orbital set as
Ni $e_g$, Ni $t_{2g}$ and O $p$ states. These orbitals will be used to construct impurity models in SEET.

We perform the SEET embedding in symmetrical orthogonal orbitals, where {\bf i)} the attribution to atomic orbital character is straightforward and {\bf ii)}
off-diagonal hybridization elements are 2-3 orders of magnitude smaller than the diagonal ones.
Note that in SEET we do not use any Wannierization procedure as it is commonly done in LDA+DMFT or GW+EDMFT.
The ability to embed multiple impurities is crucial, as non-perturbative
impurity solvers such as the ED solver used here scale exponentially in the number of impurity orbitals.

Table~\ref{tab:impurities_NiO} shows four choices of embedded orbital subsets.
Subset (a) consists of two disjoint impurities on each of the nickels, made out of two Ni $e_g$ orbitals.
Subset (b) combines those two impurities into a single four-orbital impurity.
Subset (c) builds four impurities consisting of two disjoint Ni $e_g$ and two additional disjoint Ni $t_{2g}$ orbitals (each with three
impurity orbitals).
Subset (d) supplements the four impurities of subset (c) with two additional disjoint three-orbital impurities of the oxygen $p$ orbitals.

\subsubsection{Local DOS for NiO}

Fig.~\ref{Fig:NiOLocal_SEETvsGW} shows the orbitally resolved local spectral function of NiO for the four impurity choices
of Table~\ref{tab:impurities_NiO}. Shown are also the orbitally resolved GW results corresponding to Fig.~\ref{Fig:FiniteSize},
as well as experimental local spectral functions obtained with x-ray photoemission (XPS) and bremsstrahlung-isochromat-spectroscopy
(BIS)~\cite{NiO_expt_Sawatzky84}. Note that the gap edge of XPS is shifted to zero energy and the relative height of XPS and BIS data is arbitrary.
The experimental error present in this experiment~\cite{NiO_expt_Sawatzky84} is estimated as 0.6~eV and the resulting band gap,
measured at half-maxima of both XPS and BIS peaks, is estimated to be 4.3~$\pm$~0.6~eV.

\begin{figure*}[tbh]
\includegraphics[width=0.47\textwidth]{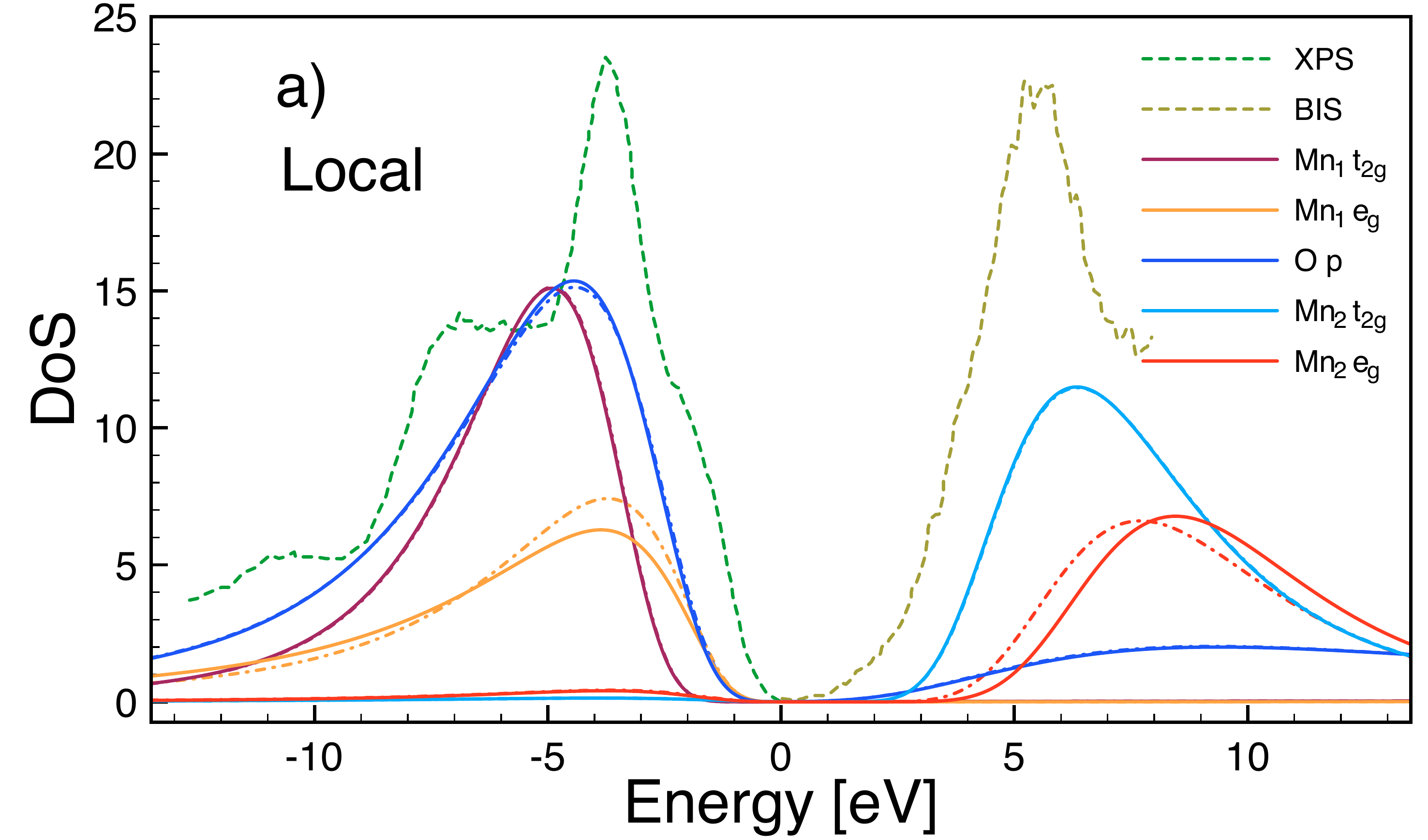}
\includegraphics[width=0.47\textwidth]{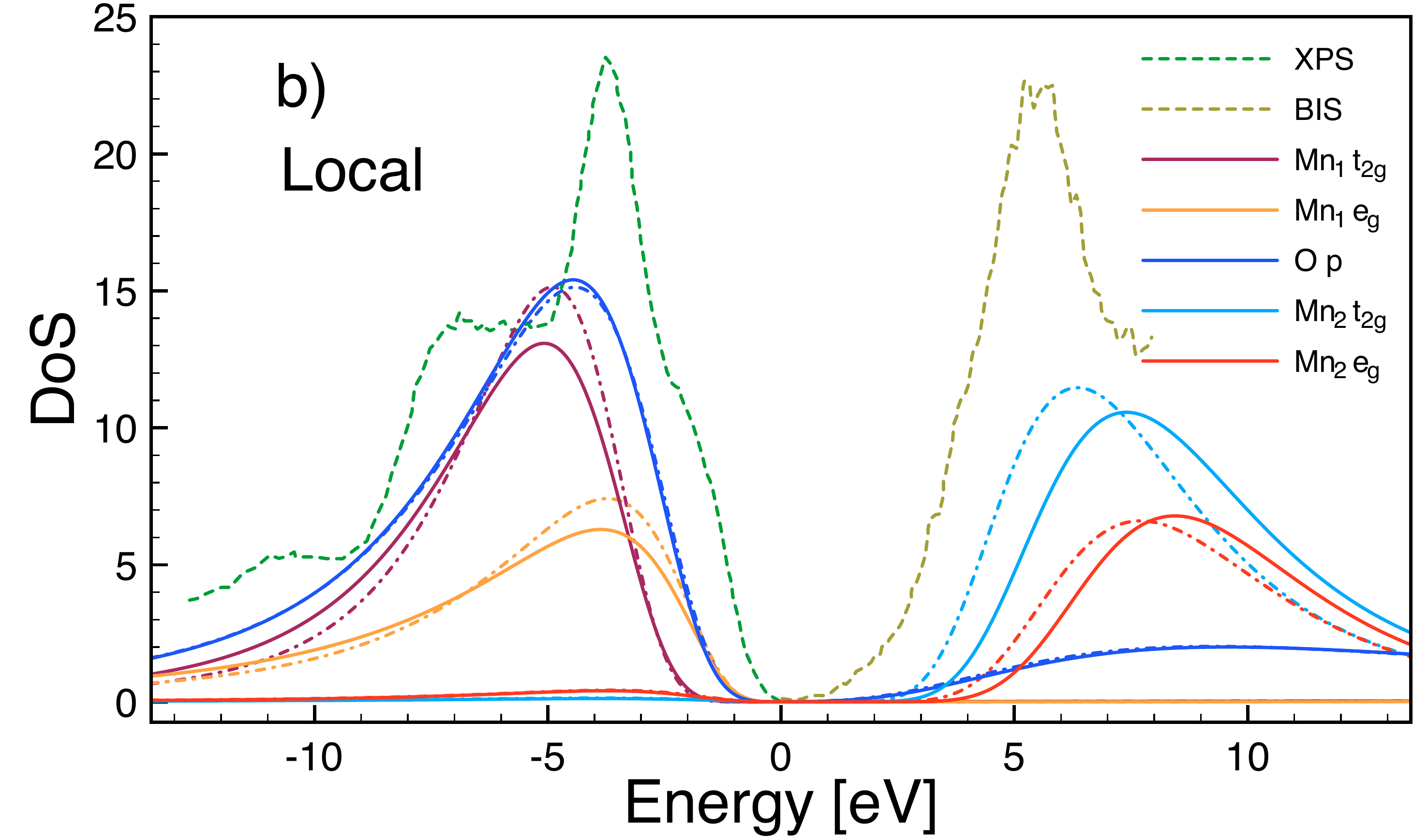}

\includegraphics[width=0.47\textwidth]{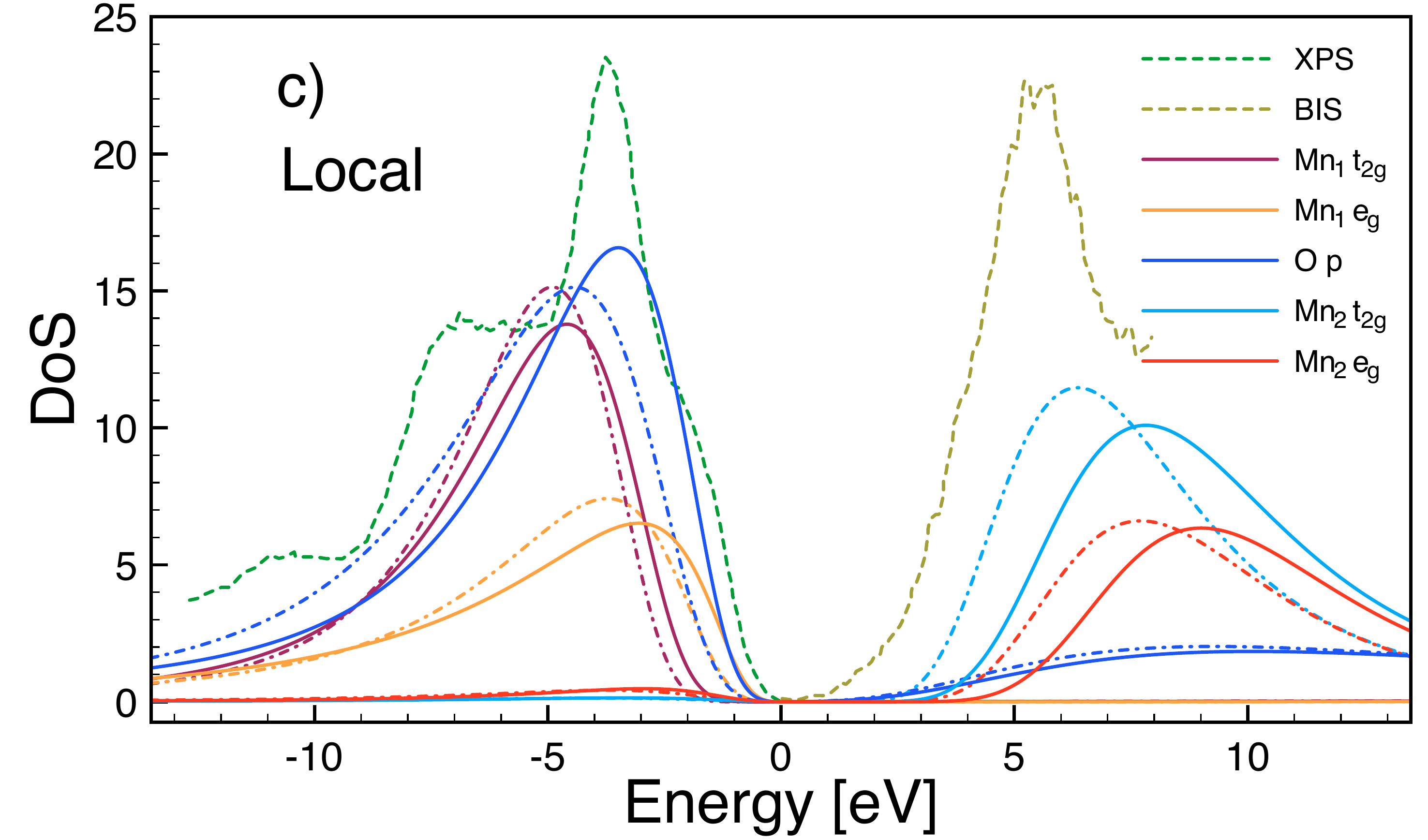}
\includegraphics[width=0.47\textwidth]{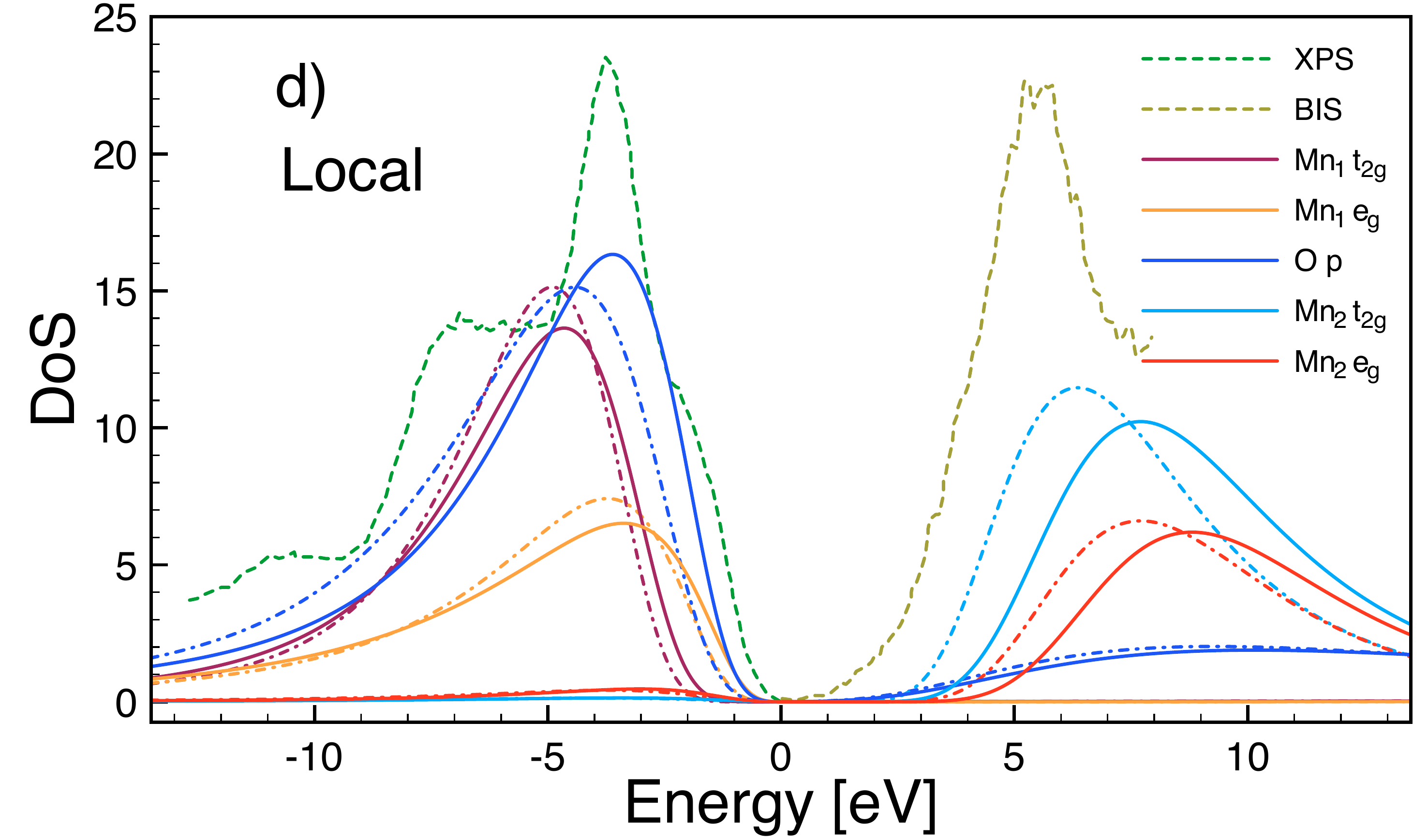}
\caption{Orbitally resolved SEET local spectral function for MnO with impurity choice of Table~\ref{tab:impurities_MnO}.
Panels in the plots correspond to impurity choices from Table~\ref{tab:impurities_MnO}. (Specifically, the panel a) corresponds to impurity choice a), etc.)
Dash-dotted lines: GW,
solid lines: SEET; Dashed lines: experimental data (see Ref.~\cite{MnO_expt_Elp91}).}
\label{Fig:MnOLocal_SEETvsGW}
\end{figure*}

Panel a) shows results from two disjoint two-orbital impurities that only consist of the nickel $e_g$ states. Substantial shifts that arise due to embedding
are evident for $e_g$ states. All other orbitals are adjusted only via the Dyson equation~\ref{Eqn:Dyson},
and these changes are small. As discussed previously, the GW results and consequently SEET results are biased by finite size effects.
Introducing a correction due to fine size effects will shrink the current GW gap (which is around $5.6$~eV for the $6$~$\times$~$6$~$\times$~$6$ lattice when measured at half peak height)
by $0.5$~$-$~$0.7$~eV resulting in the GW band gap between $5$~$-$~$5.5$~eV. SEET widens this gap to $6.5$~$-$~$6.7$~eV resulting in a band gap of $5.5$~$-$~$6.0$~eV after
accounting for finite size effects.

Results for panel b) are obtained with a single four-orbital impurity that contains the
same active orbitals as subset (a) but they are contained within a single impurity. Plots from panel b) are essentially indistinguishable from panel a), indicating that
cross-correlations at the GW level are sufficient for describing the coupling between those two disjoint impurities.

In panel c), where additional $t_{2g}$ states are considered, a small change of the magnitude but not of the overall peak
position is visible for Ni $t_{2g}$ while the $e_g$ orbitals are comparable to the ones in panels a) and b). 

Adding additional correlations on the oxygen $p$ orbitals (in panel d)) substantially shifts all states including Ni $t_{2g}$
and $e_g$ states, causing a build-up of the  shoulder density  for frequencies between -5~and~-10~eV. Here the contributions from Ni t$_{2g}$ together with oxygen p start to be responsible for this buildup.

Note that these results allows us to determine the atomic character of peaks and the size of the band gap is reasonably matching the experimental results
when accounting for finite size effects, experimental uncertainties of 0.6~eV, as well as possible inaccuracies stemming from using a Gaussian basis set.

\subsubsection{Momentum resolved DOS for NiO}

Fig.~\ref{Fig:NiOSEETvsGW} shows $\tk$-resolved spectral functions at $\Gamma$ and at $\frac{2}{3}$ distance between $\Gamma$ and $X$
for the impurity choices of Table~\ref{tab:impurities_NiO}. These calculations and plots were performed in the symmetrical orthogonal basis.
The spectral function at the $\Gamma$ point shows similar behavior to Fig.~\ref{Fig:NiOLocal_SEETvsGW}. However, a notable difference
is in the Ni $4s$ orbital. This orbital while present near the gap edge at the $\Gamma$ point, it rapidly moves to higher energies away from $\Gamma$ point thus contributing
only little to the local spectral function.
Data between $\Gamma$
and $X$ look more complicated as the nickel $t_{2g}$ and oxygen $p$ states split away from the high symmetry $\Gamma$ point.
The assignment of orbital character that arises due to SEET is largely consistent with the angle resolved photoemission experiment
presented in Fig. 2 of Ref.~\onlinecite{NiOCoO_expt_Shen90}. Only the features at low energies (lower than -8~eV) cannot be assigned
without doubt, most likely due to the deficiencies of analytical continuation and artificial broadening of existing features at these energy ranges.

\subsubsection{Local magnetic moment in NiO}

Finally, we briefly discuss the staggered magnetization of NiO. A Mulliken analysis (see e.g.~\cite{Mulliken1955}) yields
the values of Table~\ref{tab:mu_NiO}. It is evident that the most important contribution to magnetism comes from the Ni $e_g$
states, which we treat non-perturbatively in all four choices of active space. Changes between different impurities (corresponding to
the influence of strong correlations on $t_{2g}$ or oxygen $p$ orbitals) are much smaller. 

\begin{table}[tbh]
\begin{tabular}{l|ccccccc}
\hline \hline
&\multirow{2}{*}{expt}&\multirow{2}{*}{HF}&\multirow{2}{*}{GW}& \multicolumn{4}{c}{GW+SEET} \\
\cline{5-8}
&&&& a & b & c & d\\
\hline
NiO & 1.77, 1.90(6) & 1.816 &1.701&1.750&1.751&1.752&1.754\\
\hline \hline
\end{tabular}
\caption{Local magnetic moment of Ni from Mulliken analysis. Impurity choices a,b,c, and d 
correspond to the rows of Table~\ref{tab:impurities_NiO}.
The experimental data is obtained from Refs.~\cite{doi:10.1063/1.1668855} and \cite{mu_expt_Cheetham83}}\label{tab:mu_NiO}
\end{table}

\begin{figure*}[htp]
\includegraphics[width=0.47\textwidth]{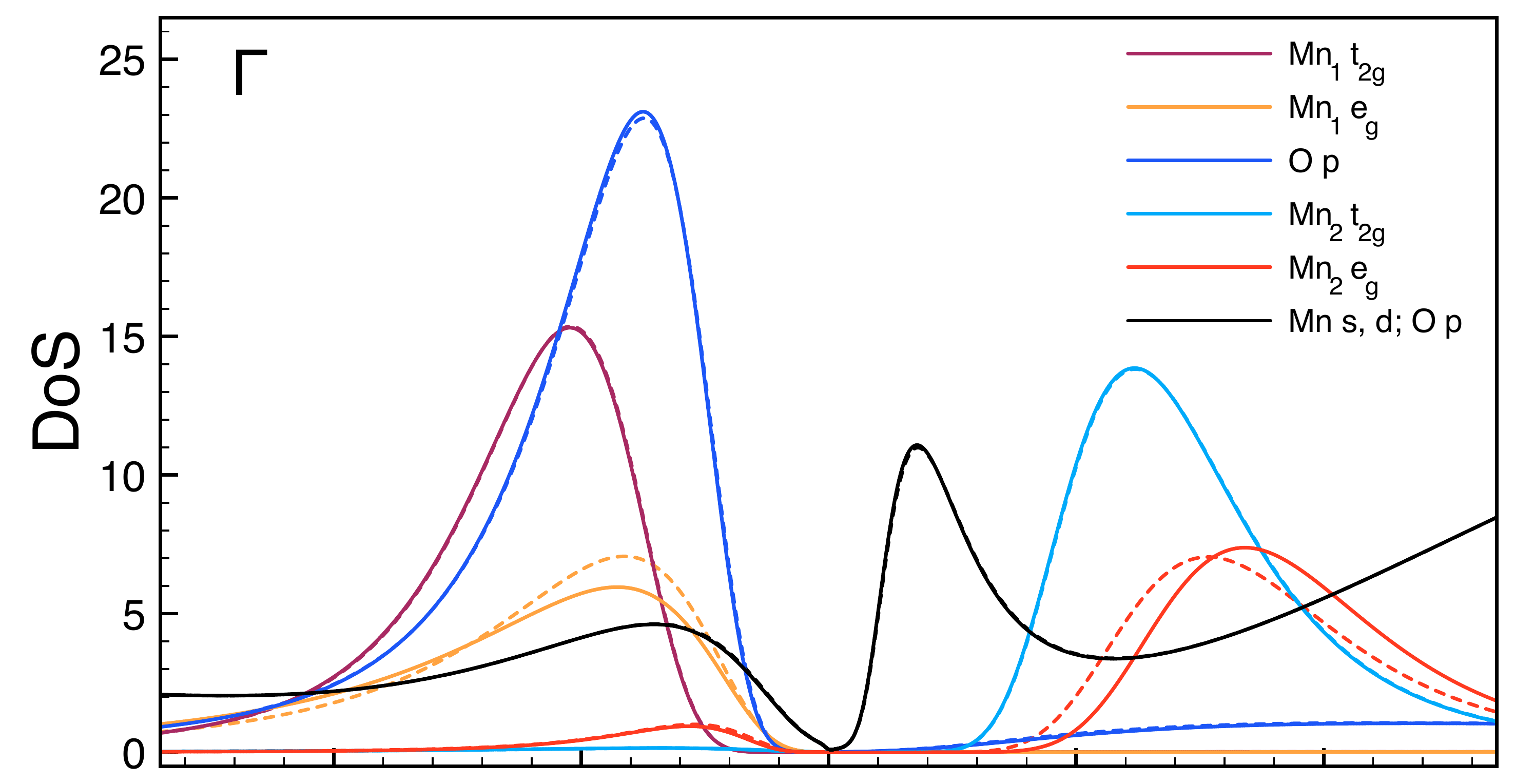}
\includegraphics[width=0.47\textwidth]{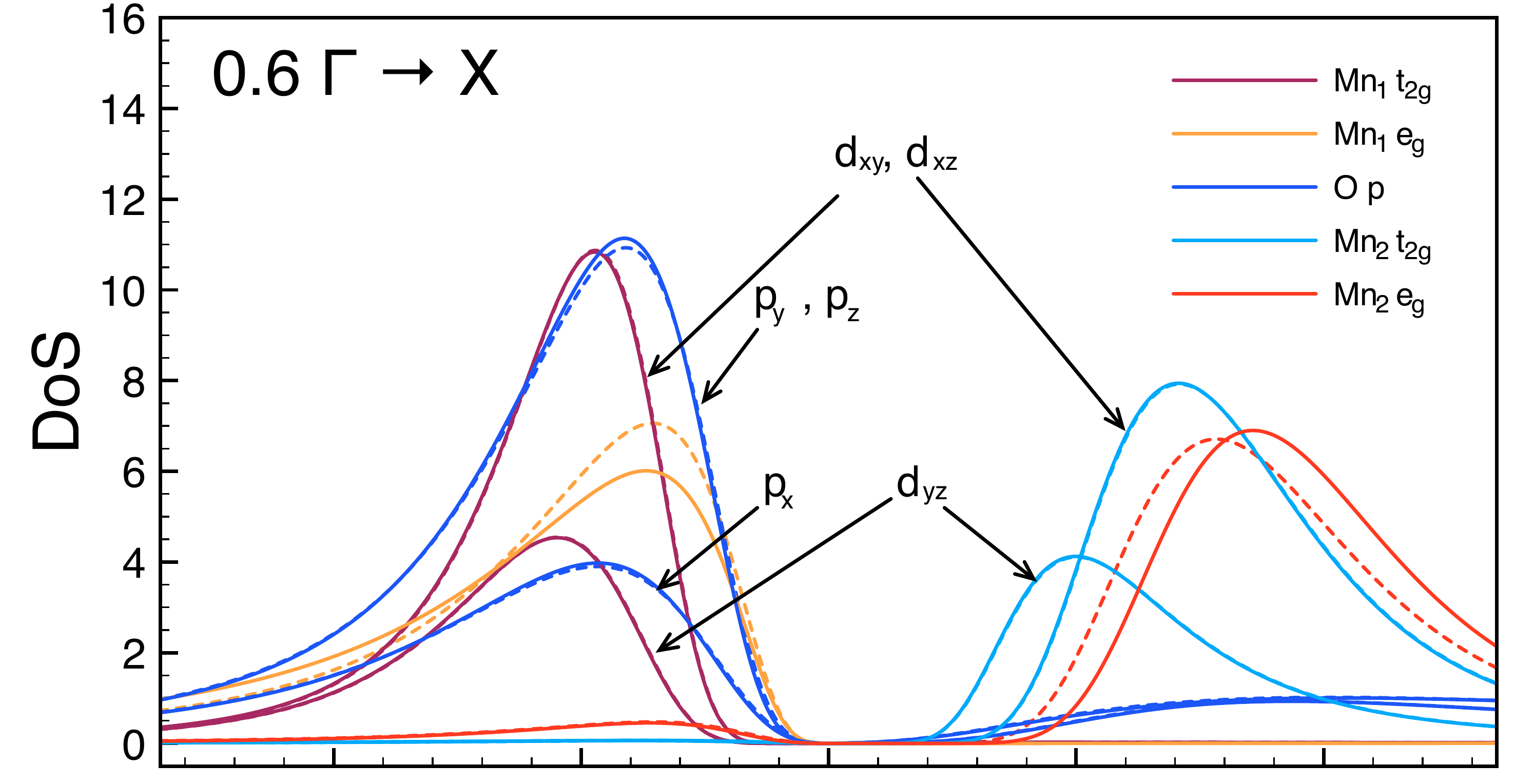}
\includegraphics[width=0.47\textwidth]{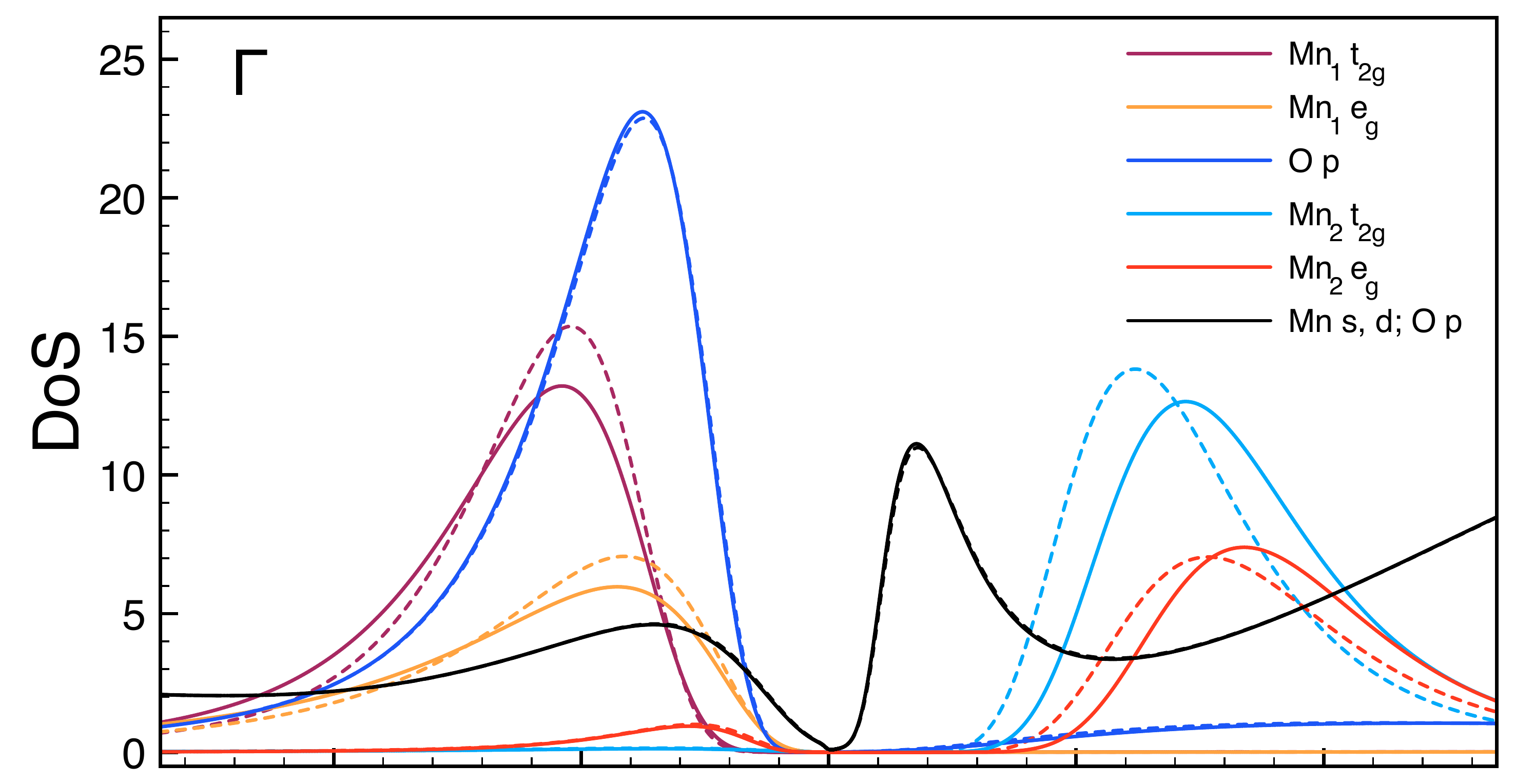}
\includegraphics[width=0.47\textwidth]{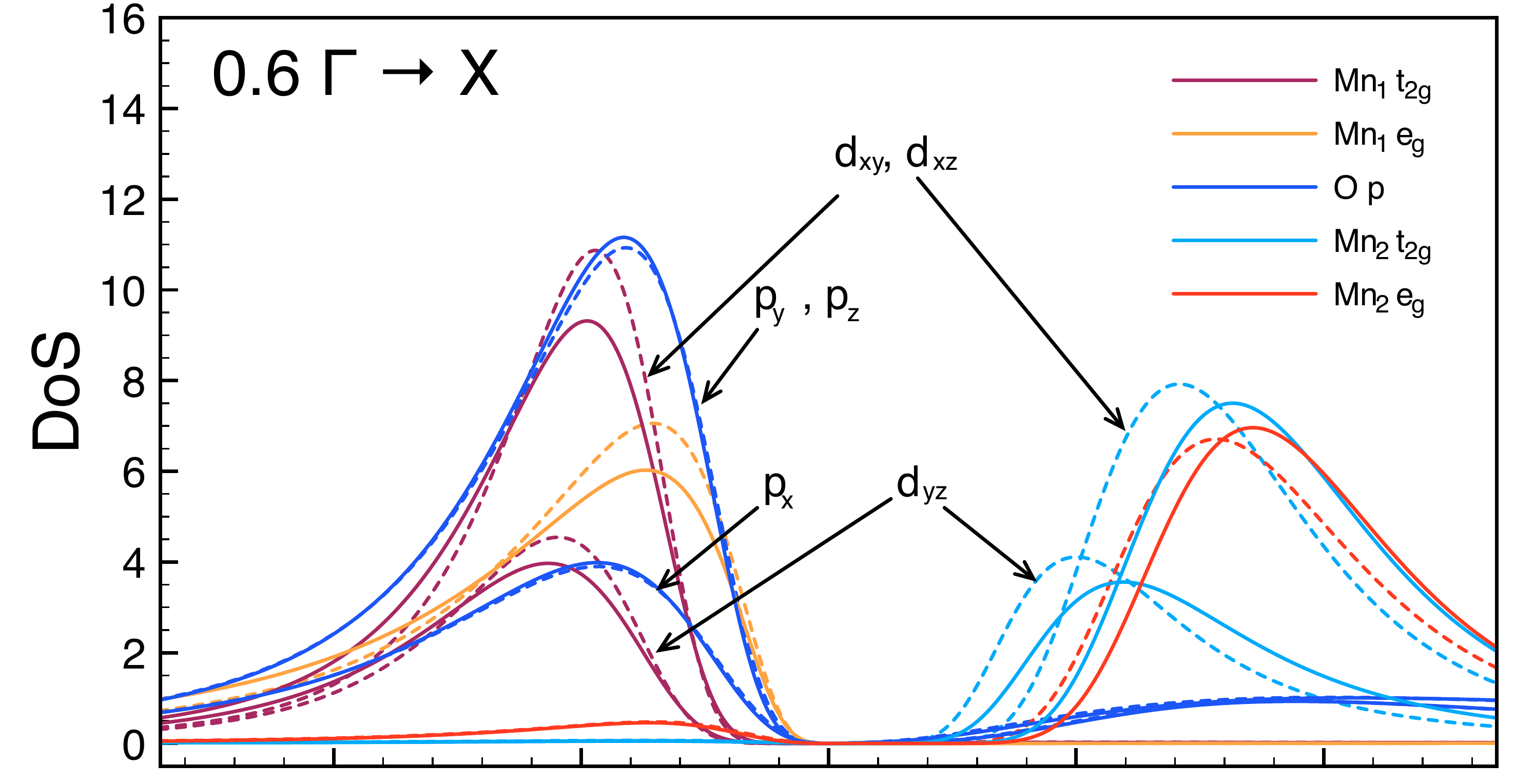}
\includegraphics[width=0.47\textwidth]{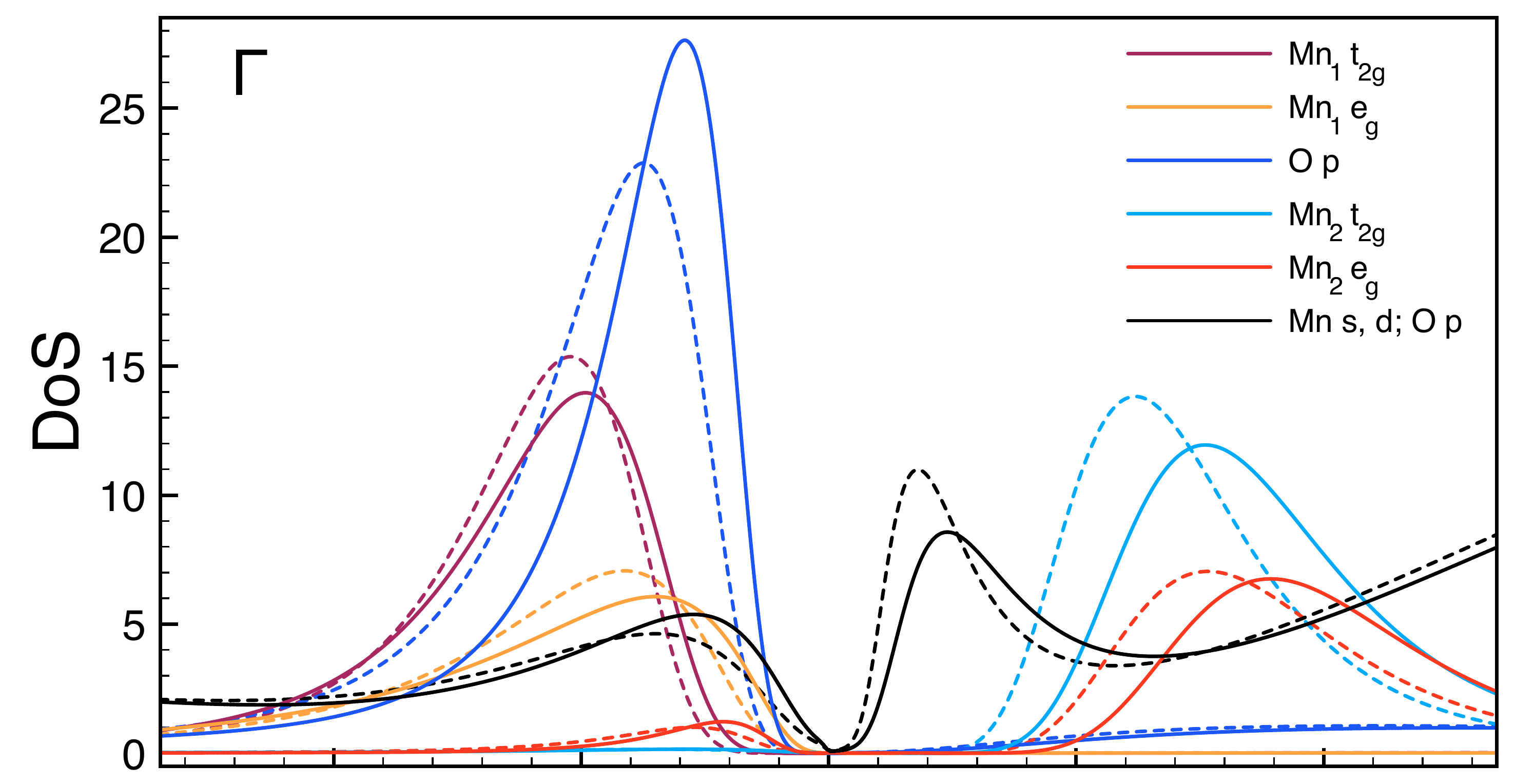}
\includegraphics[width=0.47\textwidth]{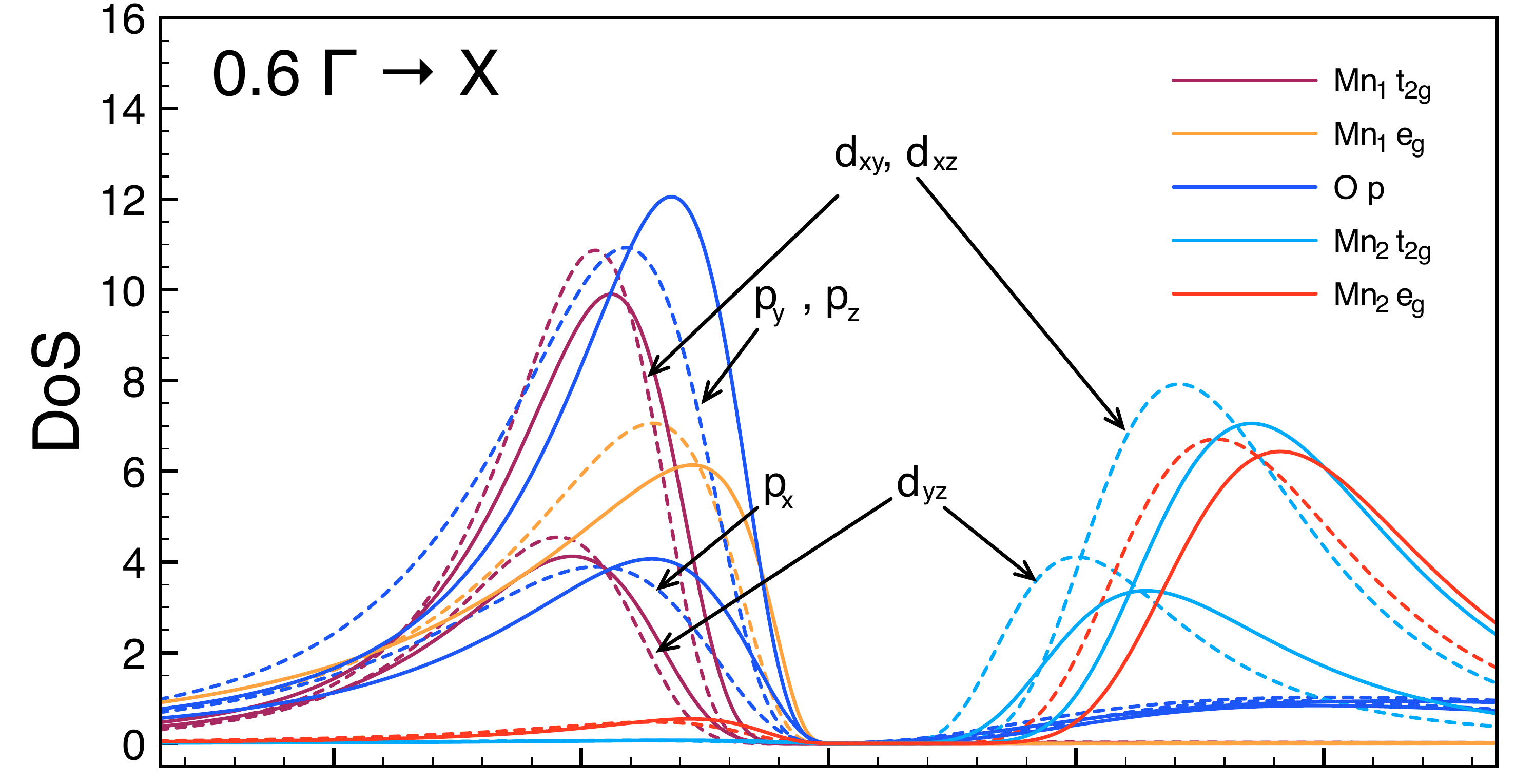}
\includegraphics[width=0.47\textwidth]{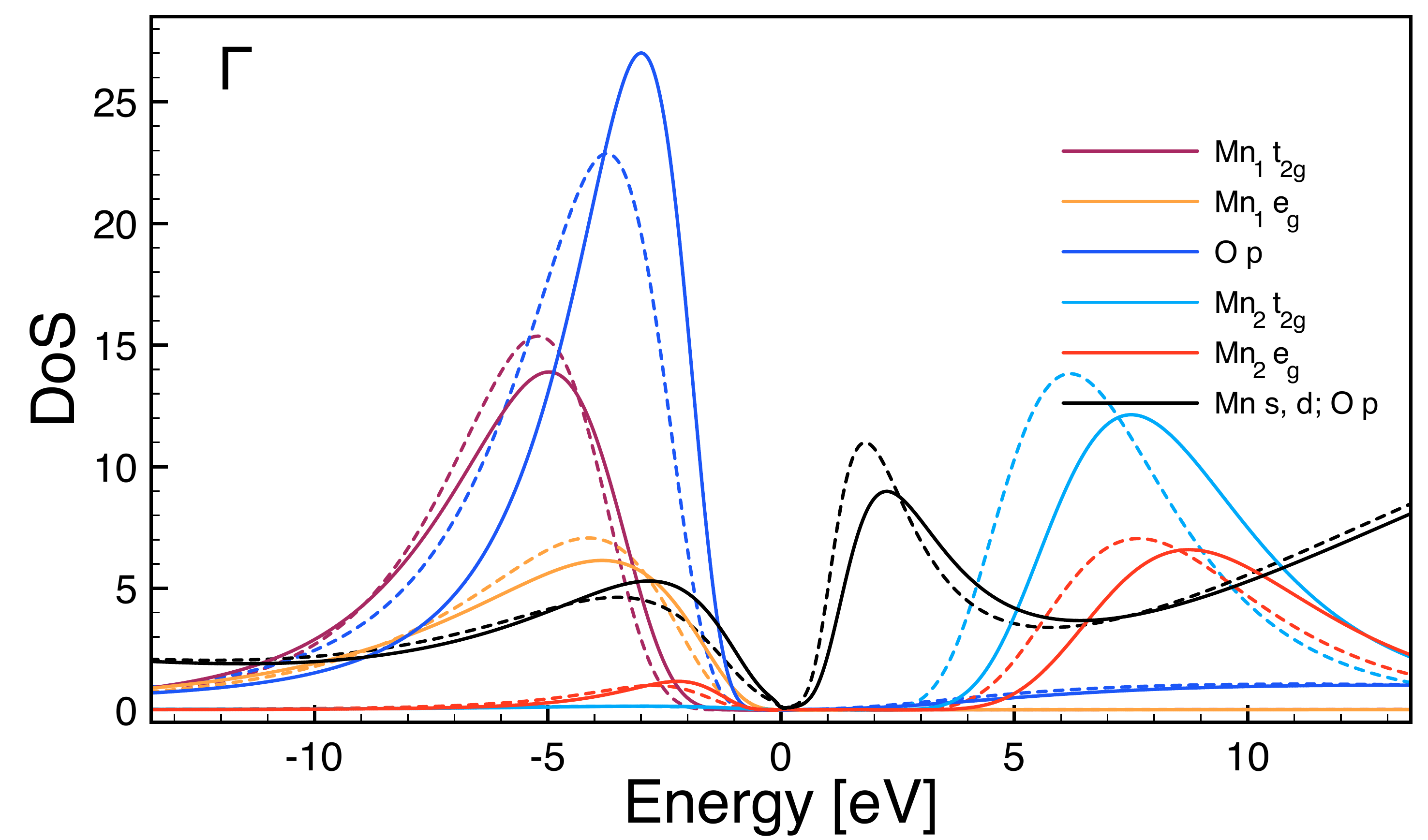}
\includegraphics[width=0.47\textwidth]{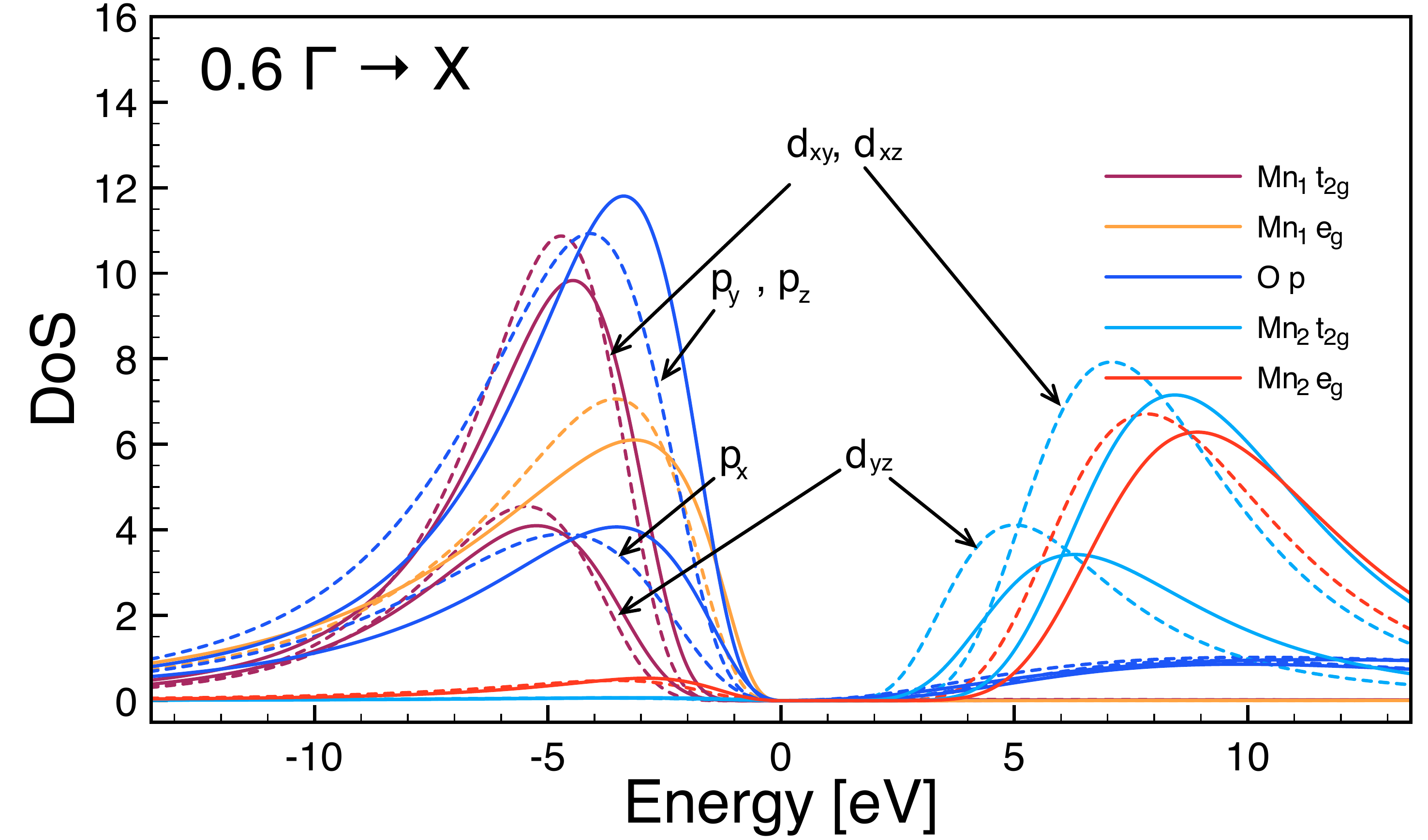}
\caption{$\tk$-resolved SEET (solid lines) and GW (dashed lines) spectral functions for MnO at $\Gamma$ (left column) and at $\frac{2}{3}$ distance
between $\Gamma$ and $X$  (right column) for the impurity choices of Table~\ref{tab:impurities_MnO}.
Different impurities were chosen in each of the rows. These impurity choices correspond to the rows
of Table~\ref{tab:impurities_MnO}, with first-to-fourth rows corresponding to impurities chosen in a-d rows of Table~\ref{tab:impurities_MnO}, respectively.}

\label{Fig:MnOSEETvsGW}
\end{figure*}

\subsection{Effect of strong correlations in MnO}

\subsubsection{Local DOS for MnO}

For MnO solid, similarly to NiO, we solve the problem in symmetrical
orthogonal orbitals and, after identification of the relevant orbitals, choose a set of disjoint impurities
for higher order treatment within SEET. Active space choices a,b and c (see Table~\ref{tab:impurities_MnO})
correspond to a, c and d in Table~\ref{tab:impurities_NiO}. For MnO, active space d combines some of the Mn $e_g$ orbitals
with neighboring oxygen $p$ states.
\begingroup
\squeezetable
\begin{table}[tbh]
\begin{ruledtabular}
\begin{tabular}{c|c|c|p{7.2cm}}
Name & Imp & Orb & Description \\
\hline
a & 2 & 2 & Mn$_1$ $e_g$; Mn$_2$ $e_g$ \\
b & 4 & 3 & Mn$_1$ $e_g$; Mn$_2$ $e_g$; Mn$_1$ $t_{2g}$; Mn$_2$ $t_{2g}$ \\
c & 6 & 3 & Mn$_1$ $e_g$; Mn$_2$ $e_g$; Mn$_1$ $t_{2g}$; Mn$_2$ $t_{2g}$; O$_1$ $p$; O$_2$ $p$ \\
\multirow{2}{*}{d}  & \multirow{2}{*}{6} & \multirow{2}{*}{3} & Mn$_1$$d_{z^2}$ + O$_1$ $p_z$; Mn$_1$$d_{x^2-y^2}$ + O$_1$ $p_x,p_y$; Mn$_1$ $t_{2g}$\\
 & & & Mn$_2$$d_{z^2}$ + O$_2$ $p_z$; Mn$_2$$d_{x^2-y^2}$ + O$_2$ $p_x,p_y$; Mn$_2$ $t_{2g}$\\
\end{tabular}
\end{ruledtabular}
\caption{Choice of the active space for MnO. Imp denotes the number of distinct disjoint impurity problems. Orb
stands for the number of impurity orbitals in the largest impurity problem.}\label{tab:impurities_MnO}
\end{table}
\endgroup

We emphasize that while an a-priori identification of the best orbital combination may in principle be possible, here we make
our orbital choice based on physical/chemical intuition and the available data from GW calculation that precedes SEET.

Fig.~\ref{Fig:MnOLocal_SEETvsGW} shows the orbitally resolved local spectral function of MnO for the four impurity choices
of Table~\ref{tab:impurities_MnO}.

The panel a) of Fig.~\ref{Fig:MnOLocal_SEETvsGW} shows that similarly to the case of NiO, there is a significant adjustment
of the Mn $e_g$ orbitals when embedding is performed. Adding only Mn $t_{2g}$, as illustrated in panel b),
keeps $e_g$ states unchanged and introduces additional renormalization in $t_{2g}$ states, as expected.
However, adding oxygen $p$ orbitals (as shown in panel c) has a large effect
and leads to adjustment of all the bands, both $e_g$ and $t_{2g}$. Embedding active orbitals from subset d (in panel d) with
combined Mn and O orbitals has little effect when compared to panel c.

In Ref.~\cite{MnO_expt_Elp91}, to define the band gap from experimental XPS and BIS data,  
the top of the valence band is taken at 50\% of the intensity of the shoulder and the end of the gap is defined at 10\% intensity
of the rising Mn 3d structure. This yields the experimental band gap of 3.9~$\pm$~0.4~eV.
The GW band gap using the same method of evaluating it as in experiment is approximately equal to around 5.0~eV. Note that
as we discussed previously, the GW result itself displays finite size effects that, on a $6$~$\times$~$6$~$\times$~$6$ lattice, we estimate to be around $0.5$~$-$~$0.7$~eV. After accounting for these effects, we estimate the GW gap to be between $4.3$~$-$~$4.5$~eV.
SEET inherits the finite size and yields the band gap of $4.8$~$-$~$5.0$~eV after accounting for them.

\subsubsection{Momentum resolved DOS for MnO}

Fig.~\ref{Fig:MnOSEETvsGW} shows $\tk$-resolved spectral functions at $\Gamma$ (left column) and at $\frac{2}{3}$ distance between $\Gamma$ and $X$
(right column) for the impurity choices of Table~\ref{tab:impurities_MnO}. 
Note that similarly to our previous discussion concerning local DOS of MnO inclusion of both $e_g$ and $t_{2g}$ orbitals into
embedding subspace (shown in the second row) leads to a large adjustments of these bands. However, adding oxygen $p$, with a
large contribution near the Fermi energy, has even a larger effect resulting in  substantially renormalized bands as shown in
the third row. This is different from the case of NiO, where the inclusion of oxygen orbitals led to much smaller changes.
As observed previously for local DOS, embedding active orbitals from the subset d (4th row) with combined Mn and O orbitals
has little effect when compared to subset c (presented in the 3rd row).

\subsubsection{Local magnetic moment in MnO}

\begin{table}[tbh]
\begin{tabular}{l|ccccccc}
\hline \hline
&\multirow{2}{*}{expt}&\multirow{2}{*}{HF}&\multirow{2}{*}{GW}& \multicolumn{4}{c}{GW+SEET} \\
\cline{5-8}
&&&& a & b & c & d\\
\hline
MnO&4.79,4.58(3)&4.937&4.870&4.887&4.897&4.897&4.897\\
\hline \hline
\end{tabular}
\caption{Local magnetic moment of Mn from Mulliken analysis. Impurity choices a,b,c, and d correspond to the rows
of Table~\ref{tab:impurities_MnO}. The experimental data is obtained from Refs.~\cite{doi:10.1063/1.1668855}~and~\cite{mu_expt_Cheetham83}}\label{tab:mu_MnO}
\end{table}

Staggered magnetic moments from a Mulliken analysis are shown in Table~\ref{tab:mu_MnO}. In contrast to NiO, where only changes due to inclusion of $e_g$ were observed,
the major correction to GW comes here from both Mn $e_g$ and Mn $t_{2g}$ orbitals to an equal degree.
This effect can be explained by the fact that in MnO both $t_{2g}$ and $e_g$ states are partially occupied,
whereas in NiO only $e_g$ states contribute to the local magnetization since the $t_{2g}$ ones are occupied. Correlations
from oxygen do not influence the population on Mn sufficiently to have an influence on the magnetization.

\section{Conclusions}\label{sec:conclusions}

We have presented results from our implementation of the self-energy embedding theory for realistic materials.
For both solid NiO and MnO, our results agreed with experimental data reasonably well and were used to assign orbital
character to the local DOS and ARPES spectra.

SEET is thermodynamically consistent and conserving and, after the selection of active orbitals that is done on the basis
of a previous weakly correlated calculation, it does not contain any ad-hoc choices of parameters such as a choice of density
functional, downfolding scheme, double counting correction, or ad-hoc truncation and re-adjustment of screened interactions.

We have shown that in SEET, we do not need to rely on quantum impurity constructions with more than a few orbitals. While
treatment of large impurity problems is possible using modern quantum chemistry impurity solvers such as zero temperature
Coupled Cluster solvers, the results of such a treatment may not be correct when correlations within the impurity are strong.
In SEET, the size of the impurity is moderately small, since much of the weakly correlated physics including screening at
the level of GW is absorbed properly when the embedding condition from Eq.~\ref{eq:seet_embedding_condition} is defined.

At present,  when running SEET, some of the methodological aspects still require physical insight (such as the choice of
active spaces), or suffer from technical limitations (such as the analytic continuation step), or are otherwise computationally
expensive  (such as the simulation with larger Gaussian bases or more momentum points); however, we believe that despite these short
term technical limitations SEET is a practicable embedding theory that can be applied to interesting correlated materials.

\acknowledgments{
    We thank Runxue Yu for performing density functional calculations for our systems. SI and EG are supported by the Simons foundation via the
    Simons Collaboration on the Many-Electron Problem, DZ and CY by the US Department of Energy (DOE) grant No. ER16391. This research used
    resources of the National Energy Research Scientific Computing Center (NERSC), a U.S. Department of Energy Office of Science
    User Facility operated under Contract No. DE-AC02-05CH11231.
}

\bibliographystyle{apsrev4-2}
\bibliography{refs}

\begin{thebibliography}{111}%
\makeatletter
\providecommand \@ifxundefined [1]{%
 \@ifx{#1\undefined}
}%
\providecommand \@ifnum [1]{%
 \ifnum #1\expandafter \@firstoftwo
 \else \expandafter \@secondoftwo
 \fi
}%
\providecommand \@ifx [1]{%
 \ifx #1\expandafter \@firstoftwo
 \else \expandafter \@secondoftwo
 \fi
}%
\providecommand \natexlab [1]{#1}%
\providecommand \enquote  [1]{``#1''}%
\providecommand \bibnamefont  [1]{#1}%
\providecommand \bibfnamefont [1]{#1}%
\providecommand \citenamefont [1]{#1}%
\providecommand \href@noop [0]{\@secondoftwo}%
\providecommand \href [0]{\begingroup \@sanitize@url \@href}%
\providecommand \@href[1]{\@@startlink{#1}\@@href}%
\providecommand \@@href[1]{\endgroup#1\@@endlink}%
\providecommand \@sanitize@url [0]{\catcode `\\12\catcode `\$12\catcode
  `\&12\catcode `\#12\catcode `\^12\catcode `\_12\catcode `\%12\relax}%
\providecommand \@@startlink[1]{}%
\providecommand \@@endlink[0]{}%
\providecommand \url  [0]{\begingroup\@sanitize@url \@url }%
\providecommand \@url [1]{\endgroup\@href {#1}{\urlprefix }}%
\providecommand \urlprefix  [0]{URL }%
\providecommand \Eprint [0]{\href }%
\providecommand \doibase [0]{https://doi.org/}%
\providecommand \selectlanguage [0]{\@gobble}%
\providecommand \bibinfo  [0]{\@secondoftwo}%
\providecommand \bibfield  [0]{\@secondoftwo}%
\providecommand \translation [1]{[#1]}%
\providecommand \BibitemOpen [0]{}%
\providecommand \bibitemStop [0]{}%
\providecommand \bibitemNoStop [0]{.\EOS\space}%
\providecommand \EOS [0]{\spacefactor3000\relax}%
\providecommand \BibitemShut  [1]{\csname bibitem#1\endcsname}%
\let\auto@bib@innerbib\@empty
\bibitem [{\citenamefont {Georges}\ and\ \citenamefont
  {Kotliar}(1992)}]{DMFT_infinite_dim_Georges_Kotliar_1992}%
  \BibitemOpen
  \bibfield  {author} {\bibinfo {author} {\bibfnamefont {A.}~\bibnamefont
  {Georges}}\ and\ \bibinfo {author} {\bibfnamefont {G.}~\bibnamefont
  {Kotliar}},\ }\href {https://doi.org/10.1103/PhysRevB.45.6479} {\bibfield
  {journal} {\bibinfo  {journal} {Phys. Rev. B}\ }\textbf {\bibinfo {volume}
  {45}},\ \bibinfo {pages} {6479} (\bibinfo {year} {1992})}\BibitemShut
  {NoStop}%
\bibitem [{\citenamefont {Metzner}\ and\ \citenamefont
  {Vollhardt}(1989)}]{Hubbard_inf_dim_Vollhardt_Metzner}%
  \BibitemOpen
  \bibfield  {author} {\bibinfo {author} {\bibfnamefont {W.}~\bibnamefont
  {Metzner}}\ and\ \bibinfo {author} {\bibfnamefont {D.}~\bibnamefont
  {Vollhardt}},\ }\href {https://doi.org/10.1103/PhysRevLett.62.324} {\bibfield
   {journal} {\bibinfo  {journal} {Phys. Rev. Lett.}\ }\textbf {\bibinfo
  {volume} {62}},\ \bibinfo {pages} {324} (\bibinfo {year} {1989})}\BibitemShut
  {NoStop}%
\bibitem [{\citenamefont {Georges}\ \emph {et~al.}(1996)\citenamefont
  {Georges}, \citenamefont {Kotliar}, \citenamefont {Krauth},\ and\
  \citenamefont {Rozenberg}}]{Georges96}%
  \BibitemOpen
  \bibfield  {author} {\bibinfo {author} {\bibfnamefont {A.}~\bibnamefont
  {Georges}}, \bibinfo {author} {\bibfnamefont {G.}~\bibnamefont {Kotliar}},
  \bibinfo {author} {\bibfnamefont {W.}~\bibnamefont {Krauth}},\ and\ \bibinfo
  {author} {\bibfnamefont {M.~J.}\ \bibnamefont {Rozenberg}},\ }\href
  {https://doi.org/10.1103/RevModPhys.68.13} {\bibfield  {journal} {\bibinfo
  {journal} {Rev. Mod. Phys.}\ }\textbf {\bibinfo {volume} {68}},\ \bibinfo
  {pages} {13} (\bibinfo {year} {1996})}\BibitemShut {NoStop}%
\bibitem [{\citenamefont {Anisimov}\ \emph
  {et~al.}(1997{\natexlab{a}})\citenamefont {Anisimov}, \citenamefont
  {Poteryaev}, \citenamefont {Korotin}, \citenamefont {Anokhin},\ and\
  \citenamefont {Kotliar}}]{LDAplusDMFT_Anisimov1997}%
  \BibitemOpen
  \bibfield  {author} {\bibinfo {author} {\bibfnamefont {V.~I.}\ \bibnamefont
  {Anisimov}}, \bibinfo {author} {\bibfnamefont {A.~I.}\ \bibnamefont
  {Poteryaev}}, \bibinfo {author} {\bibfnamefont {M.~A.}\ \bibnamefont
  {Korotin}}, \bibinfo {author} {\bibfnamefont {A.~O.}\ \bibnamefont
  {Anokhin}},\ and\ \bibinfo {author} {\bibfnamefont {G.}~\bibnamefont
  {Kotliar}},\ }\href {https://doi.org/10.1088/0953-8984/9/35/010} {\bibfield
  {journal} {\bibinfo  {journal} {Journal of Physics: Condensed Matter}\
  }\textbf {\bibinfo {volume} {9}},\ \bibinfo {pages} {7359} (\bibinfo {year}
  {1997}{\natexlab{a}})}\BibitemShut {NoStop}%
\bibitem [{\citenamefont {Lichtenstein}\ and\ \citenamefont
  {Katsnelson}(1998)}]{Lichtenstein98}%
  \BibitemOpen
  \bibfield  {author} {\bibinfo {author} {\bibfnamefont {A.~I.}\ \bibnamefont
  {Lichtenstein}}\ and\ \bibinfo {author} {\bibfnamefont {M.~I.}\ \bibnamefont
  {Katsnelson}},\ }\href {https://doi.org/10.1103/PhysRevB.57.6884} {\bibfield
  {journal} {\bibinfo  {journal} {Phys. Rev. B}\ }\textbf {\bibinfo {volume}
  {57}},\ \bibinfo {pages} {6884} (\bibinfo {year} {1998})}\BibitemShut
  {NoStop}%
\bibitem [{\citenamefont {Kotliar}\ \emph {et~al.}(2006)\citenamefont
  {Kotliar}, \citenamefont {Savrasov}, \citenamefont {Haule}, \citenamefont
  {Oudovenko}, \citenamefont {Parcollet},\ and\ \citenamefont
  {Marianetti}}]{Kotliar06}%
  \BibitemOpen
  \bibfield  {author} {\bibinfo {author} {\bibfnamefont {G.}~\bibnamefont
  {Kotliar}}, \bibinfo {author} {\bibfnamefont {S.~Y.}\ \bibnamefont
  {Savrasov}}, \bibinfo {author} {\bibfnamefont {K.}~\bibnamefont {Haule}},
  \bibinfo {author} {\bibfnamefont {V.~S.}\ \bibnamefont {Oudovenko}}, \bibinfo
  {author} {\bibfnamefont {O.}~\bibnamefont {Parcollet}},\ and\ \bibinfo
  {author} {\bibfnamefont {C.~A.}\ \bibnamefont {Marianetti}},\ }\href
  {https://doi.org/10.1103/RevModPhys.78.865} {\bibfield  {journal} {\bibinfo
  {journal} {Rev. Mod. Phys.}\ }\textbf {\bibinfo {volume} {78}},\ \bibinfo
  {pages} {865} (\bibinfo {year} {2006})}\BibitemShut {NoStop}%
\bibitem [{\citenamefont {Rusakov}\ and\ \citenamefont
  {Zgid}(2016)}]{GF2_Alexander16}%
  \BibitemOpen
  \bibfield  {author} {\bibinfo {author} {\bibfnamefont {A.~A.}\ \bibnamefont
  {Rusakov}}\ and\ \bibinfo {author} {\bibfnamefont {D.}~\bibnamefont {Zgid}},\
  }\href {https://doi.org/10.1063/1.4940900} {\bibfield  {journal} {\bibinfo
  {journal} {The Journal of Chemical Physics}\ }\textbf {\bibinfo {volume}
  {144}},\ \bibinfo {pages} {054106} (\bibinfo {year} {2016})}\BibitemShut
  {NoStop}%
\bibitem [{\citenamefont {Iskakov}\ \emph {et~al.}(2019)\citenamefont
  {Iskakov}, \citenamefont {Rusakov}, \citenamefont {Zgid},\ and\ \citenamefont
  {Gull}}]{GF2_Sergei19}%
  \BibitemOpen
  \bibfield  {author} {\bibinfo {author} {\bibfnamefont {S.}~\bibnamefont
  {Iskakov}}, \bibinfo {author} {\bibfnamefont {A.~A.}\ \bibnamefont
  {Rusakov}}, \bibinfo {author} {\bibfnamefont {D.}~\bibnamefont {Zgid}},\ and\
  \bibinfo {author} {\bibfnamefont {E.}~\bibnamefont {Gull}},\ }\href
  {https://doi.org/10.1103/PhysRevB.100.085112} {\bibfield  {journal} {\bibinfo
   {journal} {Phys. Rev. B}\ }\textbf {\bibinfo {volume} {100}},\ \bibinfo
  {pages} {085112} (\bibinfo {year} {2019})}\BibitemShut {NoStop}%
\bibitem [{\citenamefont {Pickett}\ and\ \citenamefont
  {Wang}(1984)}]{G0W0_Pickett84}%
  \BibitemOpen
  \bibfield  {author} {\bibinfo {author} {\bibfnamefont {W.~E.}\ \bibnamefont
  {Pickett}}\ and\ \bibinfo {author} {\bibfnamefont {C.~S.}\ \bibnamefont
  {Wang}},\ }\href {https://doi.org/10.1103/PhysRevB.30.4719} {\bibfield
  {journal} {\bibinfo  {journal} {Phys. Rev. B}\ }\textbf {\bibinfo {volume}
  {30}},\ \bibinfo {pages} {4719} (\bibinfo {year} {1984})}\BibitemShut
  {NoStop}%
\bibitem [{\citenamefont {Hybertsen}\ and\ \citenamefont
  {Louie}(1986)}]{G0W0_Hybertsen86}%
  \BibitemOpen
  \bibfield  {author} {\bibinfo {author} {\bibfnamefont {M.~S.}\ \bibnamefont
  {Hybertsen}}\ and\ \bibinfo {author} {\bibfnamefont {S.~G.}\ \bibnamefont
  {Louie}},\ }\href {https://doi.org/10.1103/PhysRevB.34.5390} {\bibfield
  {journal} {\bibinfo  {journal} {Phys. Rev. B}\ }\textbf {\bibinfo {volume}
  {34}},\ \bibinfo {pages} {5390} (\bibinfo {year} {1986})}\BibitemShut
  {NoStop}%
\bibitem [{\citenamefont {Aryasetiawan}\ and\ \citenamefont
  {Gunnarsson}(1998)}]{GW_Aryasetiawan98}%
  \BibitemOpen
  \bibfield  {author} {\bibinfo {author} {\bibfnamefont {F.}~\bibnamefont
  {Aryasetiawan}}\ and\ \bibinfo {author} {\bibfnamefont {O.}~\bibnamefont
  {Gunnarsson}},\ }\href {https://doi.org/10.1088/0034-4885/61/3/002}
  {\bibfield  {journal} {\bibinfo  {journal} {Reports on Progress in Physics}\
  }\textbf {\bibinfo {volume} {61}},\ \bibinfo {pages} {237} (\bibinfo {year}
  {1998})}\BibitemShut {NoStop}%
\bibitem [{\citenamefont {Kotani}\ \emph {et~al.}(2007)\citenamefont {Kotani},
  \citenamefont {van Schilfgaarde},\ and\ \citenamefont
  {Faleev}}]{QSGW_Kotani07}%
  \BibitemOpen
  \bibfield  {author} {\bibinfo {author} {\bibfnamefont {T.}~\bibnamefont
  {Kotani}}, \bibinfo {author} {\bibfnamefont {M.}~\bibnamefont {van
  Schilfgaarde}},\ and\ \bibinfo {author} {\bibfnamefont {S.~V.}\ \bibnamefont
  {Faleev}},\ }\href {https://doi.org/10.1103/PhysRevB.76.165106} {\bibfield
  {journal} {\bibinfo  {journal} {Phys. Rev. B}\ }\textbf {\bibinfo {volume}
  {76}},\ \bibinfo {pages} {165106} (\bibinfo {year} {2007})}\BibitemShut
  {NoStop}%
\bibitem [{\citenamefont {Kutepov}\ \emph {et~al.}(2009)\citenamefont
  {Kutepov}, \citenamefont {Savrasov},\ and\ \citenamefont
  {Kotliar}}]{scGW_Andrey09}%
  \BibitemOpen
  \bibfield  {author} {\bibinfo {author} {\bibfnamefont {A.}~\bibnamefont
  {Kutepov}}, \bibinfo {author} {\bibfnamefont {S.~Y.}\ \bibnamefont
  {Savrasov}},\ and\ \bibinfo {author} {\bibfnamefont {G.}~\bibnamefont
  {Kotliar}},\ }\href {https://doi.org/10.1103/PhysRevB.80.041103} {\bibfield
  {journal} {\bibinfo  {journal} {Phys. Rev. B}\ }\textbf {\bibinfo {volume}
  {80}},\ \bibinfo {pages} {041103(R)} (\bibinfo {year} {2009})}\BibitemShut
  {NoStop}%
\bibitem [{\citenamefont {Smith}\ and\ \citenamefont
  {Si}(2000)}]{EDMFT_Qimiao_2000}%
  \BibitemOpen
  \bibfield  {author} {\bibinfo {author} {\bibfnamefont {J.~L.}\ \bibnamefont
  {Smith}}\ and\ \bibinfo {author} {\bibfnamefont {Q.}~\bibnamefont {Si}},\
  }\href {https://doi.org/10.1103/PhysRevB.61.5184} {\bibfield  {journal}
  {\bibinfo  {journal} {Phys. Rev. B}\ }\textbf {\bibinfo {volume} {61}},\
  \bibinfo {pages} {5184} (\bibinfo {year} {2000})}\BibitemShut {NoStop}%
\bibitem [{\citenamefont {Chitra}\ and\ \citenamefont
  {Kotliar}(2001)}]{PhysRevB.63.115110}%
  \BibitemOpen
  \bibfield  {author} {\bibinfo {author} {\bibfnamefont {R.}~\bibnamefont
  {Chitra}}\ and\ \bibinfo {author} {\bibfnamefont {G.}~\bibnamefont
  {Kotliar}},\ }\href {https://doi.org/10.1103/PhysRevB.63.115110} {\bibfield
  {journal} {\bibinfo  {journal} {Phys. Rev. B}\ }\textbf {\bibinfo {volume}
  {63}},\ \bibinfo {pages} {115110} (\bibinfo {year} {2001})}\BibitemShut
  {NoStop}%
\bibitem [{\citenamefont {Si}\ and\ \citenamefont
  {Smith}(1996)}]{PhysRevLett.77.3391}%
  \BibitemOpen
  \bibfield  {author} {\bibinfo {author} {\bibfnamefont {Q.}~\bibnamefont
  {Si}}\ and\ \bibinfo {author} {\bibfnamefont {J.~L.}\ \bibnamefont {Smith}},\
  }\href {https://doi.org/10.1103/PhysRevLett.77.3391} {\bibfield  {journal}
  {\bibinfo  {journal} {Phys. Rev. Lett.}\ }\textbf {\bibinfo {volume} {77}},\
  \bibinfo {pages} {3391} (\bibinfo {year} {1996})}\BibitemShut {NoStop}%
\bibitem [{\citenamefont {Sun}\ and\ \citenamefont
  {Kotliar}(2002)}]{GWplusEDMFT_Sun02}%
  \BibitemOpen
  \bibfield  {author} {\bibinfo {author} {\bibfnamefont {P.}~\bibnamefont
  {Sun}}\ and\ \bibinfo {author} {\bibfnamefont {G.}~\bibnamefont {Kotliar}},\
  }\href {https://doi.org/10.1103/PhysRevB.66.085120} {\bibfield  {journal}
  {\bibinfo  {journal} {Phys. Rev. B}\ }\textbf {\bibinfo {volume} {66}},\
  \bibinfo {pages} {085120} (\bibinfo {year} {2002})}\BibitemShut {NoStop}%
\bibitem [{\citenamefont {Biermann}\ \emph {et~al.}(2003)\citenamefont
  {Biermann}, \citenamefont {Aryasetiawan},\ and\ \citenamefont
  {Georges}}]{PhysRevLett.90.086402}%
  \BibitemOpen
  \bibfield  {author} {\bibinfo {author} {\bibfnamefont {S.}~\bibnamefont
  {Biermann}}, \bibinfo {author} {\bibfnamefont {F.}~\bibnamefont
  {Aryasetiawan}},\ and\ \bibinfo {author} {\bibfnamefont {A.}~\bibnamefont
  {Georges}},\ }\href {https://doi.org/10.1103/PhysRevLett.90.086402}
  {\bibfield  {journal} {\bibinfo  {journal} {Phys. Rev. Lett.}\ }\textbf
  {\bibinfo {volume} {90}},\ \bibinfo {pages} {086402} (\bibinfo {year}
  {2003})}\BibitemShut {NoStop}%
\bibitem [{\citenamefont {Sakuma}\ \emph {et~al.}(2013)\citenamefont {Sakuma},
  \citenamefont {Werner},\ and\ \citenamefont
  {Aryasetiawan}}]{PhysRevB.88.235110}%
  \BibitemOpen
  \bibfield  {author} {\bibinfo {author} {\bibfnamefont {R.}~\bibnamefont
  {Sakuma}}, \bibinfo {author} {\bibfnamefont {P.}~\bibnamefont {Werner}},\
  and\ \bibinfo {author} {\bibfnamefont {F.}~\bibnamefont {Aryasetiawan}},\
  }\href {https://doi.org/10.1103/PhysRevB.88.235110} {\bibfield  {journal}
  {\bibinfo  {journal} {Phys. Rev. B}\ }\textbf {\bibinfo {volume} {88}},\
  \bibinfo {pages} {235110} (\bibinfo {year} {2013})}\BibitemShut {NoStop}%
\bibitem [{\citenamefont {Tomczak}\ \emph {et~al.}(2012)\citenamefont
  {Tomczak}, \citenamefont {Casula}, \citenamefont {Miyake}, \citenamefont
  {Aryasetiawan},\ and\ \citenamefont {Biermann}}]{Tomczak_2012}%
  \BibitemOpen
  \bibfield  {author} {\bibinfo {author} {\bibfnamefont {J.~M.}\ \bibnamefont
  {Tomczak}}, \bibinfo {author} {\bibfnamefont {M.}~\bibnamefont {Casula}},
  \bibinfo {author} {\bibfnamefont {T.}~\bibnamefont {Miyake}}, \bibinfo
  {author} {\bibfnamefont {F.}~\bibnamefont {Aryasetiawan}},\ and\ \bibinfo
  {author} {\bibfnamefont {S.}~\bibnamefont {Biermann}},\ }\href
  {https://doi.org/10.1209/0295-5075/100/67001} {\bibfield  {journal} {\bibinfo
   {journal} {{EPL} (Europhysics Letters)}\ }\textbf {\bibinfo {volume}
  {100}},\ \bibinfo {pages} {67001} (\bibinfo {year} {2012})}\BibitemShut
  {NoStop}%
\bibitem [{\citenamefont {Ayral}\ \emph {et~al.}(2013)\citenamefont {Ayral},
  \citenamefont {Biermann},\ and\ \citenamefont {Werner}}]{PhysRevB.87.125149}%
  \BibitemOpen
  \bibfield  {author} {\bibinfo {author} {\bibfnamefont {T.}~\bibnamefont
  {Ayral}}, \bibinfo {author} {\bibfnamefont {S.}~\bibnamefont {Biermann}},\
  and\ \bibinfo {author} {\bibfnamefont {P.}~\bibnamefont {Werner}},\ }\href
  {https://doi.org/10.1103/PhysRevB.87.125149} {\bibfield  {journal} {\bibinfo
  {journal} {Phys. Rev. B}\ }\textbf {\bibinfo {volume} {87}},\ \bibinfo
  {pages} {125149} (\bibinfo {year} {2013})}\BibitemShut {NoStop}%
\bibitem [{\citenamefont {Huang}\ \emph {et~al.}(2014)\citenamefont {Huang},
  \citenamefont {Ayral}, \citenamefont {Biermann},\ and\ \citenamefont
  {Werner}}]{PhysRevB.90.195114}%
  \BibitemOpen
  \bibfield  {author} {\bibinfo {author} {\bibfnamefont {L.}~\bibnamefont
  {Huang}}, \bibinfo {author} {\bibfnamefont {T.}~\bibnamefont {Ayral}},
  \bibinfo {author} {\bibfnamefont {S.}~\bibnamefont {Biermann}},\ and\
  \bibinfo {author} {\bibfnamefont {P.}~\bibnamefont {Werner}},\ }\href
  {https://doi.org/10.1103/PhysRevB.90.195114} {\bibfield  {journal} {\bibinfo
  {journal} {Phys. Rev. B}\ }\textbf {\bibinfo {volume} {90}},\ \bibinfo
  {pages} {195114} (\bibinfo {year} {2014})}\BibitemShut {NoStop}%
\bibitem [{\citenamefont {Leonov}\ \emph {et~al.}(2016)\citenamefont {Leonov},
  \citenamefont {Pourovskii}, \citenamefont {Georges},\ and\ \citenamefont
  {Abrikosov}}]{TMO_MIT_Leonov16}%
  \BibitemOpen
  \bibfield  {author} {\bibinfo {author} {\bibfnamefont {I.}~\bibnamefont
  {Leonov}}, \bibinfo {author} {\bibfnamefont {L.}~\bibnamefont {Pourovskii}},
  \bibinfo {author} {\bibfnamefont {A.}~\bibnamefont {Georges}},\ and\ \bibinfo
  {author} {\bibfnamefont {I.~A.}\ \bibnamefont {Abrikosov}},\ }\href
  {https://doi.org/10.1103/PhysRevB.94.155135} {\bibfield  {journal} {\bibinfo
  {journal} {Phys. Rev. B}\ }\textbf {\bibinfo {volume} {94}},\ \bibinfo
  {pages} {155135} (\bibinfo {year} {2016})}\BibitemShut {NoStop}%
\bibitem [{\citenamefont {Ayral}\ \emph {et~al.}(2017)\citenamefont {Ayral},
  \citenamefont {Biermann}, \citenamefont {Werner},\ and\ \citenamefont
  {Boehnke}}]{PhysRevB.95.245130}%
  \BibitemOpen
  \bibfield  {author} {\bibinfo {author} {\bibfnamefont {T.}~\bibnamefont
  {Ayral}}, \bibinfo {author} {\bibfnamefont {S.}~\bibnamefont {Biermann}},
  \bibinfo {author} {\bibfnamefont {P.}~\bibnamefont {Werner}},\ and\ \bibinfo
  {author} {\bibfnamefont {L.}~\bibnamefont {Boehnke}},\ }\href
  {https://doi.org/10.1103/PhysRevB.95.245130} {\bibfield  {journal} {\bibinfo
  {journal} {Phys. Rev. B}\ }\textbf {\bibinfo {volume} {95}},\ \bibinfo
  {pages} {245130} (\bibinfo {year} {2017})}\BibitemShut {NoStop}%
\bibitem [{\citenamefont {Lee}\ and\ \citenamefont
  {Haule}(2017)}]{haule_lee_h2_2017}%
  \BibitemOpen
  \bibfield  {author} {\bibinfo {author} {\bibfnamefont {J.}~\bibnamefont
  {Lee}}\ and\ \bibinfo {author} {\bibfnamefont {K.}~\bibnamefont {Haule}},\
  }\href {https://doi.org/10.1103/PhysRevB.95.155104} {\bibfield  {journal}
  {\bibinfo  {journal} {Phys. Rev. B}\ }\textbf {\bibinfo {volume} {95}},\
  \bibinfo {pages} {155104} (\bibinfo {year} {2017})}\BibitemShut {NoStop}%
\bibitem [{\citenamefont {Boehnke}\ \emph {et~al.}(2016)\citenamefont
  {Boehnke}, \citenamefont {Nilsson}, \citenamefont {Aryasetiawan},\ and\
  \citenamefont {Werner}}]{Boehnke16}%
  \BibitemOpen
  \bibfield  {author} {\bibinfo {author} {\bibfnamefont {L.}~\bibnamefont
  {Boehnke}}, \bibinfo {author} {\bibfnamefont {F.}~\bibnamefont {Nilsson}},
  \bibinfo {author} {\bibfnamefont {F.}~\bibnamefont {Aryasetiawan}},\ and\
  \bibinfo {author} {\bibfnamefont {P.}~\bibnamefont {Werner}},\ }\href
  {https://doi.org/10.1103/PhysRevB.94.201106} {\bibfield  {journal} {\bibinfo
  {journal} {Phys. Rev. B}\ }\textbf {\bibinfo {volume} {94}},\ \bibinfo
  {pages} {201106(R)} (\bibinfo {year} {2016})}\BibitemShut {NoStop}%
\bibitem [{\citenamefont {Nilsson}\ \emph {et~al.}(2017)\citenamefont
  {Nilsson}, \citenamefont {Boehnke}, \citenamefont {Werner},\ and\
  \citenamefont {Aryasetiawan}}]{multitier_GW+DMFT_werner_2017}%
  \BibitemOpen
  \bibfield  {author} {\bibinfo {author} {\bibfnamefont {F.}~\bibnamefont
  {Nilsson}}, \bibinfo {author} {\bibfnamefont {L.}~\bibnamefont {Boehnke}},
  \bibinfo {author} {\bibfnamefont {P.}~\bibnamefont {Werner}},\ and\ \bibinfo
  {author} {\bibfnamefont {F.}~\bibnamefont {Aryasetiawan}},\ }\href
  {https://doi.org/10.1103/PhysRevMaterials.1.043803} {\bibfield  {journal}
  {\bibinfo  {journal} {Phys. Rev. Materials}\ }\textbf {\bibinfo {volume}
  {1}},\ \bibinfo {pages} {043803} (\bibinfo {year} {2017})}\BibitemShut
  {NoStop}%
\bibitem [{\citenamefont {Choi}\ \emph {et~al.}(2019)\citenamefont {Choi},
  \citenamefont {Semon}, \citenamefont {Kang}, \citenamefont {Kutepov},\ and\
  \citenamefont {Kotliar}}]{ComDMFT_2019}%
  \BibitemOpen
  \bibfield  {author} {\bibinfo {author} {\bibfnamefont {S.}~\bibnamefont
  {Choi}}, \bibinfo {author} {\bibfnamefont {P.}~\bibnamefont {Semon}},
  \bibinfo {author} {\bibfnamefont {B.}~\bibnamefont {Kang}}, \bibinfo {author}
  {\bibfnamefont {A.}~\bibnamefont {Kutepov}},\ and\ \bibinfo {author}
  {\bibfnamefont {G.}~\bibnamefont {Kotliar}},\ }\href
  {https://doi.org/https://doi.org/10.1016/j.cpc.2019.07.003} {\bibfield
  {journal} {\bibinfo  {journal} {Computer Physics Communications}\ }\textbf
  {\bibinfo {volume} {244}},\ \bibinfo {pages} {277 } (\bibinfo {year}
  {2019})}\BibitemShut {NoStop}%
\bibitem [{\citenamefont {Zhu}\ and\ \citenamefont
  {Chan}(2020)}]{chan_zhu_2020_gw_dmft}%
  \BibitemOpen
  \bibfield  {author} {\bibinfo {author} {\bibfnamefont {T.}~\bibnamefont
  {Zhu}}\ and\ \bibinfo {author} {\bibfnamefont {G.~K.-L.}\ \bibnamefont
  {Chan}},\ }\href@noop {} {\bibinfo {title} {Ab initio full cell gw+dmft for
  correlated materials}} (\bibinfo {year} {2020}),\ \Eprint
  {https://arxiv.org/abs/2003.01349} {arXiv:2003.01349 [cond-mat.str-el]}
  \BibitemShut {NoStop}%
\bibitem [{\citenamefont {Chibani}\ \emph {et~al.}(2016)\citenamefont
  {Chibani}, \citenamefont {Ren}, \citenamefont {Scheffler},\ and\
  \citenamefont {Rinke}}]{PhysRevB.93.165106}%
  \BibitemOpen
  \bibfield  {author} {\bibinfo {author} {\bibfnamefont {W.}~\bibnamefont
  {Chibani}}, \bibinfo {author} {\bibfnamefont {X.}~\bibnamefont {Ren}},
  \bibinfo {author} {\bibfnamefont {M.}~\bibnamefont {Scheffler}},\ and\
  \bibinfo {author} {\bibfnamefont {P.}~\bibnamefont {Rinke}},\ }\href
  {https://doi.org/10.1103/PhysRevB.93.165106} {\bibfield  {journal} {\bibinfo
  {journal} {Phys. Rev. B}\ }\textbf {\bibinfo {volume} {93}},\ \bibinfo
  {pages} {165106} (\bibinfo {year} {2016})}\BibitemShut {NoStop}%
\bibitem [{\citenamefont {Kananenka}\ \emph {et~al.}(2015)\citenamefont
  {Kananenka}, \citenamefont {Gull},\ and\ \citenamefont {Zgid}}]{Kananenka15}%
  \BibitemOpen
  \bibfield  {author} {\bibinfo {author} {\bibfnamefont {A.~A.}\ \bibnamefont
  {Kananenka}}, \bibinfo {author} {\bibfnamefont {E.}~\bibnamefont {Gull}},\
  and\ \bibinfo {author} {\bibfnamefont {D.}~\bibnamefont {Zgid}},\ }\href
  {https://doi.org/10.1103/PhysRevB.91.121111} {\bibfield  {journal} {\bibinfo
  {journal} {Phys. Rev. B}\ }\textbf {\bibinfo {volume} {91}},\ \bibinfo
  {pages} {121111(R)} (\bibinfo {year} {2015})}\BibitemShut {NoStop}%
\bibitem [{\citenamefont {Zgid}\ and\ \citenamefont {Gull}(2017)}]{Zgid17}%
  \BibitemOpen
  \bibfield  {author} {\bibinfo {author} {\bibfnamefont {D.}~\bibnamefont
  {Zgid}}\ and\ \bibinfo {author} {\bibfnamefont {E.}~\bibnamefont {Gull}},\
  }\href {https://doi.org/10.1088/1367-2630/aa5d34} {\bibfield  {journal}
  {\bibinfo  {journal} {New Journal of Physics}\ }\textbf {\bibinfo {volume}
  {19}},\ \bibinfo {pages} {023047} (\bibinfo {year} {2017})}\BibitemShut
  {NoStop}%
\bibitem [{\citenamefont {Lan}\ and\ \citenamefont {Zgid}(2017)}]{Lan17}%
  \BibitemOpen
  \bibfield  {author} {\bibinfo {author} {\bibfnamefont {T.~N.}\ \bibnamefont
  {Lan}}\ and\ \bibinfo {author} {\bibfnamefont {D.}~\bibnamefont {Zgid}},\
  }\href {https://doi.org/10.1021/acs.jpclett.7b00689} {\bibfield  {journal}
  {\bibinfo  {journal} {The Journal of Physical Chemistry Letters}\ }\textbf
  {\bibinfo {volume} {8}},\ \bibinfo {pages} {2200} (\bibinfo {year}
  {2017})}\BibitemShut {NoStop}%
\bibitem [{\citenamefont {Hedin}(1965)}]{Hedin1965}%
  \BibitemOpen
  \bibfield  {author} {\bibinfo {author} {\bibfnamefont {L.}~\bibnamefont
  {Hedin}},\ }\href {https://doi.org/10.1103/PhysRev.139.A796} {\bibfield
  {journal} {\bibinfo  {journal} {Phys. Rev.}\ }\textbf {\bibinfo {volume}
  {139}},\ \bibinfo {pages} {A796} (\bibinfo {year} {1965})}\BibitemShut
  {NoStop}%
\bibitem [{\citenamefont {Lan}\ \emph {et~al.}(2015)\citenamefont {Lan},
  \citenamefont {Kananenka},\ and\ \citenamefont {Zgid}}]{Tran_jcp_2015}%
  \BibitemOpen
  \bibfield  {author} {\bibinfo {author} {\bibfnamefont {T.~N.}\ \bibnamefont
  {Lan}}, \bibinfo {author} {\bibfnamefont {A.~A.}\ \bibnamefont {Kananenka}},\
  and\ \bibinfo {author} {\bibfnamefont {D.}~\bibnamefont {Zgid}},\ }\href
  {https://doi.org/10.1063/1.4938562} {\bibfield  {journal} {\bibinfo
  {journal} {The Journal of Chemical Physics}\ }\textbf {\bibinfo {volume}
  {143}},\ \bibinfo {pages} {241102} (\bibinfo {year} {2015})}\BibitemShut
  {NoStop}%
\bibitem [{\citenamefont {Nguyen~Lan}\ \emph {et~al.}(2016)\citenamefont
  {Nguyen~Lan}, \citenamefont {Kananenka},\ and\ \citenamefont
  {Zgid}}]{Tran_jctc_2016}%
  \BibitemOpen
  \bibfield  {author} {\bibinfo {author} {\bibfnamefont {T.}~\bibnamefont
  {Nguyen~Lan}}, \bibinfo {author} {\bibfnamefont {A.~A.}\ \bibnamefont
  {Kananenka}},\ and\ \bibinfo {author} {\bibfnamefont {D.}~\bibnamefont
  {Zgid}},\ }\href {https://doi.org/10.1021/acs.jctc.6b00638} {\bibfield
  {journal} {\bibinfo  {journal} {Journal of Chemical Theory and Computation}\
  }\textbf {\bibinfo {volume} {12}},\ \bibinfo {pages} {4856} (\bibinfo {year}
  {2016})}\BibitemShut {NoStop}%
\bibitem [{\citenamefont {Lan}\ \emph {et~al.}(2017)\citenamefont {Lan},
  \citenamefont {Shee}, \citenamefont {Li}, \citenamefont {Gull},\ and\
  \citenamefont {Zgid}}]{Tran_Shee_2017}%
  \BibitemOpen
  \bibfield  {author} {\bibinfo {author} {\bibfnamefont {T.~N.}\ \bibnamefont
  {Lan}}, \bibinfo {author} {\bibfnamefont {A.}~\bibnamefont {Shee}}, \bibinfo
  {author} {\bibfnamefont {J.}~\bibnamefont {Li}}, \bibinfo {author}
  {\bibfnamefont {E.}~\bibnamefont {Gull}},\ and\ \bibinfo {author}
  {\bibfnamefont {D.}~\bibnamefont {Zgid}},\ }\href
  {https://doi.org/10.1103/PhysRevB.96.155106} {\bibfield  {journal} {\bibinfo
  {journal} {Phys. Rev. B}\ }\textbf {\bibinfo {volume} {96}},\ \bibinfo
  {pages} {155106} (\bibinfo {year} {2017})}\BibitemShut {NoStop}%
\bibitem [{\citenamefont {Tran}\ \emph {et~al.}(2018)\citenamefont {Tran},
  \citenamefont {Iskakov},\ and\ \citenamefont {Zgid}}]{Tran_useet}%
  \BibitemOpen
  \bibfield  {author} {\bibinfo {author} {\bibfnamefont {L.~N.}\ \bibnamefont
  {Tran}}, \bibinfo {author} {\bibfnamefont {S.}~\bibnamefont {Iskakov}},\ and\
  \bibinfo {author} {\bibfnamefont {D.}~\bibnamefont {Zgid}},\ }\href
  {https://doi.org/10.1021/acs.jpclett.8b01754} {\bibfield  {journal} {\bibinfo
   {journal} {The Journal of Physical Chemistry Letters}\ }\textbf {\bibinfo
  {volume} {9}},\ \bibinfo {pages} {4444} (\bibinfo {year} {2018})}\BibitemShut
  {NoStop}%
\bibitem [{\citenamefont {Motta}\ \emph {et~al.}(2017)\citenamefont {Motta},
  \citenamefont {Ceperley}, \citenamefont {Chan}, \citenamefont {Gomez},
  \citenamefont {Gull}, \citenamefont {Guo}, \citenamefont {Jim\'enez-Hoyos},
  \citenamefont {Lan}, \citenamefont {Li}, \citenamefont {Ma}, \citenamefont
  {Millis}, \citenamefont {Prokof'ev}, \citenamefont {Ray}, \citenamefont
  {Scuseria}, \citenamefont {Sorella}, \citenamefont {Stoudenmire},
  \citenamefont {Sun}, \citenamefont {Tupitsyn}, \citenamefont {White},
  \citenamefont {Zgid},\ and\ \citenamefont {Zhang}}]{PhysRevX.7.031059}%
  \BibitemOpen
  \bibfield  {author} {\bibinfo {author} {\bibfnamefont {M.}~\bibnamefont
  {Motta}}, \bibinfo {author} {\bibfnamefont {D.~M.}\ \bibnamefont {Ceperley}},
  \bibinfo {author} {\bibfnamefont {G.~K.-L.}\ \bibnamefont {Chan}}, \bibinfo
  {author} {\bibfnamefont {J.~A.}\ \bibnamefont {Gomez}}, \bibinfo {author}
  {\bibfnamefont {E.}~\bibnamefont {Gull}}, \bibinfo {author} {\bibfnamefont
  {S.}~\bibnamefont {Guo}}, \bibinfo {author} {\bibfnamefont {C.~A.}\
  \bibnamefont {Jim\'enez-Hoyos}}, \bibinfo {author} {\bibfnamefont {T.~N.}\
  \bibnamefont {Lan}}, \bibinfo {author} {\bibfnamefont {J.}~\bibnamefont
  {Li}}, \bibinfo {author} {\bibfnamefont {F.}~\bibnamefont {Ma}}, \bibinfo
  {author} {\bibfnamefont {A.~J.}\ \bibnamefont {Millis}}, \bibinfo {author}
  {\bibfnamefont {N.~V.}\ \bibnamefont {Prokof'ev}}, \bibinfo {author}
  {\bibfnamefont {U.}~\bibnamefont {Ray}}, \bibinfo {author} {\bibfnamefont
  {G.~E.}\ \bibnamefont {Scuseria}}, \bibinfo {author} {\bibfnamefont
  {S.}~\bibnamefont {Sorella}}, \bibinfo {author} {\bibfnamefont {E.~M.}\
  \bibnamefont {Stoudenmire}}, \bibinfo {author} {\bibfnamefont
  {Q.}~\bibnamefont {Sun}}, \bibinfo {author} {\bibfnamefont {I.~S.}\
  \bibnamefont {Tupitsyn}}, \bibinfo {author} {\bibfnamefont {S.~R.}\
  \bibnamefont {White}}, \bibinfo {author} {\bibfnamefont {D.}~\bibnamefont
  {Zgid}},\ and\ \bibinfo {author} {\bibfnamefont {S.}~\bibnamefont {Zhang}},\
  }\href {https://doi.org/10.1103/PhysRevX.7.031059} {\bibfield  {journal}
  {\bibinfo  {journal} {Phys. Rev. X}\ }\textbf {\bibinfo {volume} {7}},\
  \bibinfo {pages} {031059} (\bibinfo {year} {2017})}\BibitemShut {NoStop}%
\bibitem [{\citenamefont {Williams}\ \emph {et~al.}(2020)\citenamefont
  {Williams}, \citenamefont {Yao}, \citenamefont {Li}, \citenamefont {Chen},
  \citenamefont {Shi}, \citenamefont {Motta}, \citenamefont {Niu},
  \citenamefont {Ray}, \citenamefont {Guo}, \citenamefont {Anderson},
  \citenamefont {Li}, \citenamefont {Tran}, \citenamefont {Yeh}, \citenamefont
  {Mussard}, \citenamefont {Sharma}, \citenamefont {Bruneval}, \citenamefont
  {van Schilfgaarde}, \citenamefont {Booth}, \citenamefont {Chan},
  \citenamefont {Zhang}, \citenamefont {Gull}, \citenamefont {Zgid},
  \citenamefont {Millis}, \citenamefont {Umrigar},\ and\ \citenamefont
  {Wagner}}]{PhysRevX.10.011041}%
  \BibitemOpen
  \bibfield  {author} {\bibinfo {author} {\bibfnamefont {K.~T.}\ \bibnamefont
  {Williams}}, \bibinfo {author} {\bibfnamefont {Y.}~\bibnamefont {Yao}},
  \bibinfo {author} {\bibfnamefont {J.}~\bibnamefont {Li}}, \bibinfo {author}
  {\bibfnamefont {L.}~\bibnamefont {Chen}}, \bibinfo {author} {\bibfnamefont
  {H.}~\bibnamefont {Shi}}, \bibinfo {author} {\bibfnamefont {M.}~\bibnamefont
  {Motta}}, \bibinfo {author} {\bibfnamefont {C.}~\bibnamefont {Niu}}, \bibinfo
  {author} {\bibfnamefont {U.}~\bibnamefont {Ray}}, \bibinfo {author}
  {\bibfnamefont {S.}~\bibnamefont {Guo}}, \bibinfo {author} {\bibfnamefont
  {R.~J.}\ \bibnamefont {Anderson}}, \bibinfo {author} {\bibfnamefont
  {J.}~\bibnamefont {Li}}, \bibinfo {author} {\bibfnamefont {L.~N.}\
  \bibnamefont {Tran}}, \bibinfo {author} {\bibfnamefont {C.-N.}\ \bibnamefont
  {Yeh}}, \bibinfo {author} {\bibfnamefont {B.}~\bibnamefont {Mussard}},
  \bibinfo {author} {\bibfnamefont {S.}~\bibnamefont {Sharma}}, \bibinfo
  {author} {\bibfnamefont {F.}~\bibnamefont {Bruneval}}, \bibinfo {author}
  {\bibfnamefont {M.}~\bibnamefont {van Schilfgaarde}}, \bibinfo {author}
  {\bibfnamefont {G.~H.}\ \bibnamefont {Booth}}, \bibinfo {author}
  {\bibfnamefont {G.~K.-L.}\ \bibnamefont {Chan}}, \bibinfo {author}
  {\bibfnamefont {S.}~\bibnamefont {Zhang}}, \bibinfo {author} {\bibfnamefont
  {E.}~\bibnamefont {Gull}}, \bibinfo {author} {\bibfnamefont {D.}~\bibnamefont
  {Zgid}}, \bibinfo {author} {\bibfnamefont {A.}~\bibnamefont {Millis}},
  \bibinfo {author} {\bibfnamefont {C.~J.}\ \bibnamefont {Umrigar}},\ and\
  \bibinfo {author} {\bibfnamefont {L.~K.}\ \bibnamefont {Wagner}},\ }\href
  {https://doi.org/10.1103/PhysRevX.10.011041} {\bibfield  {journal} {\bibinfo
  {journal} {Phys. Rev. X}\ }\textbf {\bibinfo {volume} {10}},\ \bibinfo
  {pages} {011041} (\bibinfo {year} {2020})}\BibitemShut {NoStop}%
\bibitem [{\citenamefont {Rusakov}\ \emph {et~al.}(2019)\citenamefont
  {Rusakov}, \citenamefont {Iskakov}, \citenamefont {Tran},\ and\ \citenamefont
  {Zgid}}]{doi:10.1021/acs.jctc.8b00927}%
  \BibitemOpen
  \bibfield  {author} {\bibinfo {author} {\bibfnamefont {A.~A.}\ \bibnamefont
  {Rusakov}}, \bibinfo {author} {\bibfnamefont {S.}~\bibnamefont {Iskakov}},
  \bibinfo {author} {\bibfnamefont {L.~N.}\ \bibnamefont {Tran}},\ and\
  \bibinfo {author} {\bibfnamefont {D.}~\bibnamefont {Zgid}},\ }\href
  {https://doi.org/10.1021/acs.jctc.8b00927} {\bibfield  {journal} {\bibinfo
  {journal} {Journal of Chemical Theory and Computation}\ }\textbf {\bibinfo
  {volume} {15}},\ \bibinfo {pages} {229} (\bibinfo {year} {2019})}\BibitemShut
  {NoStop}%
\bibitem [{\citenamefont {Mott}(1949)}]{Mott_1949}%
  \BibitemOpen
  \bibfield  {author} {\bibinfo {author} {\bibfnamefont {N.~F.}\ \bibnamefont
  {Mott}},\ }\href {https://doi.org/10.1088/0370-1298/62/7/303} {\bibfield
  {journal} {\bibinfo  {journal} {Proceedings of the Physical Society. Section
  A}\ }\textbf {\bibinfo {volume} {62}},\ \bibinfo {pages} {416} (\bibinfo
  {year} {1949})}\BibitemShut {NoStop}%
\bibitem [{\citenamefont {Powell}\ and\ \citenamefont
  {Spicer}(1970)}]{NiO_expt_Powell70}%
  \BibitemOpen
  \bibfield  {author} {\bibinfo {author} {\bibfnamefont {R.~J.}\ \bibnamefont
  {Powell}}\ and\ \bibinfo {author} {\bibfnamefont {W.~E.}\ \bibnamefont
  {Spicer}},\ }\href {https://doi.org/10.1103/PhysRevB.2.2182} {\bibfield
  {journal} {\bibinfo  {journal} {Phys. Rev. B}\ }\textbf {\bibinfo {volume}
  {2}},\ \bibinfo {pages} {2182} (\bibinfo {year} {1970})}\BibitemShut
  {NoStop}%
\bibitem [{\citenamefont {Sawatzky}\ and\ \citenamefont
  {Allen}(1984)}]{NiO_expt_Sawatzky84}%
  \BibitemOpen
  \bibfield  {author} {\bibinfo {author} {\bibfnamefont {G.~A.}\ \bibnamefont
  {Sawatzky}}\ and\ \bibinfo {author} {\bibfnamefont {J.~W.}\ \bibnamefont
  {Allen}},\ }\href {https://doi.org/10.1103/PhysRevLett.53.2339} {\bibfield
  {journal} {\bibinfo  {journal} {Phys. Rev. Lett.}\ }\textbf {\bibinfo
  {volume} {53}},\ \bibinfo {pages} {2339} (\bibinfo {year}
  {1984})}\BibitemShut {NoStop}%
\bibitem [{\citenamefont {Shen}\ \emph {et~al.}(1990)\citenamefont {Shen},
  \citenamefont {Shih}, \citenamefont {Jepsen}, \citenamefont {Spicer},
  \citenamefont {Lindau},\ and\ \citenamefont {Allen}}]{NiOCoO_expt_Shen90}%
  \BibitemOpen
  \bibfield  {author} {\bibinfo {author} {\bibfnamefont {Z.-X.}\ \bibnamefont
  {Shen}}, \bibinfo {author} {\bibfnamefont {C.~K.}\ \bibnamefont {Shih}},
  \bibinfo {author} {\bibfnamefont {O.}~\bibnamefont {Jepsen}}, \bibinfo
  {author} {\bibfnamefont {W.~E.}\ \bibnamefont {Spicer}}, \bibinfo {author}
  {\bibfnamefont {I.}~\bibnamefont {Lindau}},\ and\ \bibinfo {author}
  {\bibfnamefont {J.~W.}\ \bibnamefont {Allen}},\ }\href
  {https://doi.org/10.1103/PhysRevLett.64.2442} {\bibfield  {journal} {\bibinfo
   {journal} {Phys. Rev. Lett.}\ }\textbf {\bibinfo {volume} {64}},\ \bibinfo
  {pages} {2442} (\bibinfo {year} {1990})}\BibitemShut {NoStop}%
\bibitem [{\citenamefont {Shen}\ \emph {et~al.}(1991)\citenamefont {Shen},
  \citenamefont {List}, \citenamefont {Dessau}, \citenamefont {Wells},
  \citenamefont {Jepsen}, \citenamefont {Arko}, \citenamefont {Barttlet},
  \citenamefont {Shih}, \citenamefont {Parmigiani}, \citenamefont {Huang},\
  and\ \citenamefont {Lindberg}}]{NiO_expt_Shen91}%
  \BibitemOpen
  \bibfield  {author} {\bibinfo {author} {\bibfnamefont {Z.-X.}\ \bibnamefont
  {Shen}}, \bibinfo {author} {\bibfnamefont {R.~S.}\ \bibnamefont {List}},
  \bibinfo {author} {\bibfnamefont {D.~S.}\ \bibnamefont {Dessau}}, \bibinfo
  {author} {\bibfnamefont {B.~O.}\ \bibnamefont {Wells}}, \bibinfo {author}
  {\bibfnamefont {O.}~\bibnamefont {Jepsen}}, \bibinfo {author} {\bibfnamefont
  {A.~J.}\ \bibnamefont {Arko}}, \bibinfo {author} {\bibfnamefont
  {R.}~\bibnamefont {Barttlet}}, \bibinfo {author} {\bibfnamefont {C.~K.}\
  \bibnamefont {Shih}}, \bibinfo {author} {\bibfnamefont {F.}~\bibnamefont
  {Parmigiani}}, \bibinfo {author} {\bibfnamefont {J.~C.}\ \bibnamefont
  {Huang}},\ and\ \bibinfo {author} {\bibfnamefont {P.~A.~P.}\ \bibnamefont
  {Lindberg}},\ }\href {https://doi.org/10.1103/PhysRevB.44.3604} {\bibfield
  {journal} {\bibinfo  {journal} {Phys. Rev. B}\ }\textbf {\bibinfo {volume}
  {44}},\ \bibinfo {pages} {3604} (\bibinfo {year} {1991})}\BibitemShut
  {NoStop}%
\bibitem [{\citenamefont {Tjernberg}\ \emph
  {et~al.}(1996{\natexlab{a}})\citenamefont {Tjernberg}, \citenamefont
  {S\"oderholm}, \citenamefont {Karlsson}, \citenamefont {Chiaia},
  \citenamefont {Qvarford}, \citenamefont {Nyl\'en},\ and\ \citenamefont
  {Lindau}}]{NiO_expt_Tjernberg96}%
  \BibitemOpen
  \bibfield  {author} {\bibinfo {author} {\bibfnamefont {O.}~\bibnamefont
  {Tjernberg}}, \bibinfo {author} {\bibfnamefont {S.}~\bibnamefont
  {S\"oderholm}}, \bibinfo {author} {\bibfnamefont {U.~O.}\ \bibnamefont
  {Karlsson}}, \bibinfo {author} {\bibfnamefont {G.}~\bibnamefont {Chiaia}},
  \bibinfo {author} {\bibfnamefont {M.}~\bibnamefont {Qvarford}}, \bibinfo
  {author} {\bibfnamefont {H.}~\bibnamefont {Nyl\'en}},\ and\ \bibinfo {author}
  {\bibfnamefont {I.}~\bibnamefont {Lindau}},\ }\href
  {https://doi.org/10.1103/PhysRevB.53.10372} {\bibfield  {journal} {\bibinfo
  {journal} {Phys. Rev. B}\ }\textbf {\bibinfo {volume} {53}},\ \bibinfo
  {pages} {10372} (\bibinfo {year} {1996}{\natexlab{a}})}\BibitemShut {NoStop}%
\bibitem [{\citenamefont {Tjernberg}\ \emph
  {et~al.}(1996{\natexlab{b}})\citenamefont {Tjernberg}, \citenamefont
  {S\"oderholm}, \citenamefont {Chiaia}, \citenamefont {Girard}, \citenamefont
  {Karlsson}, \citenamefont {Nyl\'en},\ and\ \citenamefont
  {Lindau}}]{NiO_expt_AFMandPM_Tjernberg96}%
  \BibitemOpen
  \bibfield  {author} {\bibinfo {author} {\bibfnamefont {O.}~\bibnamefont
  {Tjernberg}}, \bibinfo {author} {\bibfnamefont {S.}~\bibnamefont
  {S\"oderholm}}, \bibinfo {author} {\bibfnamefont {G.}~\bibnamefont {Chiaia}},
  \bibinfo {author} {\bibfnamefont {R.}~\bibnamefont {Girard}}, \bibinfo
  {author} {\bibfnamefont {U.~O.}\ \bibnamefont {Karlsson}}, \bibinfo {author}
  {\bibfnamefont {H.}~\bibnamefont {Nyl\'en}},\ and\ \bibinfo {author}
  {\bibfnamefont {I.}~\bibnamefont {Lindau}},\ }\href
  {https://doi.org/10.1103/PhysRevB.54.10245} {\bibfield  {journal} {\bibinfo
  {journal} {Phys. Rev. B}\ }\textbf {\bibinfo {volume} {54}},\ \bibinfo
  {pages} {10245} (\bibinfo {year} {1996}{\natexlab{b}})}\BibitemShut {NoStop}%
\bibitem [{\citenamefont {Jauch}\ and\ \citenamefont
  {Reehuis}(2004)}]{NiO_expt_Jauch04}%
  \BibitemOpen
  \bibfield  {author} {\bibinfo {author} {\bibfnamefont {W.}~\bibnamefont
  {Jauch}}\ and\ \bibinfo {author} {\bibfnamefont {M.}~\bibnamefont
  {Reehuis}},\ }\href {https://doi.org/10.1103/PhysRevB.70.195121} {\bibfield
  {journal} {\bibinfo  {journal} {Phys. Rev. B}\ }\textbf {\bibinfo {volume}
  {70}},\ \bibinfo {pages} {195121} (\bibinfo {year} {2004})}\BibitemShut
  {NoStop}%
\bibitem [{\citenamefont {Schuler}\ \emph {et~al.}(2005)\citenamefont
  {Schuler}, \citenamefont {Ederer}, \citenamefont {Itza-Ortiz}, \citenamefont
  {Woods}, \citenamefont {Callcott},\ and\ \citenamefont
  {Woicik}}]{NiO_expt_Schuler05}%
  \BibitemOpen
  \bibfield  {author} {\bibinfo {author} {\bibfnamefont {T.~M.}\ \bibnamefont
  {Schuler}}, \bibinfo {author} {\bibfnamefont {D.~L.}\ \bibnamefont {Ederer}},
  \bibinfo {author} {\bibfnamefont {S.}~\bibnamefont {Itza-Ortiz}}, \bibinfo
  {author} {\bibfnamefont {G.~T.}\ \bibnamefont {Woods}}, \bibinfo {author}
  {\bibfnamefont {T.~A.}\ \bibnamefont {Callcott}},\ and\ \bibinfo {author}
  {\bibfnamefont {J.~C.}\ \bibnamefont {Woicik}},\ }\href
  {https://doi.org/10.1103/PhysRevB.71.115113} {\bibfield  {journal} {\bibinfo
  {journal} {Phys. Rev. B}\ }\textbf {\bibinfo {volume} {71}},\ \bibinfo
  {pages} {115113} (\bibinfo {year} {2005})}\BibitemShut {NoStop}%
\bibitem [{\citenamefont {Gavriliuk}\ \emph {et~al.}(2012)\citenamefont
  {Gavriliuk}, \citenamefont {Trojan},\ and\ \citenamefont
  {Struzhkin}}]{NiO_expt_MIT_Gavriliuk12}%
  \BibitemOpen
  \bibfield  {author} {\bibinfo {author} {\bibfnamefont {A.~G.}\ \bibnamefont
  {Gavriliuk}}, \bibinfo {author} {\bibfnamefont {I.~A.}\ \bibnamefont
  {Trojan}},\ and\ \bibinfo {author} {\bibfnamefont {V.~V.}\ \bibnamefont
  {Struzhkin}},\ }\href {https://doi.org/10.1103/PhysRevLett.109.086402}
  {\bibfield  {journal} {\bibinfo  {journal} {Phys. Rev. Lett.}\ }\textbf
  {\bibinfo {volume} {109}},\ \bibinfo {pages} {086402} (\bibinfo {year}
  {2012})}\BibitemShut {NoStop}%
\bibitem [{\citenamefont {Potapkin}\ \emph {et~al.}(2016)\citenamefont
  {Potapkin}, \citenamefont {Dubrovinsky}, \citenamefont {Sergueev},
  \citenamefont {Ekholm}, \citenamefont {Kantor}, \citenamefont {Bessas},
  \citenamefont {Bykova}, \citenamefont {Prakapenka}, \citenamefont {Hermann},
  \citenamefont {R\"uffer}, \citenamefont {Cerantola}, \citenamefont
  {J\"onsson}, \citenamefont {Olovsson}, \citenamefont {Mankovsky},
  \citenamefont {Ebert},\ and\ \citenamefont
  {Abrikosov}}]{NiO_expt_MIT_Potapkin16}%
  \BibitemOpen
  \bibfield  {author} {\bibinfo {author} {\bibfnamefont {V.}~\bibnamefont
  {Potapkin}}, \bibinfo {author} {\bibfnamefont {L.}~\bibnamefont
  {Dubrovinsky}}, \bibinfo {author} {\bibfnamefont {I.}~\bibnamefont
  {Sergueev}}, \bibinfo {author} {\bibfnamefont {M.}~\bibnamefont {Ekholm}},
  \bibinfo {author} {\bibfnamefont {I.}~\bibnamefont {Kantor}}, \bibinfo
  {author} {\bibfnamefont {D.}~\bibnamefont {Bessas}}, \bibinfo {author}
  {\bibfnamefont {E.}~\bibnamefont {Bykova}}, \bibinfo {author} {\bibfnamefont
  {V.}~\bibnamefont {Prakapenka}}, \bibinfo {author} {\bibfnamefont {R.~P.}\
  \bibnamefont {Hermann}}, \bibinfo {author} {\bibfnamefont {R.}~\bibnamefont
  {R\"uffer}}, \bibinfo {author} {\bibfnamefont {V.}~\bibnamefont {Cerantola}},
  \bibinfo {author} {\bibfnamefont {H.~J.~M.}\ \bibnamefont {J\"onsson}},
  \bibinfo {author} {\bibfnamefont {W.}~\bibnamefont {Olovsson}}, \bibinfo
  {author} {\bibfnamefont {S.}~\bibnamefont {Mankovsky}}, \bibinfo {author}
  {\bibfnamefont {H.}~\bibnamefont {Ebert}},\ and\ \bibinfo {author}
  {\bibfnamefont {I.~A.}\ \bibnamefont {Abrikosov}},\ }\href
  {https://doi.org/10.1103/PhysRevB.93.201110} {\bibfield  {journal} {\bibinfo
  {journal} {Phys. Rev. B}\ }\textbf {\bibinfo {volume} {93}},\ \bibinfo
  {pages} {201110(R)} (\bibinfo {year} {2016})}\BibitemShut {NoStop}%
\bibitem [{\citenamefont {Eastman}\ and\ \citenamefont
  {Freeouf}(1975)}]{MnONiO_expt_Eastman75}%
  \BibitemOpen
  \bibfield  {author} {\bibinfo {author} {\bibfnamefont {D.~E.}\ \bibnamefont
  {Eastman}}\ and\ \bibinfo {author} {\bibfnamefont {J.~L.}\ \bibnamefont
  {Freeouf}},\ }\href {https://doi.org/10.1103/PhysRevLett.34.395} {\bibfield
  {journal} {\bibinfo  {journal} {Phys. Rev. Lett.}\ }\textbf {\bibinfo
  {volume} {34}},\ \bibinfo {pages} {395} (\bibinfo {year} {1975})}\BibitemShut
  {NoStop}%
\bibitem [{\citenamefont {Shen}\ and\ \citenamefont
  {Dessau}(1995)}]{PE_review_SHEN95}%
  \BibitemOpen
  \bibfield  {author} {\bibinfo {author} {\bibfnamefont {Z.-X.}\ \bibnamefont
  {Shen}}\ and\ \bibinfo {author} {\bibfnamefont {D.}~\bibnamefont {Dessau}},\
  }\href {https://doi.org/https://doi.org/10.1016/0370-1573(95)80001-A}
  {\bibfield  {journal} {\bibinfo  {journal} {Physics Reports}\ }\textbf
  {\bibinfo {volume} {253}},\ \bibinfo {pages} {1 } (\bibinfo {year}
  {1995})}\BibitemShut {NoStop}%
\bibitem [{\citenamefont {Lad}\ and\ \citenamefont
  {Henrich}(1988)}]{MnO_expt_Lad88}%
  \BibitemOpen
  \bibfield  {author} {\bibinfo {author} {\bibfnamefont {R.~J.}\ \bibnamefont
  {Lad}}\ and\ \bibinfo {author} {\bibfnamefont {V.~E.}\ \bibnamefont
  {Henrich}},\ }\href {https://doi.org/10.1103/PhysRevB.38.10860} {\bibfield
  {journal} {\bibinfo  {journal} {Phys. Rev. B}\ }\textbf {\bibinfo {volume}
  {38}},\ \bibinfo {pages} {10860} (\bibinfo {year} {1988})}\BibitemShut
  {NoStop}%
\bibitem [{\citenamefont {van Elp}\ \emph {et~al.}(1991)\citenamefont {van
  Elp}, \citenamefont {Potze}, \citenamefont {Eskes}, \citenamefont {Berger},\
  and\ \citenamefont {Sawatzky}}]{MnO_expt_Elp91}%
  \BibitemOpen
  \bibfield  {author} {\bibinfo {author} {\bibfnamefont {J.}~\bibnamefont {van
  Elp}}, \bibinfo {author} {\bibfnamefont {R.~H.}\ \bibnamefont {Potze}},
  \bibinfo {author} {\bibfnamefont {H.}~\bibnamefont {Eskes}}, \bibinfo
  {author} {\bibfnamefont {R.}~\bibnamefont {Berger}},\ and\ \bibinfo {author}
  {\bibfnamefont {G.~A.}\ \bibnamefont {Sawatzky}},\ }\href
  {https://doi.org/10.1103/PhysRevB.44.1530} {\bibfield  {journal} {\bibinfo
  {journal} {Phys. Rev. B}\ }\textbf {\bibinfo {volume} {44}},\ \bibinfo
  {pages} {1530} (\bibinfo {year} {1991})}\BibitemShut {NoStop}%
\bibitem [{\citenamefont {Kondo}\ \emph {et~al.}(2000)\citenamefont {Kondo},
  \citenamefont {Yagi}, \citenamefont {Syono}, \citenamefont {Noguchi},
  \citenamefont {Atou}, \citenamefont {Kikegawa},\ and\ \citenamefont
  {Shimomura}}]{MnO_PT_Kondo00}%
  \BibitemOpen
  \bibfield  {author} {\bibinfo {author} {\bibfnamefont {T.}~\bibnamefont
  {Kondo}}, \bibinfo {author} {\bibfnamefont {T.}~\bibnamefont {Yagi}},
  \bibinfo {author} {\bibfnamefont {Y.}~\bibnamefont {Syono}}, \bibinfo
  {author} {\bibfnamefont {Y.}~\bibnamefont {Noguchi}}, \bibinfo {author}
  {\bibfnamefont {T.}~\bibnamefont {Atou}}, \bibinfo {author} {\bibfnamefont
  {T.}~\bibnamefont {Kikegawa}},\ and\ \bibinfo {author} {\bibfnamefont
  {O.}~\bibnamefont {Shimomura}},\ }\href {https://doi.org/10.1063/1.373045}
  {\bibfield  {journal} {\bibinfo  {journal} {Journal of Applied Physics}\
  }\textbf {\bibinfo {volume} {87}},\ \bibinfo {pages} {4153} (\bibinfo {year}
  {2000})}\BibitemShut {NoStop}%
\bibitem [{\citenamefont {Patterson}\ \emph {et~al.}(2004)\citenamefont
  {Patterson}, \citenamefont {Aracne}, \citenamefont {Jackson}, \citenamefont
  {Malba}, \citenamefont {Weir}, \citenamefont {Baker},\ and\ \citenamefont
  {Vohra}}]{MnO_MIT_Patterson04}%
  \BibitemOpen
  \bibfield  {author} {\bibinfo {author} {\bibfnamefont {J.~R.}\ \bibnamefont
  {Patterson}}, \bibinfo {author} {\bibfnamefont {C.~M.}\ \bibnamefont
  {Aracne}}, \bibinfo {author} {\bibfnamefont {D.~D.}\ \bibnamefont {Jackson}},
  \bibinfo {author} {\bibfnamefont {V.}~\bibnamefont {Malba}}, \bibinfo
  {author} {\bibfnamefont {S.~T.}\ \bibnamefont {Weir}}, \bibinfo {author}
  {\bibfnamefont {P.~A.}\ \bibnamefont {Baker}},\ and\ \bibinfo {author}
  {\bibfnamefont {Y.~K.}\ \bibnamefont {Vohra}},\ }\href
  {https://doi.org/10.1103/PhysRevB.69.220101} {\bibfield  {journal} {\bibinfo
  {journal} {Phys. Rev. B}\ }\textbf {\bibinfo {volume} {69}},\ \bibinfo
  {pages} {220101(R)} (\bibinfo {year} {2004})}\BibitemShut {NoStop}%
\bibitem [{\citenamefont {Towler}\ \emph {et~al.}(1994)\citenamefont {Towler},
  \citenamefont {Allan}, \citenamefont {Harrison}, \citenamefont {Saunders},
  \citenamefont {Mackrodt},\ and\ \citenamefont {Apr\`a}}]{NiOMnO_HF_Towler94}%
  \BibitemOpen
  \bibfield  {author} {\bibinfo {author} {\bibfnamefont {M.~D.}\ \bibnamefont
  {Towler}}, \bibinfo {author} {\bibfnamefont {N.~L.}\ \bibnamefont {Allan}},
  \bibinfo {author} {\bibfnamefont {N.~M.}\ \bibnamefont {Harrison}}, \bibinfo
  {author} {\bibfnamefont {V.~R.}\ \bibnamefont {Saunders}}, \bibinfo {author}
  {\bibfnamefont {W.~C.}\ \bibnamefont {Mackrodt}},\ and\ \bibinfo {author}
  {\bibfnamefont {E.}~\bibnamefont {Apr\`a}},\ }\href
  {https://doi.org/10.1103/PhysRevB.50.5041} {\bibfield  {journal} {\bibinfo
  {journal} {Phys. Rev. B}\ }\textbf {\bibinfo {volume} {50}},\ \bibinfo
  {pages} {5041} (\bibinfo {year} {1994})}\BibitemShut {NoStop}%
\bibitem [{\citenamefont {Fujimori}\ and\ \citenamefont
  {Minami}(1984)}]{NiO_Fujimori84}%
  \BibitemOpen
  \bibfield  {author} {\bibinfo {author} {\bibfnamefont {A.}~\bibnamefont
  {Fujimori}}\ and\ \bibinfo {author} {\bibfnamefont {F.}~\bibnamefont
  {Minami}},\ }\href {https://doi.org/10.1103/PhysRevB.30.957} {\bibfield
  {journal} {\bibinfo  {journal} {Phys. Rev. B}\ }\textbf {\bibinfo {volume}
  {30}},\ \bibinfo {pages} {957} (\bibinfo {year} {1984})}\BibitemShut
  {NoStop}%
\bibitem [{\citenamefont {Fang}\ \emph {et~al.}(1999)\citenamefont {Fang},
  \citenamefont {Solovyev}, \citenamefont {Sawada},\ and\ \citenamefont
  {Terakura}}]{MnO_DFT_PT_Fang99}%
  \BibitemOpen
  \bibfield  {author} {\bibinfo {author} {\bibfnamefont {Z.}~\bibnamefont
  {Fang}}, \bibinfo {author} {\bibfnamefont {I.~V.}\ \bibnamefont {Solovyev}},
  \bibinfo {author} {\bibfnamefont {H.}~\bibnamefont {Sawada}},\ and\ \bibinfo
  {author} {\bibfnamefont {K.}~\bibnamefont {Terakura}},\ }\href
  {https://doi.org/10.1103/PhysRevB.59.762} {\bibfield  {journal} {\bibinfo
  {journal} {Phys. Rev. B}\ }\textbf {\bibinfo {volume} {59}},\ \bibinfo
  {pages} {762} (\bibinfo {year} {1999})}\BibitemShut {NoStop}%
\bibitem [{\citenamefont {Anisimov}\ \emph
  {et~al.}(1997{\natexlab{b}})\citenamefont {Anisimov}, \citenamefont
  {Aryasetiawan},\ and\ \citenamefont {Lichtenstein}}]{LDAplusU_Anisimov97}%
  \BibitemOpen
  \bibfield  {author} {\bibinfo {author} {\bibfnamefont {V.~I.}\ \bibnamefont
  {Anisimov}}, \bibinfo {author} {\bibfnamefont {F.}~\bibnamefont
  {Aryasetiawan}},\ and\ \bibinfo {author} {\bibfnamefont {A.~I.}\ \bibnamefont
  {Lichtenstein}},\ }\href {https://doi.org/10.1088/0953-8984/9/4/002}
  {\bibfield  {journal} {\bibinfo  {journal} {Journal of Physics: Condensed
  Matter}\ }\textbf {\bibinfo {volume} {9}},\ \bibinfo {pages} {767} (\bibinfo
  {year} {1997}{\natexlab{b}})}\BibitemShut {NoStop}%
\bibitem [{\citenamefont {Aryasetiawan}\ and\ \citenamefont
  {Gunnarsson}(1995)}]{NiO_GW_Aryasetiawan95}%
  \BibitemOpen
  \bibfield  {author} {\bibinfo {author} {\bibfnamefont {F.}~\bibnamefont
  {Aryasetiawan}}\ and\ \bibinfo {author} {\bibfnamefont {O.}~\bibnamefont
  {Gunnarsson}},\ }\href {https://doi.org/10.1103/PhysRevLett.74.3221}
  {\bibfield  {journal} {\bibinfo  {journal} {Phys. Rev. Lett.}\ }\textbf
  {\bibinfo {volume} {74}},\ \bibinfo {pages} {3221} (\bibinfo {year}
  {1995})}\BibitemShut {NoStop}%
\bibitem [{\citenamefont {Faleev}\ \emph {et~al.}(2004)\citenamefont {Faleev},
  \citenamefont {van Schilfgaarde},\ and\ \citenamefont {Kotani}}]{Sergey04}%
  \BibitemOpen
  \bibfield  {author} {\bibinfo {author} {\bibfnamefont {S.~V.}\ \bibnamefont
  {Faleev}}, \bibinfo {author} {\bibfnamefont {M.}~\bibnamefont {van
  Schilfgaarde}},\ and\ \bibinfo {author} {\bibfnamefont {T.}~\bibnamefont
  {Kotani}},\ }\href {https://doi.org/10.1103/PhysRevLett.93.126406} {\bibfield
   {journal} {\bibinfo  {journal} {Phys. Rev. Lett.}\ }\textbf {\bibinfo
  {volume} {93}},\ \bibinfo {pages} {126406} (\bibinfo {year}
  {2004})}\BibitemShut {NoStop}%
\bibitem [{\citenamefont {Li}\ \emph {et~al.}(2005)\citenamefont {Li},
  \citenamefont {Rignanese},\ and\ \citenamefont {Louie}}]{NiO_QSGW_Li05}%
  \BibitemOpen
  \bibfield  {author} {\bibinfo {author} {\bibfnamefont {J.-L.}\ \bibnamefont
  {Li}}, \bibinfo {author} {\bibfnamefont {G.-M.}\ \bibnamefont {Rignanese}},\
  and\ \bibinfo {author} {\bibfnamefont {S.~G.}\ \bibnamefont {Louie}},\ }\href
  {https://doi.org/10.1103/PhysRevB.71.193102} {\bibfield  {journal} {\bibinfo
  {journal} {Phys. Rev. B}\ }\textbf {\bibinfo {volume} {71}},\ \bibinfo
  {pages} {193102} (\bibinfo {year} {2005})}\BibitemShut {NoStop}%
\bibitem [{\citenamefont {R\"odl}\ \emph {et~al.}(2009)\citenamefont {R\"odl},
  \citenamefont {Fuchs}, \citenamefont {Furthm\"uller},\ and\ \citenamefont
  {Bechstedt}}]{Rodl09}%
  \BibitemOpen
  \bibfield  {author} {\bibinfo {author} {\bibfnamefont {C.}~\bibnamefont
  {R\"odl}}, \bibinfo {author} {\bibfnamefont {F.}~\bibnamefont {Fuchs}},
  \bibinfo {author} {\bibfnamefont {J.}~\bibnamefont {Furthm\"uller}},\ and\
  \bibinfo {author} {\bibfnamefont {F.}~\bibnamefont {Bechstedt}},\ }\href
  {https://doi.org/10.1103/PhysRevB.79.235114} {\bibfield  {journal} {\bibinfo
  {journal} {Phys. Rev. B}\ }\textbf {\bibinfo {volume} {79}},\ \bibinfo
  {pages} {235114} (\bibinfo {year} {2009})}\BibitemShut {NoStop}%
\bibitem [{\citenamefont {Eder}(2015)}]{NiO_VCA_Eder15}%
  \BibitemOpen
  \bibfield  {author} {\bibinfo {author} {\bibfnamefont {R.}~\bibnamefont
  {Eder}},\ }\href {https://doi.org/10.1103/PhysRevB.91.245146} {\bibfield
  {journal} {\bibinfo  {journal} {Phys. Rev. B}\ }\textbf {\bibinfo {volume}
  {91}},\ \bibinfo {pages} {245146} (\bibinfo {year} {2015})}\BibitemShut
  {NoStop}%
\bibitem [{\citenamefont {Ren}\ \emph {et~al.}(2006)\citenamefont {Ren},
  \citenamefont {Leonov}, \citenamefont {Keller}, \citenamefont {Kollar},
  \citenamefont {Nekrasov},\ and\ \citenamefont {Vollhardt}}]{LDADMFT_Ren06}%
  \BibitemOpen
  \bibfield  {author} {\bibinfo {author} {\bibfnamefont {X.}~\bibnamefont
  {Ren}}, \bibinfo {author} {\bibfnamefont {I.}~\bibnamefont {Leonov}},
  \bibinfo {author} {\bibfnamefont {G.}~\bibnamefont {Keller}}, \bibinfo
  {author} {\bibfnamefont {M.}~\bibnamefont {Kollar}}, \bibinfo {author}
  {\bibfnamefont {I.}~\bibnamefont {Nekrasov}},\ and\ \bibinfo {author}
  {\bibfnamefont {D.}~\bibnamefont {Vollhardt}},\ }\href
  {https://doi.org/10.1103/PhysRevB.74.195114} {\bibfield  {journal} {\bibinfo
  {journal} {Phys. Rev. B}\ }\textbf {\bibinfo {volume} {74}},\ \bibinfo
  {pages} {195114} (\bibinfo {year} {2006})}\BibitemShut {NoStop}%
\bibitem [{\citenamefont {Kune\ifmmode~\check{s}\else \v{s}\fi{}}\ \emph
  {et~al.}(2007{\natexlab{a}})\citenamefont {Kune\ifmmode~\check{s}\else
  \v{s}\fi{}}, \citenamefont {Anisimov}, \citenamefont {Lukoyanov},\ and\
  \citenamefont {Vollhardt}}]{Kunes07_prb}%
  \BibitemOpen
  \bibfield  {author} {\bibinfo {author} {\bibfnamefont {J.}~\bibnamefont
  {Kune\ifmmode~\check{s}\else \v{s}\fi{}}}, \bibinfo {author} {\bibfnamefont
  {V.~I.}\ \bibnamefont {Anisimov}}, \bibinfo {author} {\bibfnamefont {A.~V.}\
  \bibnamefont {Lukoyanov}},\ and\ \bibinfo {author} {\bibfnamefont
  {D.}~\bibnamefont {Vollhardt}},\ }\href
  {https://doi.org/10.1103/PhysRevB.75.165115} {\bibfield  {journal} {\bibinfo
  {journal} {Phys. Rev. B}\ }\textbf {\bibinfo {volume} {75}},\ \bibinfo
  {pages} {165115} (\bibinfo {year} {2007}{\natexlab{a}})}\BibitemShut
  {NoStop}%
\bibitem [{\citenamefont {Kune\ifmmode~\check{s}\else \v{s}\fi{}}\ \emph
  {et~al.}(2007{\natexlab{b}})\citenamefont {Kune\ifmmode~\check{s}\else
  \v{s}\fi{}}, \citenamefont {Anisimov}, \citenamefont {Skornyakov},
  \citenamefont {Lukoyanov},\ and\ \citenamefont {Vollhardt}}]{Kunes07_prl}%
  \BibitemOpen
  \bibfield  {author} {\bibinfo {author} {\bibfnamefont {J.}~\bibnamefont
  {Kune\ifmmode~\check{s}\else \v{s}\fi{}}}, \bibinfo {author} {\bibfnamefont
  {V.~I.}\ \bibnamefont {Anisimov}}, \bibinfo {author} {\bibfnamefont {S.~L.}\
  \bibnamefont {Skornyakov}}, \bibinfo {author} {\bibfnamefont {A.~V.}\
  \bibnamefont {Lukoyanov}},\ and\ \bibinfo {author} {\bibfnamefont
  {D.}~\bibnamefont {Vollhardt}},\ }\href
  {https://doi.org/10.1103/PhysRevLett.99.156404} {\bibfield  {journal}
  {\bibinfo  {journal} {Phys. Rev. Lett.}\ }\textbf {\bibinfo {volume} {99}},\
  \bibinfo {pages} {156404} (\bibinfo {year} {2007}{\natexlab{b}})}\BibitemShut
  {NoStop}%
\bibitem [{\citenamefont {Boman}\ \emph {et~al.}(2008)\citenamefont {Boman},
  \citenamefont {Koch},\ and\ \citenamefont {Sánchez~de
  Merás}}]{Cholesky2008}%
  \BibitemOpen
  \bibfield  {author} {\bibinfo {author} {\bibfnamefont {L.}~\bibnamefont
  {Boman}}, \bibinfo {author} {\bibfnamefont {H.}~\bibnamefont {Koch}},\ and\
  \bibinfo {author} {\bibfnamefont {A.}~\bibnamefont {Sánchez~de Merás}},\
  }\href {https://doi.org/10.1063/1.2988315} {\bibfield  {journal} {\bibinfo
  {journal} {The Journal of Chemical Physics}\ }\textbf {\bibinfo {volume}
  {129}},\ \bibinfo {pages} {134107} (\bibinfo {year} {2008})}\BibitemShut
  {NoStop}%
\bibitem [{\citenamefont {Werner}\ \emph {et~al.}(2003)\citenamefont {Werner},
  \citenamefont {Manby},\ and\ \citenamefont {Knowles}}]{Werner2003}%
  \BibitemOpen
  \bibfield  {author} {\bibinfo {author} {\bibfnamefont {H.-J.}\ \bibnamefont
  {Werner}}, \bibinfo {author} {\bibfnamefont {F.~R.}\ \bibnamefont {Manby}},\
  and\ \bibinfo {author} {\bibfnamefont {P.~J.}\ \bibnamefont {Knowles}},\
  }\href {https://doi.org/10.1063/1.1564816} {\bibfield  {journal} {\bibinfo
  {journal} {The Journal of Chemical Physics}\ }\textbf {\bibinfo {volume}
  {118}},\ \bibinfo {pages} {8149} (\bibinfo {year} {2003})}\BibitemShut
  {NoStop}%
\bibitem [{\citenamefont {Ren}\ \emph {et~al.}(2012)\citenamefont {Ren},
  \citenamefont {Rinke}, \citenamefont {Blum}, \citenamefont {Wieferink},
  \citenamefont {Tkatchenko}, \citenamefont {Sanfilippo}, \citenamefont
  {Reuter},\ and\ \citenamefont {Scheffler}}]{Ren2012}%
  \BibitemOpen
  \bibfield  {author} {\bibinfo {author} {\bibfnamefont {X.}~\bibnamefont
  {Ren}}, \bibinfo {author} {\bibfnamefont {P.}~\bibnamefont {Rinke}}, \bibinfo
  {author} {\bibfnamefont {V.}~\bibnamefont {Blum}}, \bibinfo {author}
  {\bibfnamefont {J.}~\bibnamefont {Wieferink}}, \bibinfo {author}
  {\bibfnamefont {A.}~\bibnamefont {Tkatchenko}}, \bibinfo {author}
  {\bibfnamefont {A.}~\bibnamefont {Sanfilippo}}, \bibinfo {author}
  {\bibfnamefont {K.}~\bibnamefont {Reuter}},\ and\ \bibinfo {author}
  {\bibfnamefont {M.}~\bibnamefont {Scheffler}},\ }\href
  {https://doi.org/10.1088/1367-2630/14/5/053020} {\bibfield  {journal}
  {\bibinfo  {journal} {New Journal of Physics}\ }\textbf {\bibinfo {volume}
  {14}},\ \bibinfo {pages} {053020} (\bibinfo {year} {2012})}\BibitemShut
  {NoStop}%
\bibitem [{\citenamefont {Sun}\ \emph {et~al.}(2017{\natexlab{a}})\citenamefont
  {Sun}, \citenamefont {Berkelbach}, \citenamefont {McClain},\ and\
  \citenamefont {Chan}}]{Sun2017}%
  \BibitemOpen
  \bibfield  {author} {\bibinfo {author} {\bibfnamefont {Q.}~\bibnamefont
  {Sun}}, \bibinfo {author} {\bibfnamefont {T.~C.}\ \bibnamefont {Berkelbach}},
  \bibinfo {author} {\bibfnamefont {J.~D.}\ \bibnamefont {McClain}},\ and\
  \bibinfo {author} {\bibfnamefont {G.~K.-L.}\ \bibnamefont {Chan}},\ }\href
  {https://doi.org/10.1063/1.4998644} {\bibfield  {journal} {\bibinfo
  {journal} {The Journal of Chemical Physics}\ }\textbf {\bibinfo {volume}
  {147}},\ \bibinfo {pages} {164119} (\bibinfo {year}
  {2017}{\natexlab{a}})}\BibitemShut {NoStop}%
\bibitem [{\citenamefont {Bartel}\ and\ \citenamefont
  {Morosin}(1971)}]{PhysRevB.3.1039}%
  \BibitemOpen
  \bibfield  {author} {\bibinfo {author} {\bibfnamefont {L.~C.}\ \bibnamefont
  {Bartel}}\ and\ \bibinfo {author} {\bibfnamefont {B.}~\bibnamefont
  {Morosin}},\ }\href {https://doi.org/10.1103/PhysRevB.3.1039} {\bibfield
  {journal} {\bibinfo  {journal} {Phys. Rev. B}\ }\textbf {\bibinfo {volume}
  {3}},\ \bibinfo {pages} {1039} (\bibinfo {year} {1971})}\BibitemShut
  {NoStop}%
\bibitem [{\citenamefont {Johnston}\ and\ \citenamefont
  {Heikes}(1956)}]{MnO_a}%
  \BibitemOpen
  \bibfield  {author} {\bibinfo {author} {\bibfnamefont {W.~D.}\ \bibnamefont
  {Johnston}}\ and\ \bibinfo {author} {\bibfnamefont {R.~R.}\ \bibnamefont
  {Heikes}},\ }\href {https://doi.org/10.1021/ja01595a006} {\bibfield
  {journal} {\bibinfo  {journal} {J. Am. Chem. Soc.}\ }\textbf {\bibinfo
  {volume} {78}},\ \bibinfo {pages} {3255} (\bibinfo {year}
  {1956})}\BibitemShut {NoStop}%
\bibitem [{\citenamefont {VandeVondele}\ and\ \citenamefont
  {Hutter}(2007)}]{GTHBasis}%
  \BibitemOpen
  \bibfield  {author} {\bibinfo {author} {\bibfnamefont {J.}~\bibnamefont
  {VandeVondele}}\ and\ \bibinfo {author} {\bibfnamefont {J.}~\bibnamefont
  {Hutter}},\ }\href {https://doi.org/10.1063/1.2770708} {\bibfield  {journal}
  {\bibinfo  {journal} {The Journal of Chemical Physics}\ }\textbf {\bibinfo
  {volume} {127}},\ \bibinfo {pages} {114105} (\bibinfo {year}
  {2007})}\BibitemShut {NoStop}%
\bibitem [{\citenamefont {Goedecker}\ \emph {et~al.}(1996)\citenamefont
  {Goedecker}, \citenamefont {Teter},\ and\ \citenamefont
  {Hutter}}]{GTHPseudo}%
  \BibitemOpen
  \bibfield  {author} {\bibinfo {author} {\bibfnamefont {S.}~\bibnamefont
  {Goedecker}}, \bibinfo {author} {\bibfnamefont {M.}~\bibnamefont {Teter}},\
  and\ \bibinfo {author} {\bibfnamefont {J.}~\bibnamefont {Hutter}},\ }\href
  {https://doi.org/https://doi.org/10.1103/PhysRevB.54.1703} {\bibfield
  {journal} {\bibinfo  {journal} {Phys. Rev. B}\ }\textbf {\bibinfo {volume}
  {54}},\ \bibinfo {pages} {1703} (\bibinfo {year} {1996})}\BibitemShut
  {NoStop}%
\bibitem [{\citenamefont {Hättig}(2005)}]{RI_auxbasis}%
  \BibitemOpen
  \bibfield  {author} {\bibinfo {author} {\bibfnamefont {C.}~\bibnamefont
  {Hättig}},\ }\href {https://doi.org/10.1039/B415208E} {\bibfield  {journal}
  {\bibinfo  {journal} {Phys. Chem. Chem. Phys.}\ }\textbf {\bibinfo {volume}
  {7}},\ \bibinfo {pages} {59} (\bibinfo {year} {2005})}\BibitemShut {NoStop}%
\bibitem [{\citenamefont {Paier}\ \emph {et~al.}(2005)\citenamefont {Paier},
  \citenamefont {Hirschl}, \citenamefont {Marsman},\ and\ \citenamefont
  {Kresse}}]{EwaldProbeCharge}%
  \BibitemOpen
  \bibfield  {author} {\bibinfo {author} {\bibfnamefont {J.}~\bibnamefont
  {Paier}}, \bibinfo {author} {\bibfnamefont {R.}~\bibnamefont {Hirschl}},
  \bibinfo {author} {\bibfnamefont {M.}~\bibnamefont {Marsman}},\ and\ \bibinfo
  {author} {\bibfnamefont {G.}~\bibnamefont {Kresse}},\ }\href
  {https://doi.org/10.1063/1.1926272} {\bibfield  {journal} {\bibinfo
  {journal} {The Journal of Chemical Physics}\ }\textbf {\bibinfo {volume}
  {122}},\ \bibinfo {pages} {234102} (\bibinfo {year} {2005})}\BibitemShut
  {NoStop}%
\bibitem [{\citenamefont {Sundararaman}\ and\ \citenamefont
  {Arias}(2013)}]{CoulombSingular}%
  \BibitemOpen
  \bibfield  {author} {\bibinfo {author} {\bibfnamefont {R.}~\bibnamefont
  {Sundararaman}}\ and\ \bibinfo {author} {\bibfnamefont {T.~A.}\ \bibnamefont
  {Arias}},\ }\href {https://doi.org/10.1103/PhysRevB.87.165122} {\bibfield
  {journal} {\bibinfo  {journal} {Phys. Rev. B}\ }\textbf {\bibinfo {volume}
  {87}},\ \bibinfo {pages} {165122} (\bibinfo {year} {2013})}\BibitemShut
  {NoStop}%
\bibitem [{\citenamefont {Sun}\ \emph {et~al.}(2017{\natexlab{b}})\citenamefont
  {Sun}, \citenamefont {Berkelbach}, \citenamefont {Blunt}, \citenamefont
  {Booth}, \citenamefont {Guo}, \citenamefont {Li}, \citenamefont {Liu},
  \citenamefont {McClain}, \citenamefont {Sayfutyarova}, \citenamefont
  {Sharma}, \citenamefont {Wouters},\ and\ \citenamefont {Chan}}]{PySCF}%
  \BibitemOpen
  \bibfield  {author} {\bibinfo {author} {\bibfnamefont {Q.}~\bibnamefont
  {Sun}}, \bibinfo {author} {\bibfnamefont {T.~C.}\ \bibnamefont {Berkelbach}},
  \bibinfo {author} {\bibfnamefont {N.~S.}\ \bibnamefont {Blunt}}, \bibinfo
  {author} {\bibfnamefont {G.~H.}\ \bibnamefont {Booth}}, \bibinfo {author}
  {\bibfnamefont {S.}~\bibnamefont {Guo}}, \bibinfo {author} {\bibfnamefont
  {Z.}~\bibnamefont {Li}}, \bibinfo {author} {\bibfnamefont {J.}~\bibnamefont
  {Liu}}, \bibinfo {author} {\bibfnamefont {J.~D.}\ \bibnamefont {McClain}},
  \bibinfo {author} {\bibfnamefont {E.~R.}\ \bibnamefont {Sayfutyarova}},
  \bibinfo {author} {\bibfnamefont {S.}~\bibnamefont {Sharma}}, \bibinfo
  {author} {\bibfnamefont {S.}~\bibnamefont {Wouters}},\ and\ \bibinfo {author}
  {\bibfnamefont {G.~K.}\ \bibnamefont {Chan}},\ }\href
  {https://doi.org/10.1002/wcms.1340} {\bibinfo {title} {Pyscf: the
  python-based simulations of chemistry framework}} (\bibinfo {year}
  {2017}{\natexlab{b}})\BibitemShut {NoStop}%
\bibitem [{\citenamefont {Kresse}\ and\ \citenamefont
  {Joubert}(1999)}]{Kresse1999}%
  \BibitemOpen
  \bibfield  {author} {\bibinfo {author} {\bibfnamefont {G.}~\bibnamefont
  {Kresse}}\ and\ \bibinfo {author} {\bibfnamefont {D.}~\bibnamefont
  {Joubert}},\ }\href {https://doi.org/10.1103/physrevb.59.1758} {\bibfield
  {journal} {\bibinfo  {journal} {Physical Review B}\ }\textbf {\bibinfo
  {volume} {59}},\ \bibinfo {pages} {1758} (\bibinfo {year}
  {1999})}\BibitemShut {NoStop}%
\bibitem [{\citenamefont {Shinaoka}\ \emph {et~al.}(2017)\citenamefont
  {Shinaoka}, \citenamefont {Otsuki}, \citenamefont {Ohzeki},\ and\
  \citenamefont {Yoshimi}}]{PhysRevB.96.035147}%
  \BibitemOpen
  \bibfield  {author} {\bibinfo {author} {\bibfnamefont {H.}~\bibnamefont
  {Shinaoka}}, \bibinfo {author} {\bibfnamefont {J.}~\bibnamefont {Otsuki}},
  \bibinfo {author} {\bibfnamefont {M.}~\bibnamefont {Ohzeki}},\ and\ \bibinfo
  {author} {\bibfnamefont {K.}~\bibnamefont {Yoshimi}},\ }\href
  {https://doi.org/10.1103/PhysRevB.96.035147} {\bibfield  {journal} {\bibinfo
  {journal} {Phys. Rev. B}\ }\textbf {\bibinfo {volume} {96}},\ \bibinfo
  {pages} {035147} (\bibinfo {year} {2017})}\BibitemShut {NoStop}%
\bibitem [{\citenamefont {Li}\ \emph {et~al.}(2020)\citenamefont {Li},
  \citenamefont {Wallerberger}, \citenamefont {Chikano}, \citenamefont {Yeh},
  \citenamefont {Gull},\ and\ \citenamefont {Shinaoka}}]{PhysRevB.101.035144}%
  \BibitemOpen
  \bibfield  {author} {\bibinfo {author} {\bibfnamefont {J.}~\bibnamefont
  {Li}}, \bibinfo {author} {\bibfnamefont {M.}~\bibnamefont {Wallerberger}},
  \bibinfo {author} {\bibfnamefont {N.}~\bibnamefont {Chikano}}, \bibinfo
  {author} {\bibfnamefont {C.-N.}\ \bibnamefont {Yeh}}, \bibinfo {author}
  {\bibfnamefont {E.}~\bibnamefont {Gull}},\ and\ \bibinfo {author}
  {\bibfnamefont {H.}~\bibnamefont {Shinaoka}},\ }\href
  {https://doi.org/10.1103/PhysRevB.101.035144} {\bibfield  {journal} {\bibinfo
   {journal} {Phys. Rev. B}\ }\textbf {\bibinfo {volume} {101}},\ \bibinfo
  {pages} {035144} (\bibinfo {year} {2020})}\BibitemShut {NoStop}%
\bibitem [{\citenamefont {Chikano}\ \emph {et~al.}(2019)\citenamefont
  {Chikano}, \citenamefont {Yoshimi}, \citenamefont {Otsuki},\ and\
  \citenamefont {Shinaoka}}]{CHIKANO2019181}%
  \BibitemOpen
  \bibfield  {author} {\bibinfo {author} {\bibfnamefont {N.}~\bibnamefont
  {Chikano}}, \bibinfo {author} {\bibfnamefont {K.}~\bibnamefont {Yoshimi}},
  \bibinfo {author} {\bibfnamefont {J.}~\bibnamefont {Otsuki}},\ and\ \bibinfo
  {author} {\bibfnamefont {H.}~\bibnamefont {Shinaoka}},\ }\href
  {https://doi.org/https://doi.org/10.1016/j.cpc.2019.02.006} {\bibfield
  {journal} {\bibinfo  {journal} {Computer Physics Communications}\ }\textbf
  {\bibinfo {volume} {240}},\ \bibinfo {pages} {181 } (\bibinfo {year}
  {2019})}\BibitemShut {NoStop}%
\bibitem [{\citenamefont {Boehnke}\ \emph {et~al.}(2011)\citenamefont
  {Boehnke}, \citenamefont {Hafermann}, \citenamefont {Ferrero}, \citenamefont
  {Lechermann},\ and\ \citenamefont {Parcollet}}]{Boehnke2011}%
  \BibitemOpen
  \bibfield  {author} {\bibinfo {author} {\bibfnamefont {L.}~\bibnamefont
  {Boehnke}}, \bibinfo {author} {\bibfnamefont {H.}~\bibnamefont {Hafermann}},
  \bibinfo {author} {\bibfnamefont {M.}~\bibnamefont {Ferrero}}, \bibinfo
  {author} {\bibfnamefont {F.}~\bibnamefont {Lechermann}},\ and\ \bibinfo
  {author} {\bibfnamefont {O.}~\bibnamefont {Parcollet}},\ }\href
  {https://doi.org/10.1103/PhysRevB.84.075145} {\bibfield  {journal} {\bibinfo
  {journal} {Phys. Rev. B}\ }\textbf {\bibinfo {volume} {84}},\ \bibinfo
  {pages} {075145} (\bibinfo {year} {2011})}\BibitemShut {NoStop}%
\bibitem [{\citenamefont {Dong}\ \emph {et~al.}(2020)\citenamefont {Dong},
  \citenamefont {Zgid}, \citenamefont {Gull},\ and\ \citenamefont
  {Strand}}]{dong2020legendrespectral}%
  \BibitemOpen
  \bibfield  {author} {\bibinfo {author} {\bibfnamefont {X.}~\bibnamefont
  {Dong}}, \bibinfo {author} {\bibfnamefont {D.}~\bibnamefont {Zgid}}, \bibinfo
  {author} {\bibfnamefont {E.}~\bibnamefont {Gull}},\ and\ \bibinfo {author}
  {\bibfnamefont {H.~U.~R.}\ \bibnamefont {Strand}},\ }\href@noop {} {\bibinfo
  {title} {Legendre-spectral dyson equation solver with super-exponential
  convergence}} (\bibinfo {year} {2020}),\ \Eprint
  {https://arxiv.org/abs/2001.11603} {arXiv:2001.11603 [cond-mat.str-el]}
  \BibitemShut {NoStop}%
\bibitem [{\citenamefont {Gull}\ \emph {et~al.}(2018)\citenamefont {Gull},
  \citenamefont {Iskakov}, \citenamefont {Krivenko}, \citenamefont {Rusakov},\
  and\ \citenamefont {Zgid}}]{PhysRevB.98.075127}%
  \BibitemOpen
  \bibfield  {author} {\bibinfo {author} {\bibfnamefont {E.}~\bibnamefont
  {Gull}}, \bibinfo {author} {\bibfnamefont {S.}~\bibnamefont {Iskakov}},
  \bibinfo {author} {\bibfnamefont {I.}~\bibnamefont {Krivenko}}, \bibinfo
  {author} {\bibfnamefont {A.~A.}\ \bibnamefont {Rusakov}},\ and\ \bibinfo
  {author} {\bibfnamefont {D.}~\bibnamefont {Zgid}},\ }\href
  {https://doi.org/10.1103/PhysRevB.98.075127} {\bibfield  {journal} {\bibinfo
  {journal} {Phys. Rev. B}\ }\textbf {\bibinfo {volume} {98}},\ \bibinfo
  {pages} {075127} (\bibinfo {year} {2018})}\BibitemShut {NoStop}%
\bibitem [{\citenamefont {Kaltak}\ and\ \citenamefont
  {Kresse}(2019)}]{kaltak2019minimax}%
  \BibitemOpen
  \bibfield  {author} {\bibinfo {author} {\bibfnamefont {M.}~\bibnamefont
  {Kaltak}}\ and\ \bibinfo {author} {\bibfnamefont {G.}~\bibnamefont
  {Kresse}},\ }\href@noop {} {\bibinfo {title} {Minimax isometry method}}
  (\bibinfo {year} {2019}),\ \Eprint {https://arxiv.org/abs/1909.01740}
  {arXiv:1909.01740 [cond-mat.mtrl-sci]} \BibitemShut {NoStop}%
\bibitem [{\citenamefont {Jarrell}\ and\ \citenamefont
  {Gubernatis}(1996)}]{Jarrell1996}%
  \BibitemOpen
  \bibfield  {author} {\bibinfo {author} {\bibfnamefont {M.}~\bibnamefont
  {Jarrell}}\ and\ \bibinfo {author} {\bibfnamefont {J.}~\bibnamefont
  {Gubernatis}},\ }\href {https://doi.org/10.1016/0370-1573(95)00074-7}
  {\bibfield  {journal} {\bibinfo  {journal} {Physics Reports}\ }\textbf
  {\bibinfo {volume} {269}},\ \bibinfo {pages} {133} (\bibinfo {year}
  {1996})}\BibitemShut {NoStop}%
\bibitem [{\citenamefont {Iskakov}\ and\ \citenamefont
  {Danilov}(2018)}]{ISKAKOV2018128}%
  \BibitemOpen
  \bibfield  {author} {\bibinfo {author} {\bibfnamefont {S.}~\bibnamefont
  {Iskakov}}\ and\ \bibinfo {author} {\bibfnamefont {M.}~\bibnamefont
  {Danilov}},\ }\href
  {https://doi.org/https://doi.org/10.1016/j.cpc.2017.12.016} {\bibfield
  {journal} {\bibinfo  {journal} {Computer Physics Communications}\ }\textbf
  {\bibinfo {volume} {225}},\ \bibinfo {pages} {128 } (\bibinfo {year}
  {2018})}\BibitemShut {NoStop}%
\bibitem [{\citenamefont {Lowdin}(1970)}]{LOWDIN1970185}%
  \BibitemOpen
  \bibfield  {author} {\bibinfo {author} {\bibfnamefont {P.-O.}\ \bibnamefont
  {Lowdin}}\ }(\bibinfo  {publisher} {Academic Press},\ \bibinfo {year}
  {1970})\ pp.\ \bibinfo {pages} {185 -- 199}\BibitemShut {NoStop}%
\bibitem [{\citenamefont {Gaenko}\ \emph {et~al.}(2017)\citenamefont {Gaenko},
  \citenamefont {Antipov}, \citenamefont {Carcassi}, \citenamefont {Chen},
  \citenamefont {Chen}, \citenamefont {Dong}, \citenamefont {Gamper},
  \citenamefont {Gukelberger}, \citenamefont {Igarashi}, \citenamefont
  {Iskakov}, \citenamefont {K\"{o}nz}, \citenamefont {LeBlanc}, \citenamefont
  {Levy}, \citenamefont {Ma}, \citenamefont {Paki}, \citenamefont {Shinaoka},
  \citenamefont {Todo}, \citenamefont {Troyer},\ and\ \citenamefont
  {Gull}}]{Gaenko2017}%
  \BibitemOpen
  \bibfield  {author} {\bibinfo {author} {\bibfnamefont {A.}~\bibnamefont
  {Gaenko}}, \bibinfo {author} {\bibfnamefont {A.}~\bibnamefont {Antipov}},
  \bibinfo {author} {\bibfnamefont {G.}~\bibnamefont {Carcassi}}, \bibinfo
  {author} {\bibfnamefont {T.}~\bibnamefont {Chen}}, \bibinfo {author}
  {\bibfnamefont {X.}~\bibnamefont {Chen}}, \bibinfo {author} {\bibfnamefont
  {Q.}~\bibnamefont {Dong}}, \bibinfo {author} {\bibfnamefont {L.}~\bibnamefont
  {Gamper}}, \bibinfo {author} {\bibfnamefont {J.}~\bibnamefont {Gukelberger}},
  \bibinfo {author} {\bibfnamefont {R.}~\bibnamefont {Igarashi}}, \bibinfo
  {author} {\bibfnamefont {S.}~\bibnamefont {Iskakov}}, \bibinfo {author}
  {\bibfnamefont {M.}~\bibnamefont {K\"{o}nz}}, \bibinfo {author}
  {\bibfnamefont {J.}~\bibnamefont {LeBlanc}}, \bibinfo {author} {\bibfnamefont
  {R.}~\bibnamefont {Levy}}, \bibinfo {author} {\bibfnamefont {P.}~\bibnamefont
  {Ma}}, \bibinfo {author} {\bibfnamefont {J.}~\bibnamefont {Paki}}, \bibinfo
  {author} {\bibfnamefont {H.}~\bibnamefont {Shinaoka}}, \bibinfo {author}
  {\bibfnamefont {S.}~\bibnamefont {Todo}}, \bibinfo {author} {\bibfnamefont
  {M.}~\bibnamefont {Troyer}},\ and\ \bibinfo {author} {\bibfnamefont
  {E.}~\bibnamefont {Gull}},\ }\href
  {https://doi.org/10.1016/j.cpc.2016.12.009} {\bibfield  {journal} {\bibinfo
  {journal} {Computer Physics Communications}\ }\textbf {\bibinfo {volume}
  {213}},\ \bibinfo {pages} {235} (\bibinfo {year} {2017})}\BibitemShut
  {NoStop}%
\bibitem [{\citenamefont {Levy}\ \emph {et~al.}(2017)\citenamefont {Levy},
  \citenamefont {LeBlanc},\ and\ \citenamefont {Gull}}]{Levy2017}%
  \BibitemOpen
  \bibfield  {author} {\bibinfo {author} {\bibfnamefont {R.}~\bibnamefont
  {Levy}}, \bibinfo {author} {\bibfnamefont {J.}~\bibnamefont {LeBlanc}},\ and\
  \bibinfo {author} {\bibfnamefont {E.}~\bibnamefont {Gull}},\ }\href
  {https://doi.org/10.1016/j.cpc.2017.01.018} {\bibfield  {journal} {\bibinfo
  {journal} {Computer Physics Communications}\ }\textbf {\bibinfo {volume}
  {215}},\ \bibinfo {pages} {149} (\bibinfo {year} {2017})}\BibitemShut
  {NoStop}%
\bibitem [{\citenamefont {Mishchenko}\ \emph {et~al.}(2000)\citenamefont
  {Mishchenko}, \citenamefont {Prokof'ev}, \citenamefont {Sakamoto},\ and\
  \citenamefont {Svistunov}}]{Mishchenko2000}%
  \BibitemOpen
  \bibfield  {author} {\bibinfo {author} {\bibfnamefont {A.~S.}\ \bibnamefont
  {Mishchenko}}, \bibinfo {author} {\bibfnamefont {N.~V.}\ \bibnamefont
  {Prokof'ev}}, \bibinfo {author} {\bibfnamefont {A.}~\bibnamefont
  {Sakamoto}},\ and\ \bibinfo {author} {\bibfnamefont {B.~V.}\ \bibnamefont
  {Svistunov}},\ }\href {https://doi.org/10.1103/PhysRevB.62.6317} {\bibfield
  {journal} {\bibinfo  {journal} {Phys. Rev. B}\ }\textbf {\bibinfo {volume}
  {62}},\ \bibinfo {pages} {6317} (\bibinfo {year} {2000})}\BibitemShut
  {NoStop}%
\bibitem [{\citenamefont {Otsuki}\ \emph {et~al.}(2017)\citenamefont {Otsuki},
  \citenamefont {Ohzeki}, \citenamefont {Shinaoka},\ and\ \citenamefont
  {Yoshimi}}]{Otsuki2017}%
  \BibitemOpen
  \bibfield  {author} {\bibinfo {author} {\bibfnamefont {J.}~\bibnamefont
  {Otsuki}}, \bibinfo {author} {\bibfnamefont {M.}~\bibnamefont {Ohzeki}},
  \bibinfo {author} {\bibfnamefont {H.}~\bibnamefont {Shinaoka}},\ and\
  \bibinfo {author} {\bibfnamefont {K.}~\bibnamefont {Yoshimi}},\ }\href
  {https://doi.org/10.1103/PhysRevE.95.061302} {\bibfield  {journal} {\bibinfo
  {journal} {Phys. Rev. E}\ }\textbf {\bibinfo {volume} {95}},\ \bibinfo
  {pages} {061302(R)} (\bibinfo {year} {2017})}\BibitemShut {NoStop}%
\bibitem [{\citenamefont {Stanescu}\ and\ \citenamefont
  {Kotliar}(2006)}]{Stanescu2006}%
  \BibitemOpen
  \bibfield  {author} {\bibinfo {author} {\bibfnamefont {T.~D.}\ \bibnamefont
  {Stanescu}}\ and\ \bibinfo {author} {\bibfnamefont {G.}~\bibnamefont
  {Kotliar}},\ }\href {https://doi.org/10.1103/PhysRevB.74.125110} {\bibfield
  {journal} {\bibinfo  {journal} {Phys. Rev. B}\ }\textbf {\bibinfo {volume}
  {74}},\ \bibinfo {pages} {125110} (\bibinfo {year} {2006})}\BibitemShut
  {NoStop}%
\bibitem [{\citenamefont {Wang}\ \emph {et~al.}(2009)\citenamefont {Wang},
  \citenamefont {Gull}, \citenamefont {de' Medici}, \citenamefont {Capone},\
  and\ \citenamefont {Millis}}]{Wang2009}%
  \BibitemOpen
  \bibfield  {author} {\bibinfo {author} {\bibfnamefont {X.}~\bibnamefont
  {Wang}}, \bibinfo {author} {\bibfnamefont {E.}~\bibnamefont {Gull}}, \bibinfo
  {author} {\bibfnamefont {L.}~\bibnamefont {de' Medici}}, \bibinfo {author}
  {\bibfnamefont {M.}~\bibnamefont {Capone}},\ and\ \bibinfo {author}
  {\bibfnamefont {A.~J.}\ \bibnamefont {Millis}},\ }\href
  {https://doi.org/10.1103/PhysRevB.80.045101} {\bibfield  {journal} {\bibinfo
  {journal} {Phys. Rev. B}\ }\textbf {\bibinfo {volume} {80}},\ \bibinfo
  {pages} {045101} (\bibinfo {year} {2009})}\BibitemShut {NoStop}%
\bibitem [{\citenamefont {Gull}\ \emph {et~al.}(2011)\citenamefont {Gull},
  \citenamefont {Millis}, \citenamefont {Lichtenstein}, \citenamefont
  {Rubtsov}, \citenamefont {Troyer},\ and\ \citenamefont {Werner}}]{Gull2011}%
  \BibitemOpen
  \bibfield  {author} {\bibinfo {author} {\bibfnamefont {E.}~\bibnamefont
  {Gull}}, \bibinfo {author} {\bibfnamefont {A.~J.}\ \bibnamefont {Millis}},
  \bibinfo {author} {\bibfnamefont {A.~I.}\ \bibnamefont {Lichtenstein}},
  \bibinfo {author} {\bibfnamefont {A.~N.}\ \bibnamefont {Rubtsov}}, \bibinfo
  {author} {\bibfnamefont {M.}~\bibnamefont {Troyer}},\ and\ \bibinfo {author}
  {\bibfnamefont {P.}~\bibnamefont {Werner}},\ }\href
  {https://doi.org/10.1103/revmodphys.83.349} {\bibfield  {journal} {\bibinfo
  {journal} {Reviews of Modern Physics}\ }\textbf {\bibinfo {volume} {83}},\
  \bibinfo {pages} {349} (\bibinfo {year} {2011})}\BibitemShut {NoStop}%
\bibitem [{\citenamefont {Bulla}\ \emph {et~al.}(2008)\citenamefont {Bulla},
  \citenamefont {Costi},\ and\ \citenamefont {Pruschke}}]{Bulla2008}%
  \BibitemOpen
  \bibfield  {author} {\bibinfo {author} {\bibfnamefont {R.}~\bibnamefont
  {Bulla}}, \bibinfo {author} {\bibfnamefont {T.~A.}\ \bibnamefont {Costi}},\
  and\ \bibinfo {author} {\bibfnamefont {T.}~\bibnamefont {Pruschke}},\ }\href
  {https://doi.org/10.1103/revmodphys.80.395} {\bibfield  {journal} {\bibinfo
  {journal} {Reviews of Modern Physics}\ }\textbf {\bibinfo {volume} {80}},\
  \bibinfo {pages} {395} (\bibinfo {year} {2008})}\BibitemShut {NoStop}%
\bibitem [{\citenamefont {Lin}\ \emph {et~al.}(1993)\citenamefont {Lin},
  \citenamefont {Gubernatis}, \citenamefont {Gould},\ and\ \citenamefont
  {Tobochnik}}]{doi:10.1063/1.4823192}%
  \BibitemOpen
  \bibfield  {author} {\bibinfo {author} {\bibfnamefont {H.}~\bibnamefont
  {Lin}}, \bibinfo {author} {\bibfnamefont {J.}~\bibnamefont {Gubernatis}},
  \bibinfo {author} {\bibfnamefont {H.}~\bibnamefont {Gould}},\ and\ \bibinfo
  {author} {\bibfnamefont {J.}~\bibnamefont {Tobochnik}},\ }\href
  {https://doi.org/10.1063/1.4823192} {\bibfield  {journal} {\bibinfo
  {journal} {Computers in Physics}\ }\textbf {\bibinfo {volume} {7}},\ \bibinfo
  {pages} {400} (\bibinfo {year} {1993})}\BibitemShut {NoStop}%
\bibitem [{\citenamefont {Zgid}\ and\ \citenamefont {Chan}(2011)}]{Zgid_2011}%
  \BibitemOpen
  \bibfield  {author} {\bibinfo {author} {\bibfnamefont {D.}~\bibnamefont
  {Zgid}}\ and\ \bibinfo {author} {\bibfnamefont {G.~K.-L.}\ \bibnamefont
  {Chan}},\ }\href {https://doi.org/10.1063/1.3556707} {\bibfield  {journal}
  {\bibinfo  {journal} {The Journal of Chemical Physics}\ }\textbf {\bibinfo
  {volume} {134}},\ \bibinfo {pages} {094115} (\bibinfo {year}
  {2011})}\BibitemShut {NoStop}%
\bibitem [{\citenamefont {Zgid}\ \emph {et~al.}(2012)\citenamefont {Zgid},
  \citenamefont {Gull},\ and\ \citenamefont {Chan}}]{Zgid2012}%
  \BibitemOpen
  \bibfield  {author} {\bibinfo {author} {\bibfnamefont {D.}~\bibnamefont
  {Zgid}}, \bibinfo {author} {\bibfnamefont {E.}~\bibnamefont {Gull}},\ and\
  \bibinfo {author} {\bibfnamefont {G.~K.-L.}\ \bibnamefont {Chan}},\ }\href
  {https://doi.org/10.1103/PhysRevB.86.165128} {\bibfield  {journal} {\bibinfo
  {journal} {Phys. Rev. B}\ }\textbf {\bibinfo {volume} {86}},\ \bibinfo
  {pages} {165128} (\bibinfo {year} {2012})}\BibitemShut {NoStop}%
\bibitem [{\citenamefont {Shee}\ and\ \citenamefont {Zgid}(2019)}]{Shee2019}%
  \BibitemOpen
  \bibfield  {author} {\bibinfo {author} {\bibfnamefont {A.}~\bibnamefont
  {Shee}}\ and\ \bibinfo {author} {\bibfnamefont {D.}~\bibnamefont {Zgid}},\
  }\href {https://doi.org/10.1021/acs.jctc.9b00603} {\bibfield  {journal}
  {\bibinfo  {journal} {Journal of Chemical Theory and Computation}\ }\textbf
  {\bibinfo {volume} {15}},\ \bibinfo {pages} {6010} (\bibinfo {year}
  {2019})}\BibitemShut {NoStop}%
\bibitem [{\citenamefont {Zhu}\ \emph {et~al.}(2019)\citenamefont {Zhu},
  \citenamefont {Jim\'enez-Hoyos}, \citenamefont {McClain}, \citenamefont
  {Berkelbach},\ and\ \citenamefont {Chan}}]{Zhu2019}%
  \BibitemOpen
  \bibfield  {author} {\bibinfo {author} {\bibfnamefont {T.}~\bibnamefont
  {Zhu}}, \bibinfo {author} {\bibfnamefont {C.~A.}\ \bibnamefont
  {Jim\'enez-Hoyos}}, \bibinfo {author} {\bibfnamefont {J.}~\bibnamefont
  {McClain}}, \bibinfo {author} {\bibfnamefont {T.~C.}\ \bibnamefont
  {Berkelbach}},\ and\ \bibinfo {author} {\bibfnamefont {G.~K.-L.}\
  \bibnamefont {Chan}},\ }\href {https://doi.org/10.1103/PhysRevB.100.115154}
  {\bibfield  {journal} {\bibinfo  {journal} {Phys. Rev. B}\ }\textbf {\bibinfo
  {volume} {100}},\ \bibinfo {pages} {115154} (\bibinfo {year}
  {2019})}\BibitemShut {NoStop}%
\bibitem [{\citenamefont {Voglis}\ and\ \citenamefont
  {Lagaris}(2004)}]{Voglis_arectangular}%
  \BibitemOpen
  \bibfield  {author} {\bibinfo {author} {\bibfnamefont {C.}~\bibnamefont
  {Voglis}}\ and\ \bibinfo {author} {\bibfnamefont {I.~E.}\ \bibnamefont
  {Lagaris}},\ }in\ \href@noop {} {\emph {\bibinfo {booktitle} {WSEAS
  International Conference on Applied Mathematics}}}\ (\bibinfo {year}
  {2004})\BibitemShut {NoStop}%
\bibitem [{\citenamefont {Branch}\ \emph {et~al.}(1999)\citenamefont {Branch},
  \citenamefont {Coleman},\ and\ \citenamefont {Li}}]{Branch1999}%
  \BibitemOpen
  \bibfield  {author} {\bibinfo {author} {\bibfnamefont {M.~A.}\ \bibnamefont
  {Branch}}, \bibinfo {author} {\bibfnamefont {T.~F.}\ \bibnamefont
  {Coleman}},\ and\ \bibinfo {author} {\bibfnamefont {Y.}~\bibnamefont {Li}},\
  }\href {https://doi.org/10.1137/s1064827595289108} {\bibfield  {journal}
  {\bibinfo  {journal} {{SIAM} Journal on Scientific Computing}\ }\textbf
  {\bibinfo {volume} {21}},\ \bibinfo {pages} {1} (\bibinfo {year}
  {1999})}\BibitemShut {NoStop}%
\bibitem [{\citenamefont {Mulliken}(1955)}]{Mulliken1955}%
  \BibitemOpen
  \bibfield  {author} {\bibinfo {author} {\bibfnamefont {R.~S.}\ \bibnamefont
  {Mulliken}},\ }\href {https://doi.org/10.1063/1.1740588} {\bibfield
  {journal} {\bibinfo  {journal} {The Journal of Chemical Physics}\ }\textbf
  {\bibinfo {volume} {23}},\ \bibinfo {pages} {1833} (\bibinfo {year}
  {1955})}\BibitemShut {NoStop}%
\bibitem [{\citenamefont {Fender}\ \emph {et~al.}(1968)\citenamefont {Fender},
  \citenamefont {Jacobson},\ and\ \citenamefont
  {Wedgwood}}]{doi:10.1063/1.1668855}%
  \BibitemOpen
  \bibfield  {author} {\bibinfo {author} {\bibfnamefont {B.~E.~F.}\
  \bibnamefont {Fender}}, \bibinfo {author} {\bibfnamefont {A.~J.}\
  \bibnamefont {Jacobson}},\ and\ \bibinfo {author} {\bibfnamefont {F.~A.}\
  \bibnamefont {Wedgwood}},\ }\href {https://doi.org/10.1063/1.1668855}
  {\bibfield  {journal} {\bibinfo  {journal} {The Journal of Chemical Physics}\
  }\textbf {\bibinfo {volume} {48}},\ \bibinfo {pages} {990} (\bibinfo {year}
  {1968})}\BibitemShut {NoStop}%
\bibitem [{\citenamefont {Cheetham}\ and\ \citenamefont
  {Hope}(1983)}]{mu_expt_Cheetham83}%
  \BibitemOpen
  \bibfield  {author} {\bibinfo {author} {\bibfnamefont {A.~K.}\ \bibnamefont
  {Cheetham}}\ and\ \bibinfo {author} {\bibfnamefont {D.~A.~O.}\ \bibnamefont
  {Hope}},\ }\href {https://doi.org/10.1103/PhysRevB.27.6964} {\bibfield
  {journal} {\bibinfo  {journal} {Phys. Rev. B}\ }\textbf {\bibinfo {volume}
  {27}},\ \bibinfo {pages} {6964} (\bibinfo {year} {1983})}\BibitemShut
  {NoStop}%
\end{thebibliography}%

\end{document}